\documentclass[nofootinbib,superscriptaddress,twocolumn,aps,showpacs, floatfix]{revtex4-1}
\usepackage{epsfig}
\usepackage{graphicx}
\usepackage{dcolumn}
\usepackage{bm}
\usepackage{amssymb}
\usepackage{amsthm,amsmath}
\usepackage{subfigure}
\usepackage{float}
\usepackage[utf8]{inputenc}
\usepackage{booktabs}
\usepackage{natbib}
\usepackage{comment}
\usepackage{hyperref}
\hypersetup{colorlinks=true, citecolor=blue, urlcolor=blue, linkcolor=blue}
\usepackage[dvipsnames, table, xcdraw]{xcolor}
\usepackage[referable]{threeparttablex}

\begin{document}

\title {Practices of public procurement and the risk of corrupt behavior before and after the government transition in M\'exico}

\author{Andrea Falc\'on-Cort\'es}
\email{falcon@icf.unam.mx}
\affiliation{Instituto de Ciencias F\'isicas, Universidad Nacional Aut\'onoma de M\'exico, Morelos 62210, M\'exico}

\author{Andrés Aldana}
\email{andres.aldana@c3.unam.mx}
\affiliation{Instituto de Biotecnolog\'ia, Universidad Nacional Aut\'onoma de M\'exico, Morelos 62210, M\'exico}
\affiliation{Centro de Ciencias de la Complejidad, Universidad Nacional Aut\'onoma de M\'exico, Ciudad de M\'exico 04510, M\'exico}

\author{Hernán Larralde}
\email{hernan@icf.unam.mx}
\affiliation{Instituto de Ciencias F\'isicas, Universidad Nacional Aut\'onoma de M\'exico, Morelos 62210, M\'exico}

\date{\today}

\begin{abstract}
    {\it Abstract:} Corruption has a significant impact on economic growth, democracy, and inequality. It has sever consequences at the human level. 
    Public procurement, where public resources are used to purchase goods or services from the private sector, are particularly susceptible to corrupt practices. 
    However, government turnover may bring significant changes in the way public contracting is done, and thus, in the levels and types of corruption involved in public procurement.
    In this respect, M\'exico lived a historical government transition in 2018, with the new government promising a crackdown on corruption.
    In this work, we analyze data from more than 1.5 million contracts corresponding from 2013 to 2020, to study to what extent this change of government affected the characteristics of public contracting, and we try to determine whether these changes affect how corruption takes place. 
    To do this, we propose a statistical framework to compare the characteristics of the contracting practices within each administration, separating the contracts in different classes depending on whether or not they were made with companies that have now been identified as being involved in corrupt practices. 
    We find that while the amount of resources spent with companies that turned out to be corrupt has decreased substantially, many of the patterns followed to contract these companies were maintained, and some of those in which changes did occur, are suggestive of a larger risk of corruption.\\
    
\end{abstract}

\maketitle

\section{Introduction}

Because of its impact on economic growth \cite{mauro1995corruption, hessami2014political}, democracy \cite{stockemer2013bribes}, and inequality \cite{gupta2002does}, one of the biggest challenges that a government has to deal with is corruption. Transparency International defines corruption as the abuse of public power for private benefit \cite{transparencycorruption}, {\it i.e.} it assumes that corruption involves the participation of public officials. Considering this definition, a niche in which corruption can arise naturally is in public procurement; where public and private sectors interact through contracting, mostly to purchase goods or services.
Corruption at the level of public-private contracting has high costs in many areas. For example, if buyers favor some suppliers over others through corrupt decisions, bribes or patronage (clientelism), then market competition is affected \cite{north2009violence, aidt2016rent}. This lack of competition leads to a missallocation of resources, affecting areas such as budget composition \cite{mauro1995corruption}, military and technological spending \cite{gupta2002does},  social care \cite{aidt2016rent}, and may even change the market structure and dynamics \cite{wachs2019social,fazekas2020corruption}. However, owing to the complexity of the contracts, the large sums of money involved, the number of participants, as well as the inherent complicity of public officials, this kind of corruption is difficult to identify, track, and prevent \cite{baldi2016bid, oecd2007integrity, oecd2015government}. 

A government turnover, {\it i.e.} a change in the individuals and/or parties in power, due to elections or otherwise, may bring significant changes in the way public contracting is carried out, and thus, in the types of corrupt behavior that may occur in the context of public procurement \cite{clingermayer2003governmental}. Broms {\it et al.} propose that frequent elections with uncertain outcomes may compel corrupt elites to pursue predatory strategies; however, if there is a well-established party system, regular electoral uncertainty may motivate corrupt elites to exercise restraint \cite{broms2019political}. In this context,  Fazekas {\it et al.} found that fair electoral contest and heterogeneous power-sharing may have the potential to mitigate corrupt market distortions, even in systematically corrupt places \cite{fazekas2020corruption}. On the other hand, while it seems to be true that government turnovers can diminish corruption in public procurement \cite{fazekas2020corruption, broms2019political}, there is also evidence that a change in government can maintain corrupt behaviors, only changing the favored suppliers \cite{david2019grand}. An example of this was found by the Mexican Institute for Competitiveness (IMCO by its Spanish acronym, \cite{IMCO}), which analysed public procurement data from M\'exico, finding that in the change of government that took place between 2012 and 2013, the favored suppliers also changed, but not the amount of resources and contracts given to the new suppliers. This phenomena was called \lq\lq the compadres' change'' \cite{IMCOcompadres}.

For years M\'exico has occupied low levels in the score of corruption perception index given by Transparency International \cite{TICorruption, maria2015anatomia}. Mexican citizens consider that corruption is one of the biggest problems in the country, only behind violence and insecurity \cite{maria2015anatomia}. For years, government and corporate structures have been created, maintained, and adapted to obtain private benefits from public resources, which has carried huge consequences to the economic growth and human well-being in the country \cite{meyer2019poder}. 

In 2018, M\'exico lived the largest electoral contest of its history, with more than 56.6 million voters, representing 62.62\% of citizen participation \cite{INE}. In this election, a \lq\lq leftist'' candidate won the presidency for the first time. He was the most voted winner in Mexico's history, obtaining 30.1 million votes (representing 53.1\% of the voters), and his party won 388 congress positions of the 554 available (70\% of the seats) \cite{INE}. This event not only marked a government turnover, but an ideological and structural change in the government's goals and methods \cite{meyer2019poder, munoztriunfo, hanrahan2019reflections}. In this paper, we study how this government change affected the characteristics of public contracting, and the extent to which these changes affect the forms in which corruption takes place. To do so, we analyzed data from more than 1.5 million contracts between many agencies that conform the Mexican government and private suppliers, corresponding to the period 2013-2020, which includes the first two years of the new government (2019-2020) \cite{compranet}.

Many states have taken advantage of the new technology era to record their administrative activities. This data, which can be used to analyze the projects undertaken by an administration, their success and shortfalls, mainly describes the state's procurement practices \cite{kim2014big}.  This large amount of administrative data collected by governments allows studying corruption from a new data driven perspective \cite{radermacher2018official}. A common approach to quantify corruption is by building risk factors from contract data \cite{wachs2019social, fazekas2020corruption, david2019grand, fazekas2013anatomy, fazekas2013corruption, fazekas2016objective, fazekas2017corruption, fazekas2018institutional,wachs2020corruption}. For example, single-bidder contracts have been shown to be effective in identifying and predicting corruption risk \cite{wachs2019social, wachs2020corruption, klavsnja2015corruption, charron2017careers} in various contexts; including studies on the relationship between corruption and political incumbency \cite{klavsnja2015corruption}, and the effect of campaign contributions on corruption \cite{charron2017careers}. Other approaches include the application of network theory to measure the impact of single-bidder contracts on local corruption \cite{wachs2019social}. Network theory was also used to identify corruption risk distribution among countries. For those studies, the actors (buyers and suppliers) were represented by network nodes; these nodes were linked if they entered into the same public procurement contract, and the weight of the link was the fraction of single-bidder contracts between each pair \cite{wachs2020corruption}.

Here we propose a somewhat different approach, taking advantage of specific public data that lists suppliers that were investigated and have been identified as having incurred in corrupt practices. Since 2013 M\'exico's government collects a list of companies that have been caught providing invoices for simulated operations (or EFOS for its spanish acronym \lq\lq Empresas que Facturan Operaciones Simuladas'') \cite{EFOS}. These companies sell fake receipts to buyers who use them to avoid taxes or to cover acts of embezzlement, among other things. There is also, since 2013, a list of all the suppliers that have been caught doing one or several types of corrupt activities when participating in a public contract. For example, presenting fake documentation in order to win a contract, overcharging their services, breaching contract, or diverting resources. These companies are labeled as \lq\lq sanctioned suppliers and contractors'' (or PCS by their Spanish acronym \lq\lq Proveedores y Contratistas Sancionados'') \cite{PCS}.

It should be noted that to appear in these lists, these contractors were subject to a legal investigation, and some of the contractors are appealing the decision. Thus, these lists may change slightly as time goes by as a result of companies winning their appeals, which removes them from the lists, as well as due to conclusions of long lasting investigations, which may add new companies.  Nevertheless, from the available records, we identified the contracts in which companies, which have been suspect of corrupt activities, participated.

We use this data to achieve the following goals:
\begin{enumerate}
    \item [i)] Our first goal is to develop a statistical framework to analyze whether there were changes in public contracting practices as a consequence of the government turnover in M\'exico, both in those contracts suspect of corruption (those in which EFOS or PCS companies participated) and those in which companies free of official corruption charges participated. To achieve this goal, each contract has to be described by a set of quantitative variables based on the available data.
    \item [ii)] The second goal is to use the data to build risk factors following the framework proposed by Fazekas' group, and to test whether these indicators successfully describe and identify those contracts in which companies suspect of corrupt activities participated. We then use these indicators to determine whether the change in the procurement practices due to the government turnover brought a lower or higher risk of corruption, where {\it risk of corruption}, in this work, is to be understood as the fraction of potentially corrupt contracts.
\end{enumerate}

The results in this paper are based solely on the study of the data, trying to avoid any political bias, and without use of any prior knowledge about the contracts' participants. 

\section{Methods}

\subsection{Data}
The list of public contracts from 2013 to 2020 was taken from the electronic Mexican system of public governmental information on public procurement {\it CompraNet} \cite{compranet}. This system is operated by an administrative unit designed by the Mexican federal agency for budget and public debt (SHCP for \lq\lq Secretar\'ia de Hacienda y Cr\'edito P\'ublico''). The registry of contracts in this electronic system is mandatory for all those that operate with resources from the federal budget \footnote{These include federal, state, and municipal agencies. It should also be mentioned that we have no way of knowing how complete the list is, nor whether omissions are more frequent regarding contracts in one government level or another, as well as the possibility of omissions regarding "sensitive" contracts, as could be military spending. Finally, we remark that the list only includes contracts between government agencies and {\it private} suppliers. Government agencies that act as suppliers to other government agencies are exempt from reporting in {\it CompraNet}.}. The contracts on this list have a specific set of variables or entries that describe them. The particular entries we use in this work are shown in Table \ref{datoscontratos}.

The original data lists consisted in 1.6 M of contracts. These records were curated to standardize the information available in each of them. We homogenized all the string variables to avoid issues with special characters or spacing \footnote{By law all registries in {\it CompraNet} should be made using the fiscal name of both the agency and company. This ensures that the data contains unique identifiers of buyers and suppliers.}, we also deleted all the entries in which an important variable was omitted (for example, the buyer name, the amount spent, the type of procedure, etc. ). The contracts with this kind of problems were approximately 60 K (representing $\sim$4\% of the data). Thus, from the 1.6 M contracts available in the original data lists, we consider 1.54 M. As mentioned above, from these 1.54 M of records, we took only the variables shown in Table \ref{datoscontratos}. The original set includes other variables that are not of interest for this work, such as the name of the buyers' legal representative, or the suppliers' webpage. Hereinafter, we will refer to this curated list of 1.54 M of contracts as the {\it source list}.

The list of companies identified as EFOS is available on the site of the Mexican tax agency (SAT for \lq\lq Secretar\'ia de Administraci\'on Tributaria'') \cite{EFOS} \footnote{For this study we use the lists of both definitive and presumed EFOS}, and the list of PCS contractors is available on the official Mexican open data site \cite{PCS}. The only variable of interest in these lists is the supplier's name, which we use to identify the contracts in which these companies participated.

All data sets mentioned in this section, and other tools to reproduce the results shown here are available on \cite{zenodo}.

\begin{table*}[htb!]
\begin{threeparttable}
\begin{tabular}{llll}
\hline
\hline
\textbf{Type} &
\textbf{Variable} &
  \textbf{Short name} &
  \textbf{Detail} \\ \hline
  
i) &
Buyer &
  N/A &
  {\bf Governments agency that made the contract} \\ 
  &&&{\bf  with a private company (the supplier)} \\
  \hline
  &
Supplier &
  N/A &
  {\bf Private company that was awarded a contract} \\ &&&  {\bf by a government agency (the buyer)} \tnotex{tn:a} \\
  \hline
&
Government Order &
  GO &
  \begin{tabular}[c]{@{}l@{}}{\bf Government agency (buyer) level}\\ Dummy variables: \\ GO. APF -  Federal level \\ GO.GE - State level\\ GO.GM - Municipal level \\ (This variable provides information regarding at which level\\ of government the contract was assigned, as different orders\\ have different operation rules regarding\\ duration, thresholds, etc; and different levels of scrutiny) \end{tabular}   \\
  \hline
  &
Procedure Character &
  PC &
  \begin{tabular}[c]{@{}l@{}} {\bf Legal framework in which the contract was made.}\\ Dummy variables: \\PC.N - National\\ PC.I - International\\ PC.ITLC - International under the North American Free Trade Agreement\\ (NAFTA) \\ (This variable provides information of the legal framework under which\\ the contract was assigned) \end{tabular} \\
  \hline
&
Contract Type &
  CT &
  \begin{tabular}[c]{@{}l@{}}{\bf Services or commodities contracted} \\ Dummy variables: \\ CT.OP - Public work \\ CT.S - Services\\ CT-ADQ - Acquisitions\\ CT-AR - Leases\\ CT-SLAOP - Special public work\\ (This variable provides a rough description of the market sector\\ of the contract, as again,  different sectors might, in principle,\\ have different regulations) \end{tabular} \\
  \hline
  &
Procedure Type &
  PT &
  \begin{tabular}[c]{@{}l@{}} {\bf Procedure by which the supplier won the contract}\\ Dummy variables: \\PT.AD - Single-bidder\\ PT.LP - Open public contest\\ PT.I3P - Contest between only three suppliers \\ (Describes how the contract was assigned. We recall that,\\ in principle, single-bidder contracts can be carried \\out only under \lq\lq special" circumstances)  \end{tabular} \\
  \hline
&
Size &
  S &
\begin{tabular}[c]{@{}l@{}}{\bf Size of supplier} \tnotex{tn:b}\\ Dummy variables: \\ S.MIC - Micro-supplier\\ S.PEQ - Small-supplier\\ S.MED - Medium-supplier\\ S.NOM - Large-supplier\\ S.NA - Supplier without assigned size\\ (This variable is a rough indicator of the size of the supplier,\\  which can serve as a proxy  of the capacity of the supplier\\ to carry out the project, etc) \end{tabular} \\
  \hline
  &
Year &
  N/A &
  {\bf Year in which the contract began} \\
  \hline
&
Beginning week &
  N/A & \begin{tabular}[c]{@{}l@{}}
  {\bf Week of the year in which the contract began} \\ (This variable intends to gauge whether the activity\\ is correlated to the budget calendar) \end{tabular}\\
  \hline
&
Ending-Beginning weeks &
  EBWeeks &
  {\bf Weeks that the contract lasted} \\
  \hline
  &
Spending &
  N/A &
  {\bf Amount of money spent by the buyer (in USD PPP)} \tnotex{tn:c} \\ \hline \hline
\end{tabular}
\begin{tablenotes}
\item [a] \label{tn:a} As mentioned previously, the state-owned companies are exempt to report their activities as suppliers to other government agencies \cite{leyarrendamiento, leyobra}.
\item[b] \label{tn:b} The Mexican Economic Secretariat proposed this size classification based on the amount of resources produced by the company and its number of employees \cite{EstratSE}.
\item[c] \label{tn:c} Since most of the Spending is reported in mexican currency (MXN), we converted the amounts to USD PPP using the equivalences given in \cite{oecdppp}.
\end{tablenotes}
\end{threeparttable}
\caption{Set of the variables chosen to describe the features of each contract from the source list of public contracts \cite{compranet}. In parenthesis, a short explanation of the variables that are not self explanatory is given, providing the motivation for considering each particular variable.}
\label{datoscontratos}
\end{table*}

\subsubsection{Contract classes}

Our first step was to identify those public procurement contracts won by companies labeled as having been involved in corrupt activities. Since these  companies have gone through a process to be labeled as either an EFOS or a PCS, we assume that they may be suspect of having incurred in corrupt behavior
in all the contracts they participated in, independently of whether the contracts occurred before or after the company was labeled. Thus, all the contracts in which the supplier is an EFOS or a PCS, are classified as possibly corrupt and we assign the corresponding label EFOS or PCS to them.

Following the principle of presumption of innocence, we label as NC (for Non-Corrupt) all the other contracts in the source list in which the supplier is free of official 
corruption charges. Thus, we have three classes of contracts labeled EFOS, PCS and NC,
respectively. Of course we are aware that it is very likely that corrupt contracts went undetected and end up in our NC class, however we expect that they will have little statistical weight in this class that represents the vast majority of the contracts.

\begin{table*}[htbp!]
\begin{tabular}{llll}
\hline \hline
\textbf{Type} &
  \textbf{Variable} &
  \textbf{Short name} &
  \textbf{Detail} \\ \hline
ii) &
  \begin{tabular}[c]{@{}l@{}}
  Maximum number of Contracts by a buyer
  \end{tabular}
  &
  T.Cont.Max &
  \begin{tabular}[c]{@{}l@{}}{\bf Maximum number of the contracts awarded by} \\ {\bf a buyer to a supplier}\\ $\max_j{(T.Cont_j)}$ where $j$ represents all the suppliers contracted\\ by a buyer,and $T.Cont$ the number of total contracts\\ given to each supplier \end{tabular} \\
  \hline
 &
  Maximum Spending by a buyer &
  T.Spending.Max &
  \begin{tabular}[c]{@{}l@{}}{\bf Maximum amount of money spent by a buyer}\\ {\bf in contracts with a supplier}\\ $\max_j{(T.Spending_j)}$ where $j$ represents all the suppliers\\ contracted by  the buyer, and $T.Spending$ the amount of money\\ given to each supplier  \end{tabular} \\ 
  \hline
iii)
 &
  Fraction of single-bidder contracts &
  RAD &
  \begin{tabular}[c]{@{}l@{}}{\bf Fraction of single-bidder contracts awarded}\\ {\bf by a buyer to a supplier} \\ $T.AD/T.Cont$\end{tabular} \\
  \hline
   &
  Favoritism &
  Fav &
  \begin{tabular}[c]{@{}l@{}}{\bf Favoritism of a buyer for a supplier as defined in \cite{IMCOmapeando}}
  \\ $(0.33)*(T.Cont/T.Cont.Max)$ + \\ $(0.66)*(T.Spending/T.Spending.Max)$\end{tabular} \\
  \hline
 &
  Contracts per active week &
  CPW &
  \begin{tabular}[c]{@{}l@{}}{\bf Number of contracts a supplier won} \\ {\bf  with the same buyer per active week}\\ $T.Cont/ActiveWeeks$\end{tabular} \\
  \hline
 &
  Spending per active week &
  SPW &
  \begin{tabular}[c]{@{}l@{}}{\bf Amount of money spent by a buyer in contracts} \\ {\bf with the same supplier per active week}\\ $T.Spending/ActiveWeeks$\end{tabular} \\

 \hline \hline
\end{tabular}
\caption{Set of variables designed to assess buyer features and corruption risk.  Variables of type ii) describe features of the buyers. Variables of type iii) provide information about the relationship between the buyers and suppliers. These variables are approximate versions of those proposed in \cite{fazekas2013anatomy,fazekas2013corruption, IMCOmapeando}, see text. }
\label{variables}
\end{table*}

\subsubsection{Contract description variables and risk factors}

Once we have identified the class to which each contract belongs, we use the information given in the source list to build two variables that are descriptive of the buyers: the maximum number of contracts assigned by the buyer to the same supplier, and the maximum total amount spent by the buyer with the same supplier in each year. We call these variables {\bf T.Cont.Max} and {\bf T.Spending.Max}. These variables give us an idea of the budget managed by each buyer and its activity in the public procurement market (Table \ref{variables} - Type ii)). 
Unfortunately, the available data in the source list is not detailed enough to evaluate exactly the risk factors developed in \cite{fazekas2013anatomy,fazekas2013corruption, IMCOmapeando}, thus, we propose approximate versions of the factors that give similar information about the features of the relationship between the buyers and suppliers, and we refer to these variables as Type iii) (See Table \ref{variables}). Below we explain in detail the meaning of the risk factors we use:\\

\begin{enumerate}
    \item [1.] {\bf RAD}: As we mentioned above, single-bidder contracts have been shown to be effective identifiers of corruption risk, particularly because these direct processes hinder the possibility of competition, and the selection of winners may be influenced by illicit agreements. The Mexican Constitution (in its article 134 \cite{constitucion}) establishes that all public procurement should be made by open public contest, except for exceptional cases contemplated in the Law of Acquisitions, Leases and Services of the Public Sector \cite{leyarrendamiento} and the Law of Public Works and Services Related to It \cite{leyobra} . Given these considerations, we consider as a risk factor the \lq\lq Fraction of single-bidder contracts'', proposed in \cite{wachs2020corruption} and defined as:
        \begin{equation*}
            RAD = T.AD/T.Cont
        \end{equation*}
    where {\bf T.AD} is the total number of single-bidder contracts assigned directly by a buyer to a supplier in a year, and {\bf T.Cont}, the total number of contracts between the specific buyer and supplier that year. If ${\bf RAD}\geq 0.5$, that is, if more than half of the contracts between a supplier and a buyer are single-bidder, we consider that the risk of corruption is high \cite{wachs2020corruption}. Strictly speaking, the variable is descriptive of the relation between each buyer and supplier each year. However, in what follows, we assign to each contract the value of {\bf RAD} corresponding to the relation between the buyer and supplier celebrating the contract. 
    
    \item [2.] {\bf Fav:} Clientelism, or favoritism, is also a red flag for corruption. This practice refers to the situation in which the buyer favors a particular supplier by giving it an atypically high volume of contracts and money. While it is hard to detect other than in very obvious cases, the IMCO proposed a quantitative way to measure favoritism as follows \cite{IMCOmapeando}:
        \begin{align*}
            Fav &= (0.33)\frac{T.Cont}{T.Cont.Max}& \\
            &+(0.66)\frac{T.Spending}{T.Spending.Max}&
        \end{align*}
    with, {\bf T.Cont} and {\bf T.Spending} the number of total contracts and total spending made by the buyer with the supplier in a year, and {\bf T.Cont.Max} and {\bf T.Spending.Max} the variables that are descriptive of the buyer as explained above. This factor would give a score of 1 to the supplier that is the most favored by the buyer in both the number of contracts and money spent. The rest of the suppliers are then scored with respect to this most favored possible supplier. If ${\bf Fav}\geq 0.9$, then we consider that that relationship buyer-supplier has a large risk of corruption \cite{IMCOmapeando}. As above, we assign to each contract the value of the {\bf Fav} corresponding to the relation between the buyer and suppler celebrating the contract.

    \item [3.] {\bf CPW and SPW:} The IMCO also identifies a particular form of corruption in which a buyer and a company manipulate the conditions of a contract procedure, dividing an expensive contract into multiple smaller ones that are assigned to the company in a very short time interval (sometimes the same day). These smaller contracts are easier to assign directly to a single-bidder, and are less likely to be scrutinized  \cite{IMCOmapeando}. To identify this behavior, we consider two risk factors: \lq\lq Contracts per Active Week'' and \lq\lq Spending per Active Week'' defined as:
        \begin{eqnarray*}
            CPW &=& T.Cont/ActiveWeeks \\
            SPW &=& T.Spending/ActiveWeeks
        \end{eqnarray*}
    where {\bf T.Cont} and {\bf T.Spending} are the same variables explained before, and {\bf ActiveWeeks} is the number of weeks in which a supplier was assigned contracts by a buyer in each year. Then, if a large contract was fractioned into many small ones, we expect that {\bf CPW} would be large. We consider that there is a large risk of corruption if ${\bf CPW}\geq 5$ ({\it i.e.} that the company received more than one contract per day) and ${\bf SPW}\geq$ 10K USD PPP \footnote{We established the threshold for {\bf SPW} after checking the behavior of contracts in the EFOS and PCS classes with ${\bf CPW}\geq 5$. Further, the precision/recall curves (Fig. S14 of SM) show that moving these thresholds does not make a difference on the results.}. As above, we assign to each contract the value of the {\bf CPW} and {\bf SPW} corresponding to the relation between the buyer and suppler celebrating the contract.
\end{enumerate}

Therefore, considering the variables available in the source list, and the variables related to buyers' features and risk factors, there are three types of items in the final data set: 

\begin{enumerate}
    \item [i)] Items that describe the features of the contracts. For example, the procedure by which the supplier won the contract (single-bidder, public contest, etc.), the amount allocated to the contract, or the week in which the contract began (Table \ref{datoscontratos}).
    \item [ii)] Items that describe the  features of the buyers. These include the maximum amount spent by a buyer with a supplier, or the maximum number of contracts carried out by a buyer with a single supplier. 
    \item [iii)] Items that give information about the relationship between the buyers and suppliers. These items are expected to work as risk factors for corruption. Examples of these are the fraction of single-bidder contracts awarded by a buyer to supplier, or the favoritism of a buyer for a supplier.
\end{enumerate}

In what follows, the metrics computed for variables of type i) and iii) will be the fraction of {\bf\emph{contracts}} that satisfy certain property ({\it e.g.} the fraction of contracts in which a small company participated, or the fraction of contracts between buyers and suppliers characterized by a given value of RAD), and the metrics related to variables of type ii) will be related to the fraction of {\bf\emph{buyers}} with certain features ({\it e.g.} the fraction of buyers with a maximum spending larger than 5K USD PPP). Table S1 of Supplementary Material (SM) shows the most common descriptive statistics of all the variables for each contract class.  

\subsection{Statistical analysis}

One of the central goals of this work is to determine whether the government turnover brought along a methodological change in public procurement. To achieve this goal we first need to analyze whether, within the same government period, there exist significant differences between the three contract classes we defined. Specifically, we aim to verify that each contract class presents statistical differences in their characteristic variables ({\it i.e.} variables of types i) and ii)), when compared with the other classes. This will provide further justification for separating contracts into the three classes, and will help associate each class with certain characteristic description variables. After that, we compare the statistical profile of each class between the different governments, which will tell us whether there were significant changes in public procurement practices within the classes with the change of government. And finally, we  test the risk factors proposed above, and determine if the government transition resulted in a higher or lower risk of corruption in public procurement.        

\subsubsection{Binomial Test and Kolmogorov-Smirnov Test}

Since each contract has two kinds of descriptive variables, dummy variables and non-dummy variables, we use different tests to compare each kind of variable between classes and periods. To measure differences between classes we take the following procedure: for the dummy variables set we use the Binomial Test (B-Test) \cite{conover98, hollander13}, and for the non-dummy variables set we use the Kolmogorov-Smirnov Test (KS-Test) \cite{smirnov1939estimate} \footnote{We choose these particular tests because a balance between samples sizes is not required to obtain reliable comparison results.}. The specific steps to do this are the following:

\begin{enumerate}
    \item [1.] We separate the data corresponding to the two different government periods. The 1st period covers from 2013 to 2018, the 2nd period from 2019 to 2020. 

    \item [2.] For each contract class we extract the data for each variable for all the contracts belonging to the class in each period. 

    \item [3.] We compare classes by pairs in the same period performing the two-sample B-Test or the two-sample KS-Test for each variable \footnote{To compute the B-Test we follow the standard procedure to compute the $z-score$ and the $p-value$ given in \cite{conover98, hollander13, oakley21} and for the KS-Test we use the R function {\it ks.test} from the {\it dgof} package \cite{R, dgof}.}. 

    \item [4.] We consider that there are significant statistical differences between contract classes in those variables for which:
        \begin{enumerate}
            \item [a.] The B-Test results in a p-value $p_v\leq0.05$ and where the difference between fractions of the dummy variables is $\geq0.1$.
            \item [b.] The KS-Test results in a statistic $D\geq0.1$ and a p-value $p_v\leq0.05$.
        \end{enumerate}
With these values we ensure that at least 10\% of the contracts of one class present, in these variables, a different behavior from the contracts of the other class. The sample sizes for each variable type in each class and period, are in Table S2 of the SM. 
\end{enumerate}  

The results of these comparisons are shown in sub-section 3.2. 

\subsubsection{Measuring differences between government periods}

To measure the differences between government periods, we take a slightly different path than that taken above since there is a natural variability in the distribution functions within a period, and we seek to detect differences beyond this variability. Thus we perform the comparison as follows:

\begin{enumerate}

    \item [1.] We separate the data of each government period by year.

    \item [2.] For each contract class we compute:
    \begin{enumerate}
        \item [a.] For each dummy variable: the fraction of contracts in which the variable is present, over all the contracts belonging to the class in every year.
        \item [b.] For each non-dummy variable: the cumulative distribution function (CDF) of the variable \footnote{To make this computation we use the {\it ecdf} R function from the {\it stats} package \cite{R}.} over all the contracts belonging to the class in each year. 
    \end{enumerate}

    \item [3.] Then, for non-dummy variables, we compute the confidence interval (CI) at 99\% \footnote{We considered this value for the confidence interval because of the small amount of data we have to compute it. For the computation of the CI we use the {\it CI} R function from the {\it Rmisc} package \cite{R, Rmisc}} for each distribution, using the data from each year of the 1st period. For dummy variables we use a boxplot of the data from 2013-2018 for comparison \footnote{To generate the boxplot we use the R function {\it boxplot} from the {\it stats} package \cite{R}.}.
    
    \item [4.] Now, using the data from the contracts under the new government, those dummy variables whose fractions are outside of the minimum and maximum ranges of the boxplot for the corresponding data from the first period, and those non-dummy variables with at least 25\% of their cumulative distribution curve lying outside of the CI of the corresponding data of the first period, are considered to have significant statistical differences beyond the natural variability of the distributions. Conversely, those dummy variables whose fractions are inside the corresponding boxplot's ranges, or the non-dummy variables for which at least 75\% of their cumulative distribution curves are within the corresponding CI, will be considered statistically equivalent, meaning that the behavior corresponding to this variable is similar between periods. 
\end{enumerate}

This method provides a quantitative tool to identify the differences (and similarities) of the different contract classes between periods.

The results of these comparisons are shown in sub-section 3.3.

\begin{table*}[htbp!]
\begin{tabular}{@{}l|l|l|l|l|l|l|l|l@{}}
\hline \hline
\textbf{Year}           & 
\textbf{2013} & \textbf{2014} &
\textbf{2015} & \textbf{2016} &
\textbf{2017} & \textbf{2018} &
\textbf{2019} & \textbf{2020} \\ \hline
\textbf{Federal Budget (FB)}  &
$5.02\cdot 10^{11}$       & $5.55\cdot 10^{11}$       &
$5.64\cdot 10^{11}$       & $5.64\cdot 10^{11}$       & 
$5.48\cdot 10^{11}$       & $5.75\cdot 10^{11}$       & 
$6.28\cdot 10^{11}$       & $6.51\cdot 10^{11}$       \\
\textbf{Total Spending (TS)}  &
$6.43\cdot 10^{10}$       & $8.97\cdot 10^{10}$       &
$7.86\cdot 10^{10}$       & $6.78\cdot 10^{10}$       & 
$8.68\cdot 10^{10}$       & $5.52\cdot 10^{10}$       &
$3.03\cdot 10^{10}$       & $4.08\cdot 10^{10}$       \\
\textbf{Total Contracts (TC)} & 
173,253       & 190,995       & 
215,260       & 225,955       & 
224,723       & 188,826       & 
182,483       & 138,821       \\
\textbf{TS/FB}           &  0.13          & 0.16          & 
0.14          & 0.12          &  0.16          & 0.10          &  0.05          & 0.06 \\ \hline \hline
\end{tabular}
\caption{{\it Public data per year in the period 2013-2020.} Approved federal budget (FB - extracted from \cite{fb13, fb14, fb15, fb16, fb17, fb18, fb19, fb20}), total spending made on public procurement (TS - extracted from the source list), total number of contracts made (TC- extracted from the source list), and the ratio TS/FB. FB and TS are reported in USD PPP using the equivalences given in \cite{oecdppp}.}
\label{tabcomp1}
\end{table*}

\begin{table*}[htbp!]
\begin{tabular}{l |lll|lll|lll}
\hline \hline
{\bf Year}      & \multicolumn{3}{c|}{{\bf 2013}} & \multicolumn{3}{c|}{{\bf 2014}}  & \multicolumn{3}{c}{{\bf 2015}}  \\
\hline
          & EFOS           & PCS            & NC             & EFOS   & PCS    & NC      & EFOS           & PCS            & NC       \\
{\bf Contracts} & 332            & 3,429          & 169,429        & 505    & 3,826  & 186,664 & 542            & 4,285          & 210,433  \\
{\bf Spending}  & $1.01\cdot 10^{8}$         & $7.05\cdot 10^{9}$         & $5.72\cdot 10^{10}$        & $2.18\cdot 10^{8}$ & $6.76\cdot 10^{9}$ & $8.27\cdot 10^{10}$ & $1.24\cdot 10^{8}$         & $5.96\cdot 10^{9}$         & $7.25\cdot 10^{10}$  \\
{\bf \%TC}      & 0.19           & 1.98           & 97.83          & 0.26   & 2.00   & 97.74   & 0.25           & 1.99           & 97.76   \\
{\bf \%TS}      & 0.15           & 10.95          & 88.90          & 0.24   & 7.54   & 92.22   & 0.16           & 7.59           & 92.25 \\ 
\hline \hline
\end{tabular}
\end{table*}

\begin{table*}[htbp!]
\begin{tabular}{l |lll|lll|lll}
\hline \hline
{\bf Year}      & \multicolumn{3}{c}{{\bf 2016}}  &  \multicolumn{3}{c|}{{\bf 2017}} & \multicolumn{3}{c}{{\bf2018}}   \\
\hline
          & EFOS           & PCS            & NC             & EFOS   & PCS    & NC      & EFOS           & PCS            & NC    \\
{\bf Contracts}  & 433    & 4,630  & 220,892 &  314            & 6,019          & 218,390        & 217    & 3,842  & 184,767 \\
{\bf Spending}  & $1.06\cdot 10^{8}$ & $5.75\cdot 10^{9}$ & $6.20\cdot 10^{10}$ &  $9.63\cdot 10^{7}$         & $6.17\cdot 10^{9}$         & $8.05\cdot 10^{10}$        & $3.48\cdot 10^{7}$ & $6.11\cdot 10^{9}$ & $4.90\cdot 10^{10}$  \\
{\bf \%TC}      & 0.19   & 2.04   & 97.77 &  0.14           & 2.67           & 97.19          & 0.11   & 2.03   & 97.68     \\
{\bf \%TS}      & 0.16   & 8.47   & 91.37&  0.11           & 7.11           & 92.78          & 0.06   & 11.08  & 88.86   \\ 
\hline \hline
\end{tabular}
\end{table*}

\begin{table*}[htbp!]
\begin{tabular}{l |lll|lll}
\hline \hline
{\bf Year}      & \multicolumn{3}{c|}{{\bf 2019}} & \multicolumn{3}{c}{{\bf 2020}}  \\
\hline
          & EFOS           & PCS            & NC             & EFOS   & PCS    & NC   \\
{\bf Contracts}     & 55             & 3,635          & 178,793        & 13     & 1,417  & 137,461 \\
{\bf Spending}   & $1.17\cdot 10^{7}$         & $2.78\cdot 10^{9}$         & $2.75\cdot 10^{10}$        
& $9.63\cdot 10^{6}$ & $7.25\cdot 10^{8}$ & $4.00\cdot 10^{10}$ \\
{\bf \%TC}    & 0.03           & 1.99           & 97.98          & 0.01   & 1.02   & 98.97   \\
{\bf \%TS}     & 0.03           & 9.15           & 90.82          & 0.02   & 1.78   & 98.20\\
\hline \hline
\end{tabular}
\caption{{\it Data for each contract class per year in the period 2013-2020.} Number of contracts, spending (indicated in USD PPP), percentage of total contracts (\%TC) and percentage of total spending (\%TS) made in each contract class. This data were extract from the source list.}
\label{tabcompii}
\end{table*}

\section{Results}

\subsection{Context for comparison}

\begin{figure*}[htbp!]
  \centerline{\includegraphics*[width=0.7\textwidth]{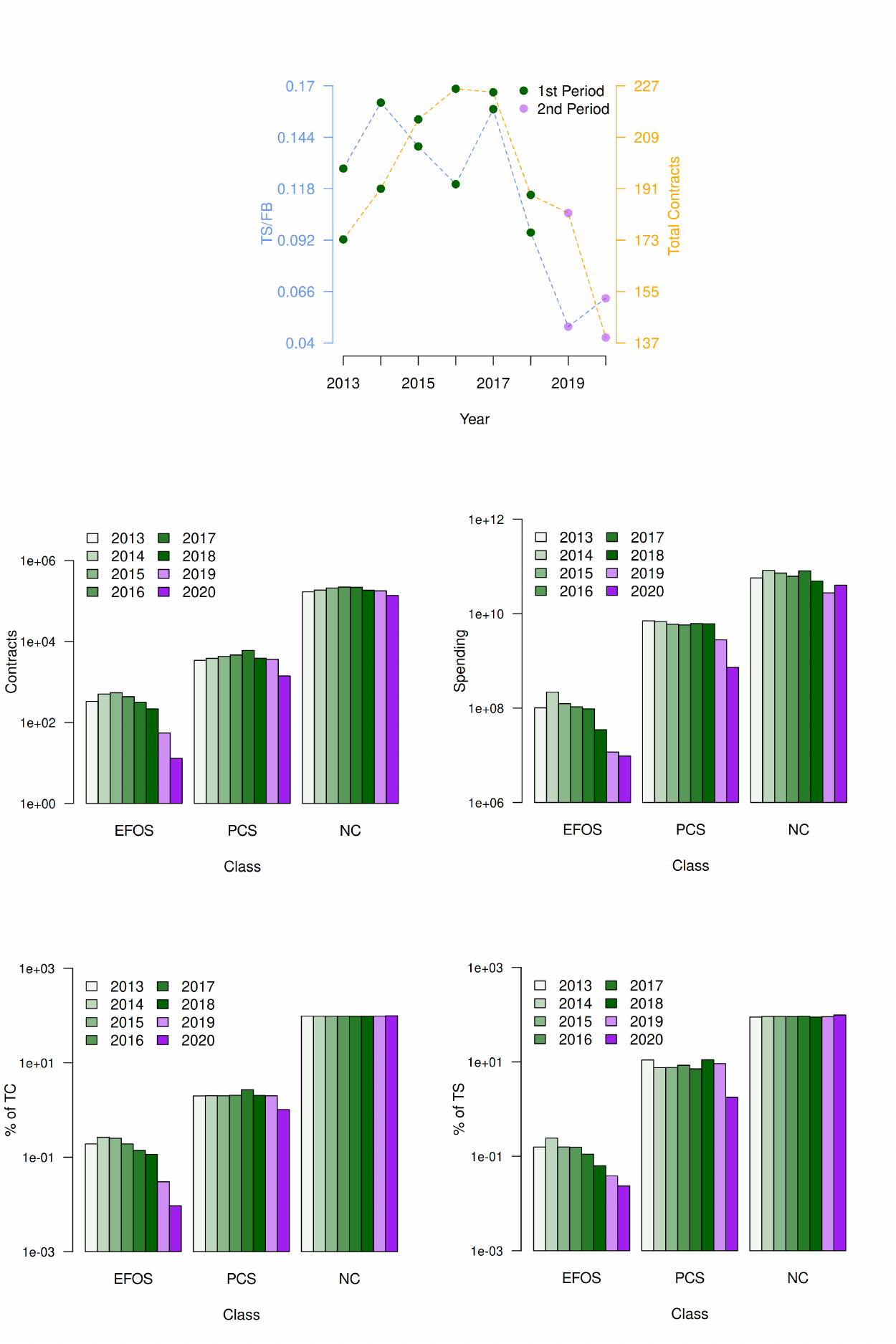}}
    \caption{{\it Comparison of the number of contracts and spending per year}. {\bf Top:} Ratio between the total spending reported in the contract source list and federal budget (blue dashed line - in USD PPP), and total number of contracts reported (orange dashed line - in thousands). The green circles represent the 1st government period (2013-2018), whereas the purple circles correspond to the 2nd government period (2019-2020). {\bf Center Left:} Logarithmic plot of the number of contracts in each class. {\bf Center Right:} Logarithmic plot of the resources spent in each class. {\bf Bottom Left:} Logarithmic plot of the percentage of the total contracts in each class. {\bf Bottom Right:} Logarithmic plot of the percentage of the total spending made in each class. 
    Green and purple bars correspond to the 1st and 2nd government periods respectively.}
    \label{Context1}
\end{figure*}

To make a fair comparison between government periods and between different contract classes, we present how many resources were spent in each class, as well
as how many contracts belonged to each class per year. Table \ref{tabcomp1} shows the 
approved federal budget (FB) per year (in USD PPP)  \cite{fb13, fb14, fb15, fb16, fb17, fb18, fb19, fb20} , the Total Spending (TS) on public procurement according to the source list (in USD PPP), the Total number of Contracts (TC) made in each year, and the ratio TS/FB. We observe that the ratio TS/FB varies from 0.10 to 0.16 in the 1st government period (2013-2018, Fig. \ref{Context1} - {\bf Top}, blue dashed line, green circles), but after the
change of government, the ratio fell to 0.05 and 0.06 for 2019 and 2020 (Fig. \ref{Context1} - {\bf Top}, blue dashed line, purple circles),
{\it i.e.} even when the budget increased, the spending in public procurement fell by approximately a 
half. Fig. \ref{Context1} - {\bf Top}  (orange dashed line) shows that the number of contracts also fell 
at the end of the 1st period, then continued to fall at the beginning of the 2nd period, and 
remained low during the second year of the 2nd period. This decrease may be a consequence of the politics of \lq\lq republican austerity'' imposed by the new government \cite{amlocorrupcion}.

Considering these data, we compared the number of contracts and the amount of money spent on each contract class for every year (Table \ref{tabcompii} - Rows 2 and 3, and Fig. \ref{Context1}- {\bf Center}).  It is noticeable that the number of contracts in the EFOS class in 2019 and 2020, and the amount of money spent on them, fell by an order of magnitude. In contrast, the number of contracts and the total spending in contracts in the PCS class remained roughly in the same scale over several years, and only suffered a significant decrease in 2020. For contracts in the NC class, the numbers are similar between periods.

We normalized the absolute numbers of contracts in each class by the corresponding TC and TS to compare the percentage of the Total Contracts (\%TC) and of the Total Spending (\%TS) in each year (Table \ref{tabcompii} - Rows 4 and 5, and Fig. \ref{Context1} - {\bf Bottom}). We see that the NC class represents approximately 90\% of the Total Spending and Total Contracts for all years. Also for the contracts in the PCS class there were no major changes between years in \%TC and \%TS up to 2020, where \%TC decayed roughly by half to 1.02\% and \%TS fell to 1.78\%. However, for the contracts in the EFOS class, there was a large decay in the fraction of contracts assigned, and a corresponding decrease in resources spent when the government changed.

Then, the first important change in public procurement due the government transition, was a marked decrease in public spending and in the number of contracts made. This decrease was especially noticeable in the EFOS class.

\subsection{Comparison between contract classes}

To verify whether the three different contract classes have intrinsic, statistically significant, procedural differences between them, we computed a B-Test (for the set of dummy variables) and a KS-Test (for the non-dummy variables) comparing by pairs each variable that describes the contracts. This comparison was made between all the contract classes in the same government period.

\begin{figure}[htbp!]
  \centerline{\includegraphics*[width=0.4\textwidth]{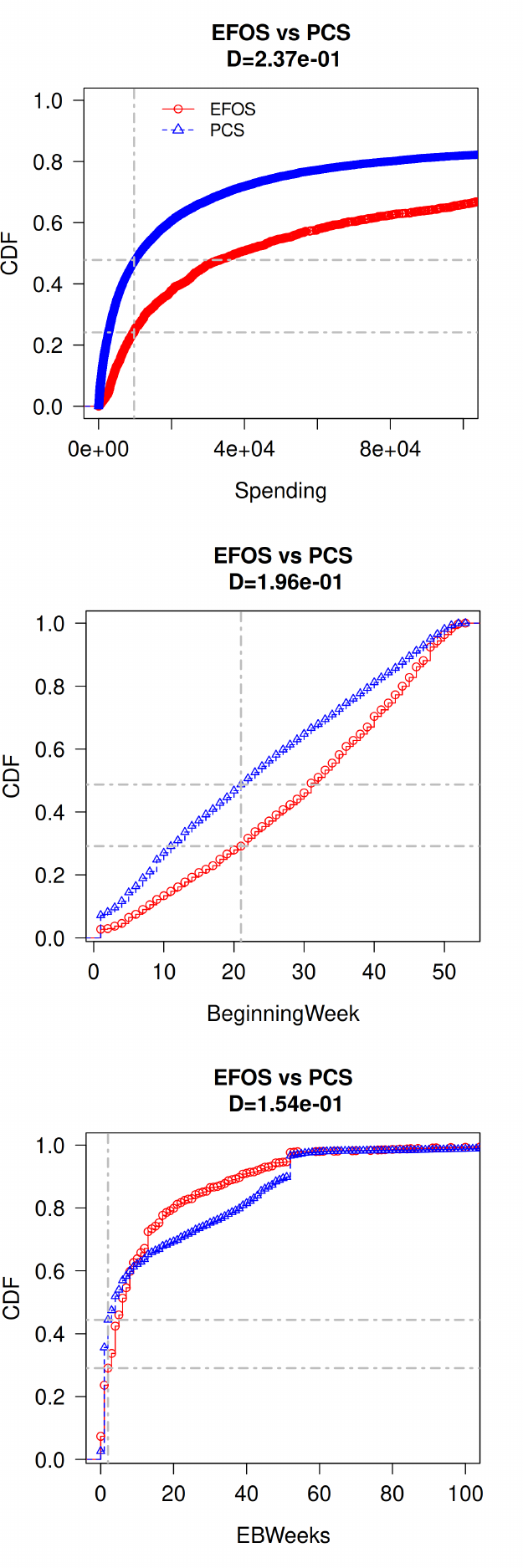}}
\caption{{\it An example of a set of variables with significant differences between EFOS and PCS classes within the first government period}.
{\bf Top:} Cumulative distribution function (CDF) for the variable {\bf Spending} in the contracts of the EFOS class (red circles solid line) and in the PCS class (blue triangles dashed line). {\bf Center:} CDF for the week of the year in which the contract began ({\bf BeginningWeek}). {\bf Bottom:} CDF for the duration of the contract ({\bf EBWeeks}). The grey dashed line shows the value at which the CDFs showed the maximum difference. }
\label{KS-T1stFig}
\end{figure}

First we compared the EFOS class vs. the PCS class. For the 1st period, Table S3 and Fig. S1 of the Supplementary Material show that there were 17 variables for which the statistical tests showed significant differences. These variables were of type i), {\it i.e.} variables that describe the features of the contracts. For example, we found that most of the contracts in the PCS class were made with companies classified as \lq\lq large supplier" ({\bf S.NOM}), while the contracts in the EFOS class were mostly done with micro, small, or medium companies \footnote{To test if this difference in the type of companies contracted was beyond the null hypothesis of randomly choosing a contractor, we analyzed the distribution of each company size in each class, finding that the choice of supplier indeed favored one particular company size for each different class.}. Fig. \ref{KS-T1stFig} shows a subset of those variables for which contracts carried out with companies labeled as EFOS and PCS were significantly different in the 1st government period (2013-2018). For the variable called {\bf Spending} ({\bf Top}), we found that the contracts in the EFOS class (red circles solid line) tend to be more expensive than the contracts in PCS class (blue triangles dashed line). We can also see in Fig. \ref{KS-T1stFig} - {\bf Bottom} that the PCS companies obtain contracts of short duration (less than 3 weeks) more frequently than EFOS companies. However, in both classes there are a few contracts of large duration (more than one year). Finally, Fig. \ref{KS-T1stFig}  - {\bf Center} shows that the 1st period government made contracts with EFOS mainly during the 2nd half of the year ({\bf BeginningWeek}), while companies labeled as PCS were contracted mostly during the 1st half of the year.
For the 2nd government period, the comparison EFOS vs. PCS showed 14 variables with significant differences, all of them of type i) as well (Table S4 and Fig. S2 of SM). Interestingly, 11 of these variables were also present in the set of variables that displayed differences in the 1st government period. Variables such as fraction of contracts with large suppliers ({\bf S.NOM}) and {\bf Spending} appear with similar behavior as in the 1st period. Also, even when the PCS contracts tend to be shorter than EFOS contracts, some of the PCS contracts reached duration times of up to two years, versus one year for the longest EFOS contracts (Fig. S2 of SM). The remaining variables with significant differences are related to procedure type, procedure character, or contract type. For example, almost all the PCS contracts were for acquisitions ({\bf CT.ADQ}), while less than half of the EFOS contracts were of this type (Table S4 of SM). Also, the distribution of the variable for the  single-bidder procedure ({\bf PT.AD}) shows that this procedure is more frequent in the PCS class than in the EFOS class (Table S4 of SM). These results highlight the fact that each of the corrupt contract classes has a different statistical profile, which in turn justifies keeping the corrupt classes separated. Further, given that the PCS contracts far outnumber the EFOS contracts every year (see Table \ref{tabcompii}), if we join both classes in a single \lq\lq corrupt" class, the statistics of this \lq\lq joint'' class would mostly reflect the statistics of the PCS class.

Regarding the comparison between EFOS and NC contracts, 10 variables presented significant differences for the 1st period, these variables were of types i) and type ii) (Table S5 and Fig. S3 of SM). It is noticeable that the two variables that showed the most prominent differences were those that characterize buyers' features ({\it i.e.} variables of type ii)); these were the maximum number of contracts awarded by a buyer to a supplier (labeled {\bf T.Cont.Max}), and the maximum amount of money spent by a buyer in contracts with a supplier ({\bf T.Spending.Max}). These variables show that the EFOS class had a higher proportion of buyers with a proportionally stronger market activity ({\it i.e.} buyers characterized by having awarded a maximum number of contracts to a supplier {\bf T.Cont.Max} $\geq$ 5) and larger budget ({\bf T.Spending.Max}$\geq$1.9M USD PPP) than the buyers in the NC class (Fig. S3 of SM). For the 2nd period the statistical tests gave 16 variables (again of the two first types) with significant differences (Table S6 and Fig. S4 of SM). Here, also the variables {\bf T.Cont.Max} and {\bf T.Spending.Max}, were those that showed the most prominent differences. The behavior of these variables was similar to that observed in the 1st government period, giving the same differences between EFOS and NC classes for the buyers' features.

The PCS vs. NC contract classes' comparison for the 1st government period (Table S7 and Fig. S5 of SM) showed significant differences in 10 variables, again distributed among the variable types i) and ii). For example, the variable for national procedure character (labeled {\bf PC.N}) shows that the NC contract's fraction made under national regulations, {\it i.e.} the contracting protocols thet followed Mexican laws, was much higher than in the PCS class. The variable for large suppliers ({\bf S.NOM}) shows that this kind of company was proportionally contracted with higher frequency in the PCS class than in the NC class. The variables for buyers' features ({\bf T.Cont.Max} and {\bf T.Spending.Max}) presented differences similar to those discussed above: the PCS class had a higher proportion of buyers with a proportionally stronger market activity and larger budget than the buyers in the NC class. For the 2nd government period, the tests showed 13 variables with significant differences, again within the variable types i) and ii) (Table S8 and Fig. S6 of SM). For example, the variable for large suppliers ({\bf S.NOM}) has a similar behavior as in the 1st government period: large suppliers were more frequently contracted in PCS class than in the NC class. Finally, Fig. S6 of the  shows that the PCS contracts tend to last more than the contracts in the NC class.

Thus, we see that our three contract classes do present significant differences when compared among each other in the same government period, and that these differences are similar between periods. These results provide support to the separation of the contracts in these three classes, and also to the hypothesis that the corruption that occurs through EFOS and PCS have different procedural patterns. At the same time, these two contract classes have differences with the NC contracts.  The next step is to identify whether each class presents differences between government periods. 

\begin{figure*}[htbp!]
  \centerline{\includegraphics*[width=1.0\textwidth]{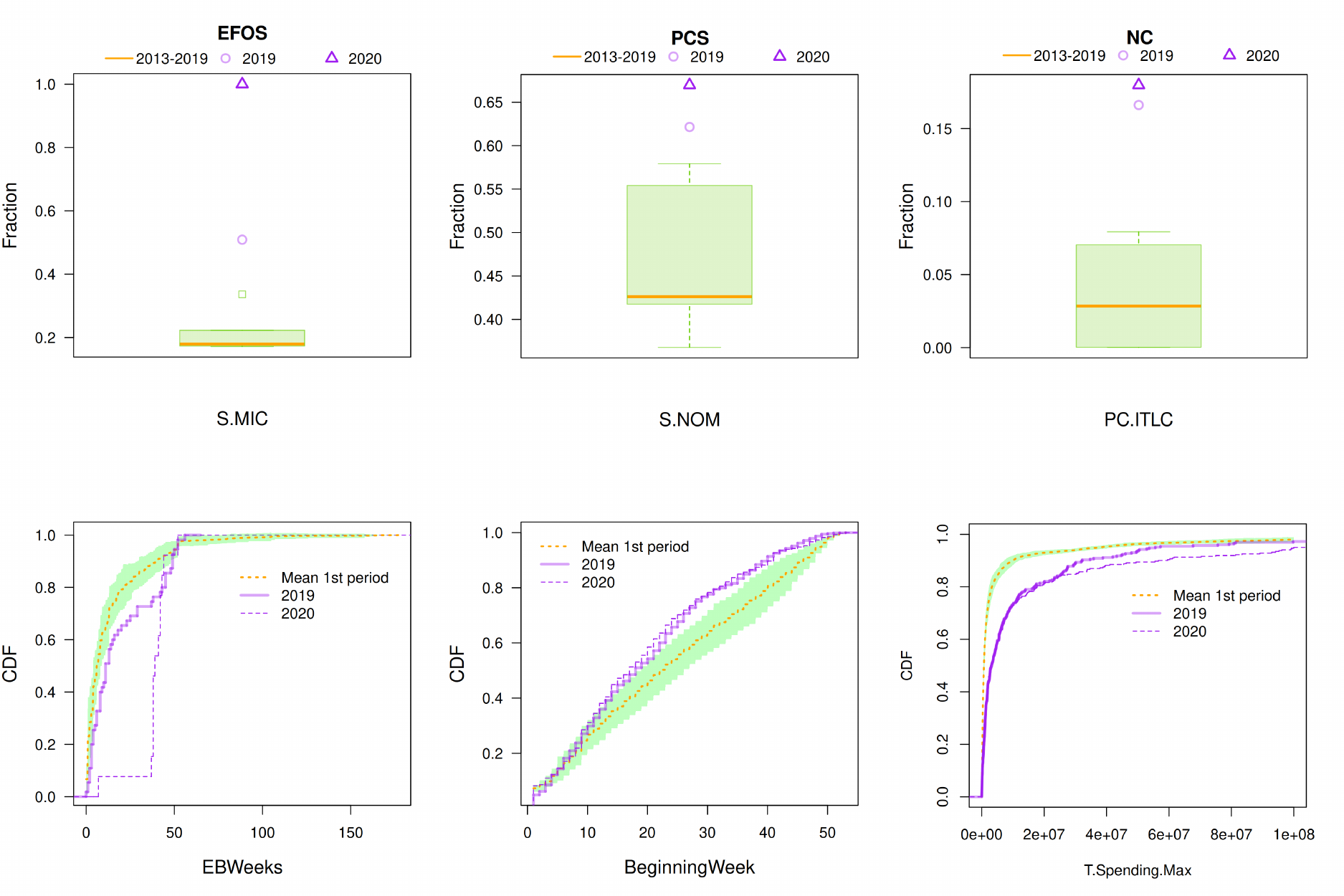}}
    \caption{{\it Example variables illustrating the main differences between both government periods}. {\bf Left:} Differences between contracts in the EFOS class. {\it Top Left:} Fraction of \lq\lq micro companies''  ({\bf S.MIC}) for the contracts in the EFOS class. In green, boxplot of the 6 years comprising the 1st government period. {\it Bottom Left:} Cumulative distribution function (CDF) for the number of weeks that the contract lasted ({\bf EBWeeks}). The green area represents the CI at 99\% generated by the data of the 1st period. {\bf Center:} Differences between contracts in the PCS class. {\it Top Center:} Fraction of contracts in the PCS class with \lq\lq large'' companies ({\bf S.NOM}). {\it Bottom Center:} CDF for week that the contract began ({\bf BeginningWeek}). {\bf Right:} Differences between contracts in the NC class. {\it Top Right:} Fraction of contracts following the NAFTA procedures ({\bf PC.ITLC}) in the NC class. {\it Bottom Right:} CDF for the maximum spending made by a buyer with a single supplier ({\bf T.Spending.Max}). In all graphs, the orange line corresponds to the mean of the 1st government period, and the two purple colored curves correspond to the first two years of the second government period.}
    \label{CI-Dif}
\end{figure*}

\subsection{Comparison between government periods}

As mentioned before, to identify differences and similarities in public procurement between government periods, we compared the variables that define the contracts either by computing the confidence interval (CI) generated by the years of the 1st period (2013-2018) and verifying whether or not both curves corresponding to each year of the 2nd period (2019-2020) lie within the CI, or, for the dummy variables, checking if the results for the second period fall inside the box plot of the data of the first period. Since one of our goals is to try to determine whether the government transition that occurred in M\'exico at the end of 2018 brought a change in the procurement practices, we compared the data for each period in the three different contract classes. Our analysis of the data sets produced the following results.

\subsubsection{Main differences between both periods}

Fig. \ref{CI-Dif} shows the most significant differences between both periods for the three contract classes: EFOS, PCS, and NC. First, we observe that in the 1st period, $\sim$ 20\% (on average) of the suppliers classified as EFOS were micro-companies (Fig.\ref{CI-Dif} - {\bf Top Left}). In contrast, in the 2nd period, this fraction grew to 50\% in 2019, and to 100\% in 2020. This signals an important change in the way interactions with EFOS were carried out between both periods. Another important difference in this class was the contract duration (Fig.\ref{CI-Dif} - {\bf Bottom Left}). In the 1st period, only 20\% (on average) of the contracts with companies labeled as EFOS lasted more than ten weeks, while for the 2nd period, 40\% of these contracts had a duration of more than ten weeks in 2019, and in 2020 this fraction grew to almost 90\%.
Even when the government of 2nd period invested less money on EFOS, and granted them fewer contracts (Fig. \ref{Context1}), the majority of these contracts were of long duration. Fig. S7 of SM shows the remaining variables in which there were significant differences between periods. For example, we notice that the new government tends to contract EFOS at the beginning of the year ({\bf BeginningWeek}), while the previous government did so mostly during the second half of the year.

For the PCS class we observe that in the 1st government period, the PCS contracts were made mostly (57\%) with small and medium companies (Fig.\ref{CI-Dif} - {{\bf Top Center}}). In contrast, in the 2nd government period, the majority of these contracts passed to large suppliers. On the other hand, with the new government, PCS contracting decreased in the middle of the year; {\it i.e}. the 2nd period government had a slightly lower tendency to contract PCS between weeks 20 and 40 than the 1st period government (Fig. \ref{CI-Dif} - {\bf Bottom Center}). Fig. S8 of the Supplementary Material shows the remaining variables for which contracts in the PCS class presented significant differences between periods. For example, in the 1st period, 1\% of contracts in the PCS class were leases (labeled {\bf CT.AR}), this fraction grew to almost 1.5\% in 2019, and to 2\% in 2020.

Next, for the NC class, we notice that the 2nd period had an increase of 15\% in the number of contracts made under the rules and regulations of the North American Free Trade Agreement (NAFTA) (Fig.\ref{CI-Dif} - {\bf Top Right}). On the other hand, the two variables associated with buyer features had differences between periods. Fig.\ref{CI-Dif} - {\bf Bottom Right} shows that almost 80\% of the buyers in the 1st period had a maximum spending ({\bf T.Spending.Max}) less than 2K USD PPP, while in the 2nd period this percentage fell to almost 60\%. Fig. S9 of the SM shows that the maximum of contacts made by a buyer with the same supplier in the 2nd period remains relatively close to the CI generated by the 1st period.
Fig. S9 of SM shows the remaining variables for which NC contracts presented significant differences between periods. For example, the variable for single-bidder procedure type ({\bf PT.AD}) shows that there was also a slight increase in the percentage of contracts won through this procedure from 75\% to 80\% from one period to another.  

\begin{figure*}[htbp!]
  \centerline{\includegraphics*[width=1.0\textwidth]{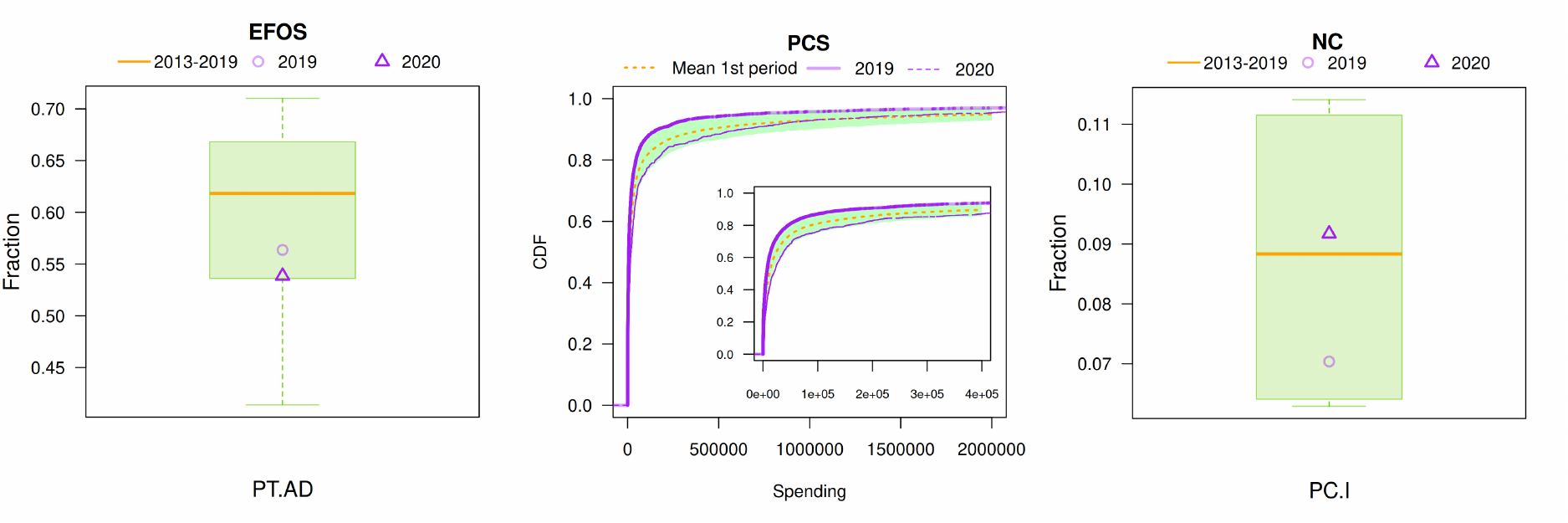}}
    \caption{{\it Example variables illustrating the main similarities between both government periods}. {\bf Left:} Similarities between contracts in the EFOS class. Fraction of contracts assigned to single-bidders ({\bf PT.AD}) in the EFOS class. In green, boxplot of the 6 years comprising the 1st government period. {\bf Center:} Similarities between contracts in the PCS class. CDF for the {\bf Spending} variable. Inset: A close up to the behavior of the CDF for small values of the {\bf Spending} variable. The green area represents the CI at 99\% generated by the data of the 1st period. {\bf Right:} Similarities between contracts in the NC class. Fraction of contracts in the NC class made under an international legal framework ({\bf PC.I}). In all graphs, the orange line corresponds to the mean of the 1st government period, and two purple colored curves correspond to the first two years of the second government period.}
    \label{CI-Sim}
\end{figure*}

\subsubsection{Main similarities between both periods}

Having identified the main differences between the contract classes in both government periods, it may also be helpful to identify the similarities, {\it i.e.} identify those variables that did not experience significant changes due to government transition. This will give  an idea of which aspects of the practices related to EFOS and PCS contracts may have remained. 

Fig. \ref{CI-Sim} shows the main similarities between governments for the three contract classes. We notice, for example, that for the EFOS class, the percentage of contracts assigned to single-bidders ({\bf PT.AD}) did not experience significant changes between periods, staying within the range of 50\% to 60\% (Fig. \ref{CI-Sim} {\bf Top Left}). There are also similarities in the percentage of contracts for leases ({\bf CT.AR}, Fig. S10 of SM).\\
For the PCS class, we found similarities in only one variable. For the {\bf Spending} variable we found that 80\% of the contracts were for less than 1M USD PPP in both periods, but in both periods there were a few contracts that reached 4M USD PPP (Fig. \ref{CI-Sim} - {\bf Center}).

For the NC class, we found similarities in the 
procedure character, in which 7-11\% of the contracts were made under international rules for both periods (Fig. \ref{CI-Sim} - {\bf Right}). In Fig. S12 of SM we can observe that there are also similarities in the contract type. We find that almost 35\% of the contracts were for services in the 1st period, and in the 2nd period this fraction remains at 38\% in 2019, and 32\% in 2020. It is also noticeable that the NC contracts for both periods had the same distribution for the duration variable (labeled {\bf EBWeeks}), where, while most of them lasted less than 40 weeks, there were some that reached more than 200 weeks of duration.

These results show that corruption related with EFOS and PCS suffered important changes in specific features, such as the kind of contracted supplier, the duration of the contracts, and the time of the year in which the corrupt companies won the contracts; and there were other features that did not change between governments, such as the fraction of contracts in the EFOS class won by single-bidder, or the distribution of resources spent on contracts of the PCS class. Also, while the NC class is in principle not legally related to corruption, the differences and similarities of the features that characterize this class of contracts from one period to the other, provide an idea of how the practices of public procurement changed during the government transition.

\begin{figure*}[htbp!]
 \centerline{\includegraphics*[width=1.0\textwidth]{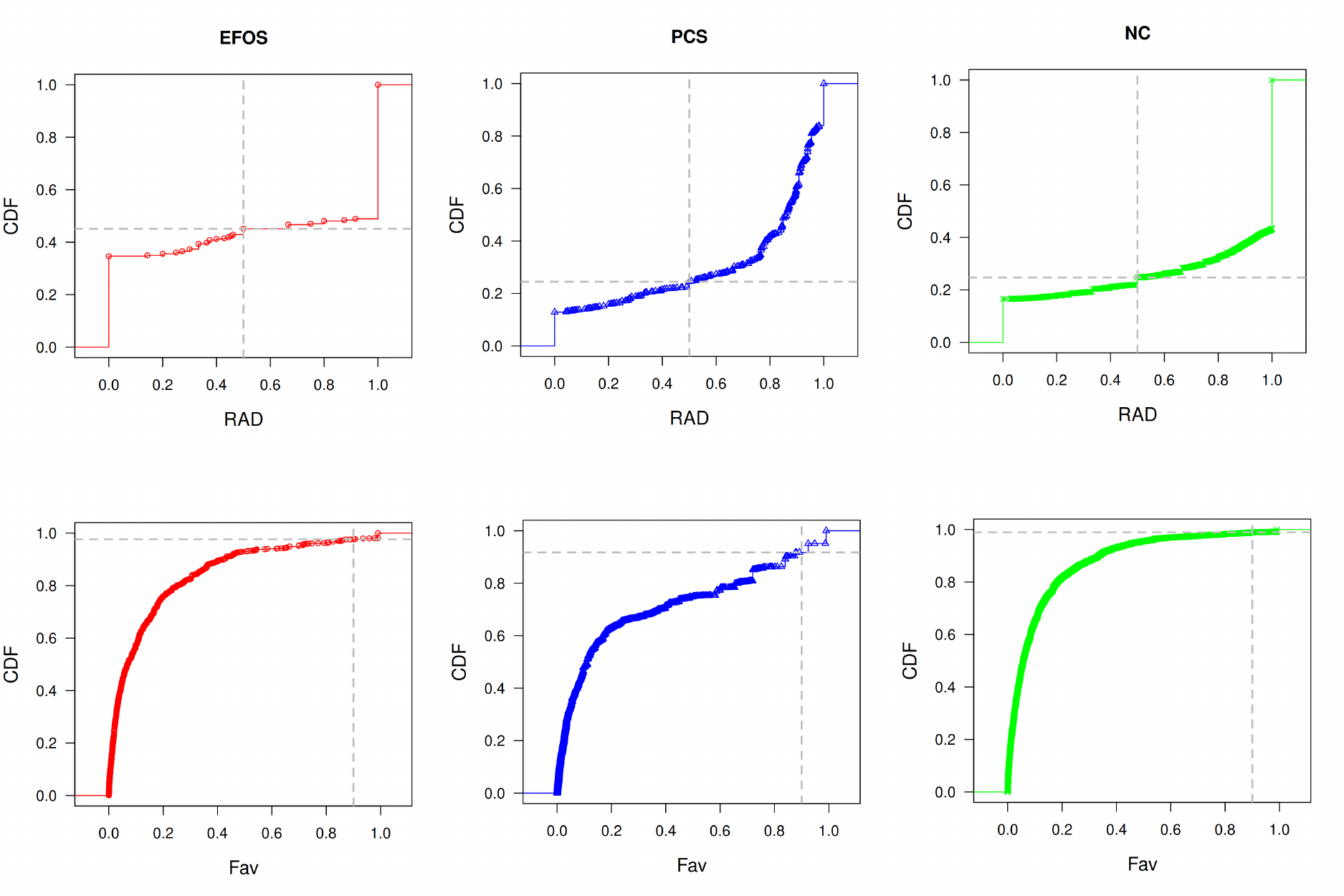}}
\caption{{\it First experiment.} CDF for {\bf RAD} and {\bf Fav} variables ({\bf Up} and {\bf Bottom}, respectively) for the three different classes EFOS ({\bf Left}), PCS ({\bf Center}), and NC ({\bf Right}). The dashed grey lines indicate the threshold for each risk factor and the value of the CDF in which it was crossed. This value in the CDF indicates the probability to find a contract with the risk factor above the marked threshold given that the contract belongs to the corresponding class.}
\label{Test_RF}
\end{figure*}

\subsection{Testing the risk factors}

To reach our second goal, we tested the risk factors proposed in section 2.1.2 and presented in Table \ref{variables} (labeled Type iii)). To do this we performed two experiments:

    \begin{enumerate}
        \item [1.] First we measured the accuracy of each risk factor as a descriptor of the contracts corresponding to the corrupt classes (EFOS and PCS). This was done by computing the fractions of contracts within each corrupt class for which the value of the variables proposed as risk factors exceed the corruption thresholds. If either fraction is larger than 0.5, then the risk factor may be considered as a useful descriptor of the contracts in the corresponding corrupt class. These fractions can be interpreted as the conditional probabilities that a contract exceeds the corruption threshold of the variables proposed as risk factors, given that it belongs to the corresponding corrupt class.
        \item [2.] Second, we measured the accuracy of each risk factor to identify those contracts in which EFOS or PCS participated. To do this we compute the fractions of contracts that belong to either the EFOS or to the PCS class respectively, within the contracts for which the values of the variables proposed as risk factors exceed the corruption threshold. If either fraction is larger than the probability to find a contract of the corresponding corrupt class at random from the complete set of contracts, we consider that the risk factor is helpful identifying contracts in the corresponding class. Again, these fractions are the conditional probabilities that a contract belongs to either the EFOS or the PCS class, given that it exceeds the corresponding corruption threshold of the variables proposed as risk factors.
    \end{enumerate}

These experiments are done without separating the contracts between both government periods, {\it i.e.} we considered all data from 2013-2020
to test each risk factor. Once we determine the risk factors that most accurately
describe and identify the corrupt contract classes, we determine whether the
government turnover brought along a higher or lower risk of corruption using the same framework proposed in section 2.2.2 for non-dummy variables. 

Fig. \ref{Test_RF} shows the results of the first experiment, namely, the probabilities that a contract exceeds the corruption threshold ({\it i.e.} that the value of the variable {\bf RAD}, which we have assigned as a descriptor of each contract, exceeds 0.5; and favoritism {\bf Fav}, exceeds 0.9) given that it belongs to each corrupt class respectively. We observe that the risk factor {\bf RAD} has a nearly 60\% accuracy describing contracts in the EFOS class and a nearly 80\% accuracy describing contracts in the PCS class. On the other hand, less than 1\% of the contracts in both classes presented a favoritism greater than 0.9. Fig. S13 of SM shows the accuracy for the remaining risk factors.  It is noticeable that the presumably corrupt behavior which combines more than 5 contracts and 10K USD PPP spending per active week, which should describe the practice of dividing expensive contracts into smaller ones, showed an accuracy of less than 2\% for contracts in the EFOS class and less than 30\% for the PCS class. Fig. S14 of Supplementary Material shows the results of this 1st experiment from the perspective of a {\it recall curve} \footnote{In the context of binary classifiers, {\it recall} is defined as the ratio between the correctly detected instances by the classifier and all positive instances.} over a larger range of thresholds.  

\begin{figure}[htbp!]
 \centerline{\includegraphics*[width=0.4\textwidth]{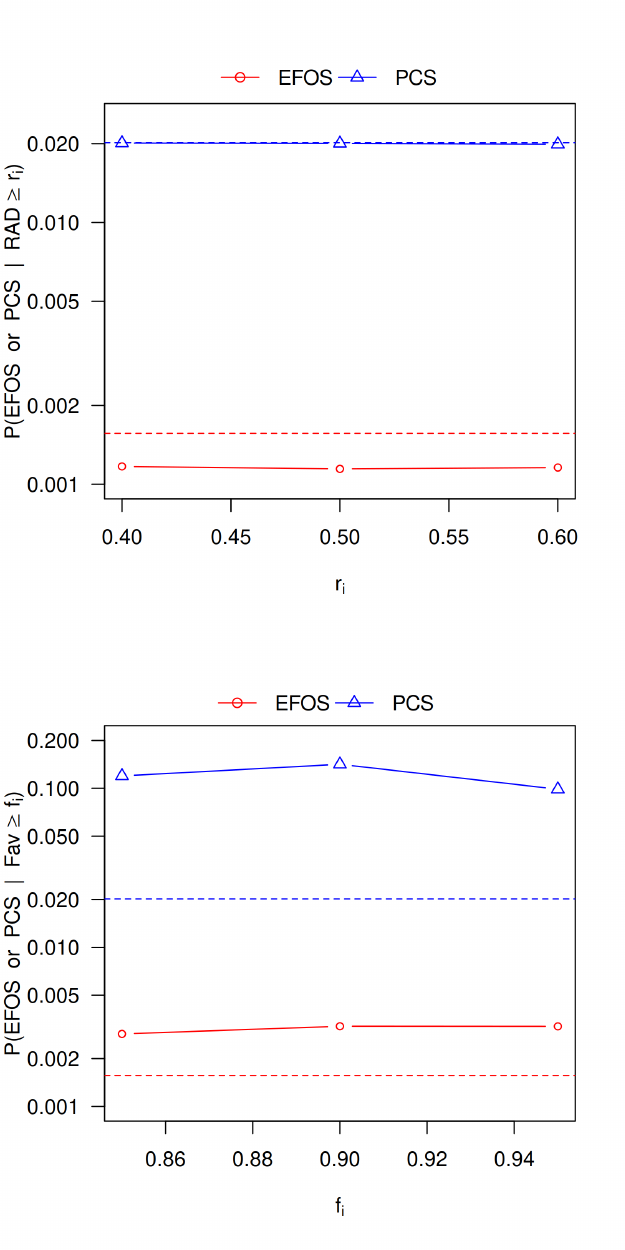}}
\caption{{\it Second experiment.} Probability for a contract to be corrupt (EFOS or PCS, solid lines) given it has an {\bf RAD}$\geq r_i$ ({\bf Top}) or a {\bf Fav} $\geq f_i$ ({\bf Bottom}), for several values of the thresholds, $r_i=\{0.4, 0.5, 0.6\}$ and $f_i=\{0.85, 0.9, 0.95\}$. The dashed lines correspond to the probability to find and EFOS (red) or PCS (blue) contract in the source list.}
\label{Test_RF_1}
\end{figure}

Fig. \ref{Test_RF_1} shows the results of experiment 2, these are the fraction of EFOS (solid red line) and PCS (solid blue line) contracts within all those contracts that exceed a certain threshold for a range of possible threshold values for the {\bf RAD} and {\bf Fav} variables ({\bf Top} and {\bf Bottom}, respectively). The dashed lines represent the probability to randomly find in the source list an EFOS (red) or PCS (blue) contract. We can observe that the {\bf RAD} risk factor is not a good identifier for neither of the corrupt classes, since none of the solid lines are above their respective dashed line. The opposite happened for the {\bf Fav} risk factor. Here we found that the probability to find an EFOS or PCS contract within those with a favoritism larger than the threshold (solid lines) is indeed higher than the probability to randomly find a corrupt contract in the source list (dashed lines). Fig. S13 of the Supplementary Material shows that {\bf CPW} and {\bf SPW} variables are good identifying PCS class but not EFOS class. As above, Fig. S14 of Supplementary Material shows the results of the 2nd experiment from the perspective of a {\it precision curve} \footnote{In the context of binary classifiers, {\it precision} is defined as the ratio between the actual positive instances and all the cases predicted as positive by the classifier.} over a larger range of different thresholds.  

Thus, even though {\bf Fav} is not a good descriptor of corrupt behavior since only 1\% of the EFOS and PCS contracts have a favoritism larger than 0.9; it is a good identifier since the fraction of both classes of corrupt contracts, among the contracts with a value of {\bf Fav} greater than 0.9, is significantly higher than their concentration in the complete sample.
In contrast, the {\bf RAD} variable is good as a descriptor of EFOS and PCS contracts, since more than 50\% of these contracts occur in buyer-supplier relationships for which {\bf RAD}$\geq$0.5. However, {\bf RAD} is not a good identifier since the probability to find an EFOS or PCS contract within all of those that exceed a given threshold for {\bf RAD} is never larger than choosing at random from the complete sample. Surprisingly, the fraction of contracts in the EFOS class among the contracts with {\bf RAD}$\geq$0.5 is actually lower than their concentration in the complete source list. 

\begin{figure}[htbp!]
 \centerline{\includegraphics*[width=0.4\textwidth]{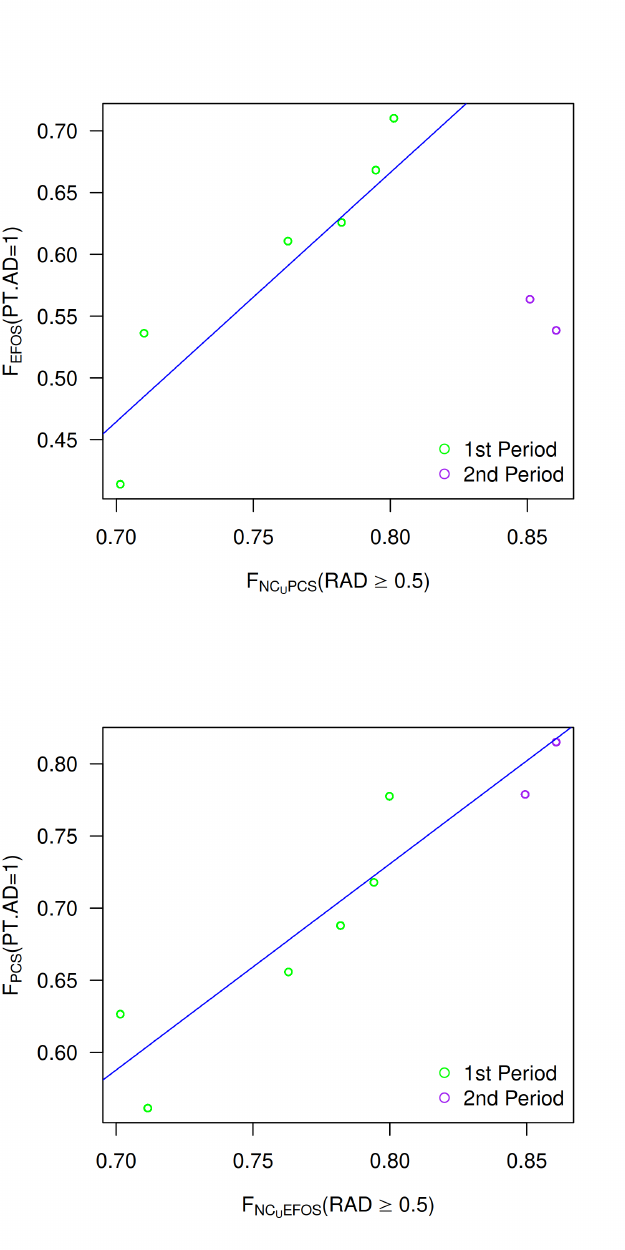}}
\caption{{\it Correlation between {\bf RAD} and corruption risk}. Linear model to correlate the fraction of contracts with {\bf RAD}$\geq$0.5 per year (as independent variable) with the fraction of contracts by single-bidder per year ({\it i.e.} contracts for which {\bf PT.AD}=1), as dependent variable, for the EFOS and PCS classes ({\bf Top} and {\bf Bottom}, respectively). To avoid trivial auto-correlations,  we exclude from the set corresponding to the independent variable the data of the dependent variables. The years corresponding to the 1st period are marked in green, those corresponding to the 2nd period in purple. Since the number of contracts varies widely from one year to another, the data was weighted by the number of contracts in each class per year to perform the least square fit. For the first model (EFOS class) we obtained an $R^2 = 0.72$, an intercept of -0.94, and a slope of 2.01. For the second model (PCS class) we obtained an $R^2 = 0.82$, an intercept of -0.41, and a slope of 1.42. All the parameters with  $p_v\ll 0.01$.}
\label{Corr_PTADvsRAD}
\end{figure}

\begin{figure}[htbp!]
 \centerline{\includegraphics*[width=0.4\textwidth]{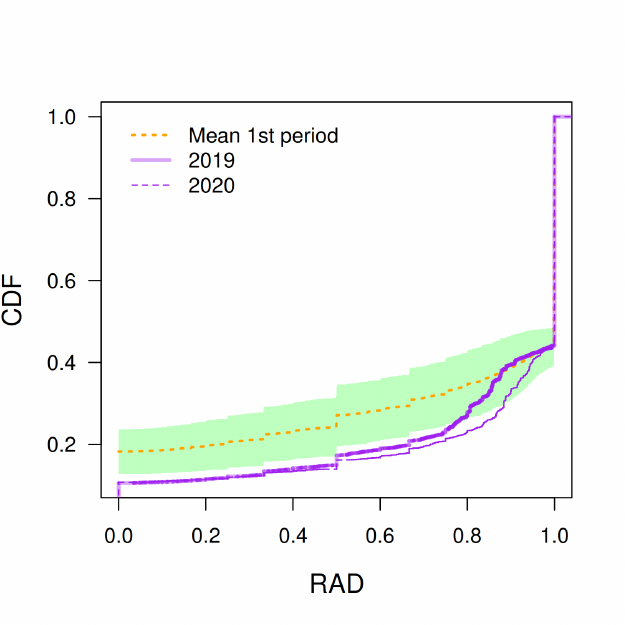}}
\caption{{\it Corruption risk between periods.} CDF for the {\bf RAD} variable. The orange line corresponds to the mean of the 1st government period, the green area represents the CI at 99\% generated by the data of the 1st period, and two purple colored curves correspond to the first two years of the second government period.}
\label{RAD_betwperiods}
\end{figure}

Now, to determine whether the government turnover brought a change of procurement practices associated to corruption, we need to analyse which of the risk factors is correlated to an increase in the fraction of potentially corrupt contracts. To do this, we compute linear models between the total fraction of contracts between a buyer and a supplier for which {\bf RAD}$\geq$ 0.5 per year as independent variable, and the fraction of contracts by single-bidder in each of the corrupt classes per year as the dependent variables. A similar procedure was done with the contracts with {\bf Fav} $\geq$ 0.9 \footnote{To avoid trivial auto-correlations, in all cases we exclude from the set corresponding to the independent variable the data of the dependent variables.}. To take into account the heterogeneity between the number of contracts in each class per year, we weighted each observation by the number of contracts assigned to each class per year. Fig. \ref{Corr_PTADvsRAD} shows the results of these analyses for {\bf RAD} risk factor. We observe that the increase in the fraction of contracts characterized by a {\bf RAD}$\geq$0.5 is linearly correlated with an increase of the fraction of potentially corrupt contracts in both classes, with a $R^2=0.72$ for contracts in the EFOS class and $R^2 = 0.82$ for those in the PCS class. Both $R^2$ are statistically significant with a $p_v\ll 0.01$. For the {\bf Fav} risk factor the analyses show that there is no significant correlation between the fraction of contracts with {\bf Fav}$\geq$0.9 and the fraction of potentially corrupt contracts (Fig. S15 of SM). To confirm that {\bf RAD} variable is the best (among our variables) to predict corruption risk related to EFOS and PCS, we compute several multivariate regression models considering, one by one, variables such as procedure character {\bf PC}, contract type {\bf CT}, supplier's size {\bf S}, and the remaining risk factors. This experiment shows that the impact of {\bf RAD} in the prediction of corruption risk is always significantly larger than the other variables (Tables S9 and S10 of SM). Thus, we conclude that {\bf RAD} is a good indicator of corruption risk, which is in line with previous results found in related works \cite{wachs2020corruption}.

Then, considering the variable {\bf RAD} as the best we have to describe corruption related to EFOS and PCS, and given its correlation with the risk of corruption, we can use the framework used in section 2.2.2 to test how the risk of corruption changed with the government turnover. Fig. \ref{RAD_betwperiods} shows the CDF of the {\bf RAD} variable, separated by periods. We can observe that the {\bf RAD} variable in the 2nd period (purple colored lines) lies outside of the CI of the distributions of the old government for {\bf RAD} values from 0.5 to 0.75, indicating that though less numerous, due to the large amount of contracts assigned to single bidders, the practices in the 2nd period present a higher risk of corruption than in the first period.

\section{Discussion}

M\'exico lived a historical government turnover in 2018 when, for the first time, a leftist candidate won the presidency, being the most voted candidate in M\'exico's democratic history. The new government appeared to represent a complete change of regime, credibly promising to break with the widespread corrupt practices of previous governments. In this work we presented a statistical framework to identify to what extent public procurement practices underwent significant changes due to the government transition, focusing on those contracts related to companies labeled as EFOS (companies that provide invoices for simulated operations) or as PCS (sanctioned suppliers and contractors) due to having incurred in some other kind of corrupt behavior. To do this, we analyzed data from more than 1.5 million government contracts corresponding from 2013 to 2020. We classified each contract in one of three different classes: EFOS or PCS if the contract had been carried out with a company identified as incurring in corrupt practices by the Mexican tax agency (SAT), or Non-Corrupt (NC) if it had not been identified as such. Once classified, we characterized each contract class by a set of variables that give information about the contract itself ({\it i.e.} the contract type, the supplier company's size, the amount spent, etc) or the buyers' features (for example the maximum number of contracts given by the buyer to a single company in a year). We also proposed risk factors constructed considering the framework developed in \cite{fazekas2016objective, wachs2020corruption}. We tested which of these risk factors had the best accuracy to describe and identify the companies labeled as EFOS or PCS, and then compared between government periods to determine whether the risk of corruption decreased, increased or remained the same after the government transition.

Before discussing the detailed comparison between the two governments, a first important difference between them concerns the number of presumably corrupt contracts per year carried out during each government, as well as the total amount of resources spent in these contracts. In the first year of the new government, the number of EFOS contracts fell by a factor of four from the last year of the previous government (or a factor of eight with respect to the average of the six years of the previous government), and another factor of four, to a total of only 13 cases, in the second year. A decrease in the number of EFOS contracts represented a reduction by a factor of three with respect to the previous year (or a factor of approximately 11 with respect to the average of the previous government) in the amount of resources spent in this kind of contracts. 
Also, and representing many more resources, the number of contracts carried out with companies identified as PCS during the first year of the new government was comparable to those of the previous government, however, the amount spent on these contracts during the first year of the new government fell by a factor of two with respect to the last year of the previous government. In the second year, the number of contracts fell by a factor larger than 2, and the resources spent fell by another factor of 4. Thus, the data showed that there has been a significant reduction in both the total number as well as the relative fraction of corrupt contracts, and the corresponding resources spent on such contracts. 
Whether this reduction was due to an effective crackdown on corruption, a consequence of the austerity program undertaken by the new government, or some other reason, is a question the data cannot answer.

However, our purpose in this work was to try to go beyond the analysis of the total amounts of resources spent in contracts that were suspect of being corrupt. We attempted to establish whether the practices and warning flags regarding suspect contracts actually changed from one government period to the other. To do so, using the variables characterizing each contract, we began by verifying that each contract class presented statistically significant differences between them in the same government period. Thus, we were able to assign a rough statistical profile to each class of contracts in each government period. For example, we found that contracts in the EFOS class were characteristically not carried out with large companies, and that this class had a large proportion of contracts for public work in comparison with the PCS and NC classes. On the other hand, the PCS class profile was distinguished by having very few contracts with micro, small, and medium size companies, and that the main activity in this class was for acquisitions. Finally, both corrupt classes were identified by having a higher proportion of buyers with a proportionally stronger market activity and larger budget than the buyers in the NC class. These characteristic statistical profiles were maintained between periods. Having verified that the various classes ---defined by whether or not the supplier had participated in corrupt activities, and if so, what kind of activities--- also had significant statistical differences between them, the next step was to analyze whether, as a consequence of the government transition, the practices in public procurement suffered changes within each class.

Regarding contracts with EFOS, we identified that the new government tended to favor micro-suppliers more than the older government. Companies this size obtained only 20\% of EFOS contracts (on average) in the 1st government period. This percentage grew to 50\% in 2019 and to almost 100\% in 2020. The reduction in the fraction of large EFOS suppliers may be due to government action targeting these companies as they represent a large drain of resources. Also, EFOS contracts' duration suffered a significant change: the proportion of EFOS contracts longer than ten weeks grew from 20\% to 40\% in 2019, and to almost 90\% in 2020. 
On the other hand, there were also similarities between periods for this class of contracts. For example, the percentage of the contracts assigned to single-bidders remained between 55\%-65\% in both periods.

In regards to contracts in the PCS class, these showed differences in the kind of suppliers that were favored, exhibiting a relative increase (from 40\% to 65\%) in the fraction of large companies that won contracts between both government periods. 
Moreover, the fraction of single-bidders in this class also increased from one government to another. On the other hand, the variable {\bf Spending} showed that the way resources were distributed in PCS contracts remained roughly the same between periods.

The NC class presented differences mainly in the contract procedures and in the features of the buyers involved. In the 2nd government period, there was an increase of $\sim$ 20\% of contracts made under NAFTA procedures compared to the 1st period. We also found that the kind of buyers that participated in the contracts in this class, tend to spend more money with a single company in the 2nd period than in the 1st period. 

Our framework to analyze a set of contracts through certain characteristic variables allowed us to identify specific changes and similarities between contract classes and between government periods. But this approach did not tell us whether the risk of corruption increased or decreased after the government transition. To do this we considered specific indicators, namely {\bf RAD} (fraction of single-bidder contracts), {\bf Fav} (favoritism), {\bf CPW} (contracts per active week) and {\bf SPW} (spending per active week), based on previously proposed risk factors of corruption \cite{fazekas2013corruption, fazekas2013anatomy, IMCOmapeando}. These can be used as red flags for those contracts for which these variables exceed a certain threshold. For example, a buyer-supplier relation (and all contracts in it) should be red flagged as risky if, say, {\bf RAD}$\geq$0.5 \cite{wachs2020corruption}, or if {\bf Fav}$\geq$0.9 \cite{IMCOmapeando}. Our study showed that for the EFOS and PCS classes, the risk factor {\bf Fav} performed well identifying corrupt contracts among those that exceeded the threshold of 0.9. Actually, it was twice as probable to find EFOS contracts in the subset of contracts with {\bf Fav}$\geq$0.9, than in the whole source list, and almost five times more probable to find PCS contracts. However, it had little accuracy as a descriptor since {\bf Fav} exceeded the threshold 0.9 in less than 1\% of companies labeled as having participated in corrupt activities. 
The variables {\bf CPW} and {\bf SPW} were not useful risk factors, since only a few EFOS and PCS contracts had more than five contracts per active week and expenditures for more than 10K USD PPP per active week, thus we were unable to identify the corruption scheme in which a buyer assigned many small contracts in a short period of time to the same supplier \cite{IMCOmapeando}. This may also suggest an important limitation of the effectiveness of these variables to predict corruption with the available data. Actually, having access to more specific information about the contracts and their participants, would be helpful to re-compute these risk factors and to re-test their effectiveness to describe and identify corruption.

On the other hand, the risk factor {\bf RAD} was accurate in describing (on average) more than 50\% and 80\% of the companies labeled as EFOS and PCS, respectively. However, this result also implies that many corrupt contracts were won in open contest. This suggests that in our corrupt classes there may be two kinds of illegal schemes; one in which public officials colluded with the supplier and frequently assigned them contracts as single-bidders ({\it i.e.} {\bf RAD} $\geq$ 0.5), and another scheme in which the contracted companies committed a fraudulent action without the government being involved. This second scheme is likely to occur in EFOS and PCS contracts that are mostly assigned through open contest (so {\bf RAD} $<$ 0.5). Here we focus in schemes of the first kind, as corruption implies government complicity \cite{transparencycorruption}. 
We also found that the {\bf RAD} variable was not efficient in identifying corrupt contracts among those with {\bf RAD}$\geq$0.5. Actually, the probability to find a PCS contract in this subset is the same as the probability to find it choosing randomly in the complete source list, and the probability to find an EFOS contract among those with {\bf RAD}$\geq$0.5 is significantly less than for the random search in the complete source list. This result is explained by the fact that the single-bidder contracts are very common in the NC class (Fig.\ref{Test_RF} {\bf Right}), as common as in the PCS class and even more common than in the EFOS class. Thus, even when the single-bidder contracts should be only exceptional \cite{constitucion, leyarrendamiento, leyobra}, they have become an extremely common procedure in M\'exico's public procurement practices. 
Further, in line with previous findings \cite{wachs2020corruption}, we found that the {\bf RAD} variable is positively correlated with an increase of the risk of corruption.
Given this correlation,  the picture that emerges is that even though the absolute number of corrupt contracts has been reduced, the risk of corruption related to EFOS and PCS contracts has not decreased with the new government, since the fraction of single-bidder contracts, {\bf RAD}, actually increased from one government to another.

Overall, our study provides a framework to identify the properties and behavior of corruption in public procurement, as well as to evaluate how corrupt practices change during a government turnover. We used these tools to analyze the impact of government transition in M\'exico in the practices of public procurement. The methodology may be a stepping stone to build new methods involving the analysis of the variables displaying differences between EFOS, PCS and NC, in order to better describe and identify corrupt contracts among big data sets.

\section{Acknowledgements}
AFC thanks PostDoctoral Scholarship DGAPA-UNAM for financial support. AFC thanks to D Cervantes-Filoteo and JR Nicol\'as-Carlock for fruitful discussion.

\end{document}


\section{\large{Supplementary Material - Practices of public procurement and the risk of corrupt behavior before and after the government transition in M\'exico}}

\vspace{1.5cm}

\begin{table*}[htb!]
\begin{tabular}{l|clll|clll|clll}
\hline \hline
                     & \multicolumn{4}{c|}{\textbf{EFOS}}                                                                     & \multicolumn{4}{c|}{\textbf{PCS}}                                                                      & \multicolumn{4}{c}{\textbf{NC}}                                                              \\
                     \hline
Variable (Dummy)     & \multicolumn{4}{c|}{Fraction}                                                                          & \multicolumn{4}{c|}{Fraction}                                                                          & \multicolumn{4}{c}{Fraction}                                                                 \\
\hline
GO.APF               & \multicolumn{4}{c|}{0.6483}                                                                            & \multicolumn{4}{c|}{0.9623}                                                                            & \multicolumn{4}{c}{0.8942}                                                                   \\
GO.GE                & \multicolumn{4}{c|}{0.2335}                                                                            & \multicolumn{4}{c|}{0.0349}                                                                            & \multicolumn{4}{c}{0.0731}                                                                   \\
GO.GM                & \multicolumn{4}{c|}{0.1182}                                                                            & \multicolumn{4}{c|}{0.0027}                                                                            & \multicolumn{4}{c}{0.0326}                                                                   \\
PC.N                 & \multicolumn{4}{c|}{0.9229}                                                                            & \multicolumn{4}{c|}{0.4651}                                                                            & \multicolumn{4}{c}{0.8498}                                                                   \\
PC.I                 & \multicolumn{4}{c|}{0.0630}                                                                            & \multicolumn{4}{c|}{0.2852}                                                                            & \multicolumn{4}{c}{0.0857}                                                                   \\
PC.ITLC              & \multicolumn{4}{c|}{0.0141}                                                                            & \multicolumn{4}{c|}{0.2497}                                                                            & \multicolumn{4}{c}{0.0644}                                                                   \\
CT.OP                & \multicolumn{4}{c|}{0.2572}                                                                            & \multicolumn{4}{c|}{0.0202}                                                                            & \multicolumn{4}{c}{0.1017}                                                                   \\
CT.S                 & \multicolumn{4}{c|}{0.321}                                                                             & \multicolumn{4}{c|}{0.117}                                                                             & \multicolumn{4}{c}{0.3444}                                                                   \\
CT.ADQ               & \multicolumn{4}{c|}{0.3737}                                                                            & \multicolumn{4}{c|}{0.8469}                                                                            & \multicolumn{4}{c}{0.5285}                                                                   \\
CT.AR                & \multicolumn{4}{c|}{0.0228}                                                                            & \multicolumn{4}{c|}{0.0108}                                                                            & \multicolumn{4}{c}{0.0056}                                                                   \\
CT.SLAOP             & \multicolumn{4}{c|}{0.0253}                                                                            & \multicolumn{4}{c|}{0.0049}                                                                            & \multicolumn{4}{c}{0.0197}                                                                   \\
PT.AD                & \multicolumn{4}{c|}{0.5786}                                                                            & \multicolumn{4}{c|}{0.6994}                                                                            & \multicolumn{4}{c}{0.748}                                                                    \\
PT.I3P               & \multicolumn{4}{c|}{0.2173}                                                                            & \multicolumn{4}{c|}{0.0844}                                                                            & \multicolumn{4}{c}{0.1078}                                                                   \\
PT.LP                & \multicolumn{4}{c|}{0.2041}                                                                            & \multicolumn{4}{c|}{0.2161}                                                                            & \multicolumn{4}{c}{0.1442}                                                                   \\
S.NOM                & \multicolumn{4}{c|}{0.0319}                                                                            & \multicolumn{4}{c|}{0.4979}                                                                            & \multicolumn{4}{c}{0.1069}                                                                   \\
S.MED                & \multicolumn{4}{c|}{0.2107}                                                                            & \multicolumn{4}{c|}{0.0714}                                                                            & \multicolumn{4}{c}{0.1448}                                                                   \\
S.PEQ                & \multicolumn{4}{c|}{0.275}                                                                             & \multicolumn{4}{c|}{0.1737}                                                                            & \multicolumn{4}{c}{0.2121}                                                                   \\
S.MIC                & \multicolumn{4}{c|}{0.2082}                                                                            & \multicolumn{4}{c|}{0.1882}                                                                            & \multicolumn{4}{c}{0.2214}                                                                   \\
S.NA                 & \multicolumn{4}{c|}{0.2742}                                                                            & \multicolumn{4}{c|}{0.0687}                                                                            & \multicolumn{4}{c}{0.3149}                                                                   \\
\hline
Variable (Non-Dummy) & \multicolumn{1}{|l}{Mean}               & SD                 & 1st Qu.            & 3rd Qu.            & \multicolumn{1}{|l}{Mean}               & SD                 & 1st Qu.            & 3rd Qu.            & \multicolumn{1}{|l}{Mean}               & SD                 & 1st Qu.   & 3rd Qu.            \\
\hline
BeginningWeek        & \multicolumn{1}{l}{29.72}              & 14.58              & 18.00              & 42.00              & \multicolumn{1}{l}{22.89}              & 15.06              & 9.00               & 36.00              & \multicolumn{1}{l}{25.18}              & 14.72              & 13.00     & 38.00              \\
EBWeeks              & \multicolumn{1}{l}{13.57}              & 18.89              & 2.00               & 17.00              & \multicolumn{1}{l}{17.36}              & 24.18              & 1.00               & 32.00              & \multicolumn{1}{l}{15.51}              & 22.65              & 1.00      & 23.00              \\
Spending             & \multicolumn{1}{l}{$2.9\cdot 10^{5}$}  & 928,311            & 10,059             & $1.8\cdot 10^{5}$  & \multicolumn{1}{l}{$1.3 \cdot 10^{7}$} & $1.4 \cdot 10^{7}$ & 2,616              & 48,894             & \multicolumn{1}{l}{$3.1 \cdot 10^{5}$} & $1.3 \cdot 10^{7}$ & 4,760     & 55,370             \\
T.Cont.Max           & \multicolumn{1}{l}{32.73}              & 165.98             & 4.00               & 14.75              & \multicolumn{1}{l}{26.17}              & 163.08             & 3.00               & 12.00              & \multicolumn{1}{l}{7.34}               & 71.76              & 1.00      & 4.00               \\
T.Spending.Max       & \multicolumn{1}{l}{$7.9 \cdot 10^{7}$} & $2.9 \cdot 10^{8}$ & $1.4 \cdot 10^{6}$ & $2.7 \cdot 10^{7}$ & \multicolumn{1}{l}{$6.4 \cdot 10^{7}$} & $5.0 \cdot 10^{8}$ & $1.5 \cdot 10^{6}$ & $2.6 \cdot 10^{7}$ & \multicolumn{1}{l}{$1.6 \cdot 10^{7}$} & $2.2 \cdot 10^{8}$ & $256,300$ & $2.6 \cdot 10^{6}$ \\
CPW                  & \multicolumn{1}{l}{1.59}               & 1.73               & 1.00               & 2.00               & \multicolumn{1}{l}{5.79}               & 7.64               & 1.00               & 6.00               & \multicolumn{1}{l}{3.45}               & 6.70               & 1.00      & 2.00               \\
SPW                  & \multicolumn{1}{l}{$3.9\cdot 10^{5}$}  & $1.2 \cdot 10^{6}$ & 18,688             & $2.7\cdot 10^{5}$  & \multicolumn{1}{l}{$8.6 \cdot 10^{6}$} & $1.7 \cdot 10^{7}$ & 31,587             & $1.4 \cdot 10^{7}$ & \multicolumn{1}{l}{$5.7 \cdot 10^{5}$} & $2.4 \cdot 10^7$   & 12,180    & $1.9 \cdot 10^{5}$ \\
Fav                  & \multicolumn{1}{l}{0.15}               & 0.21               & 0.02               & 0.19               & \multicolumn{1}{l}{0.28}               & 0.32               & 0.03               & 0.50               & \multicolumn{1}{l}{0.12}               & 0.17               & 0.01      & 0.14               \\
RAD                  & \multicolumn{1}{l}{0.57}               & 0.46               & 0.00               & 1.00               & \multicolumn{1}{l}{0.69}               & 0.34               & 0.52               & 0.94               & \multicolumn{1}{l}{0.74}               & 0.38               & 0.53      & 1.00       \\
\hline \hline
\end{tabular}
\caption{Descriptive statistics for all the variables types (i,ii,and iii) without making distinction between government periods. For dummy variables, we compute the fraction of contracts in which the variable is equal to one; this is equivalent to the variable's mean. For non-dummy variables, the table shows the mean, the standard deviation (SD), and the 1st and 3rd quartiles.}
\end{table*}

\begin{table*}[htb!]
\begin{tabular}{lllll}
\hline \hline
{\bf Government Period}            & {\bf Variables}  & {\bf EFOS}  & {\bf PCS}    & {\bf NC}        \\
\hline
                             &  &  &   &    \\
      & Type i)   & 2,343 & 26,031 & 1,190,638 \\
      & Type ii)  & 371   & 450    & 2,417     \\
      & Type iii)   & 2,343 & 26,031 & 1,190,638 \\
      &           &       &        & \\ 
      \hline
\multirow{-7}{*}{1st Period} &  &  &   &    \\
      & Type i)   & 68    & 5,052  & 316,254   \\
     & Type ii)  & 27    & 173    & 330       \\
     & Type iii)   & 68    & 5,052  & 316,254   \\
\multirow{-5}{*}{2nd Period} &  &     &     &    \\
\hline \hline
\end{tabular}
\caption{Sample size for each variable type in each class for both government periods.}
\end{table*}

\begin{table*}[htb!]
\begin{tabular}{llllllllll}
\hline \hline
\multicolumn{5}{c}{\textbf{Dummy Variables}}                                                                              &  &  & \multicolumn{3}{c}{\textbf{Non Dummy Variables}}                                                      \\
\hline
Type & Variable & $f_{EFOS}$ & $f_{PCS}$ & $|z|$ &  &  & Type & Variable      & $D$    \\
\hline
i)                           & CT.ADQ                           & 0.3713                             & 0.8424                            & 54.97                         &  &  & i)                           & Spending                              & 0.2365                         \\
     & PC.N     & 0.9253     & 0.4846    & 40.89 &  &  &      & BeginningWeek & 0.1957 \\
                             & S.NOM                            & 0.0328                             & 0.4712                            & 40.99                         &  &  &                              & EBWeeks                               & 0.1553                         \\
     & GO.APF   & 0.6448     & 0.9576    & 57.51 &  &  &      &     & \\
                             & PC.I                             & 0.0606                             & 0.3011                            & 24.80                         &  &  &                           &                                   &                          \\
     & CT.OP    & 0.2582     & 0.0230    & 54.09 &  &  &      &   &  \\
                             & CT.S                             & 0.3218                             & 0.1189                            & 27.46                         &  &  &                              &                                 &                       \\
     & S.NA     & 0.2787     & 0.0782    & 31.71 &  &  &                              &                                       &                                \\
                             & PC.ITLC                          & 0.0140                             & 0.2141                            & 23.29                         &  &  &                              &                                       &                                \\
     & GO.GE    & 0.2338     & 0.0393    & 39.42 &  &  &                              &                                       &                                \\
                             & S.MED                            & 0.2129                             & 0.0703                            & 24.09                         &  &  &                              &                                       &                                \\
     & PT.I3P   & 0.2176     & 0.0843    & 21.03 &  &  &                              &                                       &                                \\
                             & GO.GM                            & 0.1212                             & 0.0030                            & 48.75                         &  &  &                              &                                       &                                \\
     & PT.AD    & 0.5791     & 0.6820    & 10.17  &  &  &                              &                                       & \\
\hline \hline
\end{tabular}

\caption{{\it B-Test and KS-Test results for the 1st period (EFOS vs PCS):} Set of Dummy and Non Dummy Variables for which were significant differences between EFOS and PCS classes. We specified the variable type (Type), its name (Variable), for dummy variables: the fraction in EFOS class ($f_{EFOS}$), the fraction in PCS class ($f_{PCS}$), and z-score ($|z|$) of the B-Test, and for non dummy variables: the statistic $D$ of the KS-Test. All $p_v\ll 0.01$.}
\end{table*}

\begin{table*}[htb!]
\begin{tabular}{llllllllll}
\hline \hline
\multicolumn{5}{c}{\textbf{Dummy Variables}}                                                                                                                             &  &  & \multicolumn{3}{c}{\textbf{Non Dummy Variables}}                                                 \\
\hline
Type & Variable & $f_{EFOS}$ & $f_{PCS}$ & $|z|$ &  &  & Type & Variable & $D$    \\
\hline
i)                           & S.NOM                            & 0.0000                             & 0.6349                            & 10.75                         &  &  & i)                           & Spending                         & 0.3736                         \\
     & PC.N     & 0.8382     & 0.3640    & 8.04  &  &  &      & EBWeeks  & 0.3139 \\
                             & PC.ITLC                          & 0.0147                             & 0.4330                            & 6.92                          &  &  &                              & &                         \\
     & S.MIC    & 0.6029     & 0.1854    & 8.69  &  &  &                              &                                  &                                \\
                             & CT.ADQ                           & 0.4558                             & 0.8695                            & 9.88                          &  &  &                              &                                  &                                \\
     & PT.AD    & 0.5588     & 0.7889    & 4.59  &  &  &                              &                                  &                                \\
                             & GO.APF                           & 0.7647                             & 0.9868                            & 14.20                         &  &  &                              &                                  &                                \\
     & CT.OP    & 0.2205     & 0.0059    & 18.83 &  &  &                              &                                  &                                \\
                             & GO.GE                            & 0.2205                             & 0.0124                            & 13.91                         &  &  &                              &                                  &                                \\
     & CT.S     & 0.2941     & 0.1072    & 4.89  &  &  &                              &                                  &        \\
     & PT.I3P     & 0.2058     & 0.084    & 3.53  &  &  &                              &                                  &        \\
     & PT.LP     & 0.2352     & 0.1262    & 2.67  &  &  &                              &                                  &        \\
\hline \hline
\end{tabular}
\caption{{\it B-Test and KS-Test results for the 2nd period (EFOS vs PCS):} Set of Dummy and Non Dummy Variables for which were significant differences between EFOS and PCS classes. We specified the variable type (Type), its name (Variable), for dummy variables: the fraction in EFOS class ($f_{EFOS}$), the fraction in PCS class ($f_{PCS}$), and z-score ($|z|$) of the B-Test, and for non dummy variables: the statistic $D$ of the KS-Test. All $p_v\ll 0.01$.}
\end{table*}

\begin{table*}[htb!]
\begin{tabular}{llllllllll}
\hline \hline
\multicolumn{5}{c}{\textbf{Dummy Variables}}                                                                                                                            &  &  & \multicolumn{3}{c}{\textbf{Non Dummy Variables}}                                                       \\
\hline
Type & Variable & $f_{EFOS}$ & $f_{NC}$ & $|z|$ &  &  & Type & Variable       & $D$    \\
\hline
i)                           & GO.APF                           & 0.6448                             & 0.8809                           & 35.18                         &  &  & i)                           & Spending                               & 0.1850                         \\
     & GO.GE    & 0.2338     & 0.0805   & 27.18 &  &  &      & BeginningWeek  & 0.1500 \\
                             & PT.AD                            & 0.5791                             & 0.7318                           & 16.66                         &  &  & ii)                          & T.Cont.Max                                    & 0.4467                       \\
     & CT.ADQ   & 0.3713     & 0.5141   & 13.82 &  &  &      & T.Sepending.Max    & 0.4190 \\
                             & CT.OP                            & 0.2582                             & 0.1167                           & 21.28                         &  &  &                              &                                    &                       \\
     & PT.I3P   & 0.2176     & 0.1166   & 15.20 &  &  &      &           &  \\
\hline\hline                             
\end{tabular}
\caption{{\it B-Test and KS-Test results for the 1st period (EFOS vs NC):} Set of Dummy and Non Dummy Variables for which were significant differences between EFOS and NC classes. We specified the variable type (Type), its name (Variable), for dummy variables: the fraction in EFOS class ($f_{EFOS}$), the fraction in NC class ($f_{NC}$), and z-score ($|z|$) of the B-Test, and for non dummy variables: the statistic $D$ of the KS-Test. All $p_v\ll 0.01$.}
\end{table*}

\begin{table*}[hbt!]
\begin{tabular}{llllllllll}
\hline \hline
\multicolumn{5}{c}{\textbf{Dummy Variables}}                                                                                                                            &  &  & \multicolumn{3}{c}{\textbf{Non Dummy Variables}}                                                       \\
\hline
Type & Variable & $f_{EFOS}$ & $f_{NC}$ & $|z|$ &  &  & Type & Variable       & $D$    \\
\hline
i)                           & S.MIC                            & 0.6029                             & 0.2682                           & 6.22                          &  &  & i)                           & Spending                               & 0.3560                         \\
     & PT.AD    & 0.5588     & 0.8088   & 5.24  &  &  &      & EBWeeks        & 0.3048 \\
                             & GO.APF                           & 0.7647                             & 0.9440                           & 6.42                          &  &  &                              & BeginningWeek                            & 0.1886                         \\
     & GO.GE    & 0.2205     & 0.0450   & 6.97  &  &  &   ii)   & T.Cont.Max  & 0.3643 \\
                             & CT.OP                            & 0.2205                             & 0.0451                           & 6.96                          &  &  &                           & T.Spending.Max                                   & 0.3498                          \\
                             & PC.ITLC                            & 0.014                             & 0.1720                           & 3.43                          &  &  &                           &                                    &                           \\
        & S.NOM                            & 0.000                             & 0.1417                           & 3.35                          &  &  &                           &                                    &                           \\
        & S.NA                           & 0.1176                             & 0.2569                           & 2.62                          &  &  &                           &                                    &                           \\
        & PT.I3P                            & 0.2058                             & 0.074                           & 4.13                          &  &  &                           &                                    &                           \\
        & CT.ADQ                            & 0.4558                             & 0.5824                           & 2.11                          &  &  &                           &                                    &                           \\
        & PT.LP                            & 0.2352                             & 0.1167                           & 3.04                          &  &  &                           &                                    &                           \\

\hline \hline                             
\end{tabular}
\caption{{\it B-Test and KS-Test results for the 2nd period (EFOS vs NC):} Set of Dummy and Non Dummy Variables for which were significant differences between EFOS and NC classes. We specified the variable type (Type), its name (Variable), for dummy variables: the fraction in EFOS class ($f_{EFOS}$), the fraction in NC class ($f_{NC}$), and z-score ($|z|$) of the B-Test, and for non dummy variables: the statistic $D$ of the KS-Test. All $p_v\ll 0.01$.}
\end{table*}

\begin{table*}[htb!]
\begin{tabular}{llllllllll}
\hline \hline
\multicolumn{5}{c}{\textbf{Dummy Variables}}                                                                                                                            &  &  & \multicolumn{3}{c}{\textbf{Non Dummy Variables}}                                                   \\
\hline
Type & Variable & $f_{PCS}$ & $f_{NC}$ & $|z|$  &  &  & Type & Variable   & $D$    \\
\hline
i)                           & PC.N                             & 0.4846                            & 0.8767                           & 185.13                         &  &  & i)                           & Spending                           & 0.1214                         \\
     & S.NOM    & 0.4712    & 0.0975   & 194.06 &  &  & ii)  & T.Cont.Max        & 0.4412 \\
                             & CT.ADQ                           & 0.8424                            & 0.5141                           & 104.88                         &  &  &                              & T.Spending.Max                               & 0.4258                         \\
     & S.NA     & 0.0782    & 0.3302   & 85.86  &  &  &      &  &  \\
                             & CT.S                             & 0.1189                            & 0.3418                           & 75.27                          &  &  &                              &                                 &                        \\
     & PC.I     & 0.3011    & 0.0873   & 118.09 &  &  &  &  &  \\
                             & PC.ITLC                          & 0.2141                            & 0.0358                           & 145.84                         &  &  &                              &                     & \\
\hline\hline                             
\end{tabular}
\caption{{\it B-Test and KS-Test results for the 1st period (PCS vs NC):} Set of Dummy and Non Dummy Variables for which ere significant differences between PCS and NC classes. We specified the variable type (Type), its name (Variable), for dummy variables: the fraction in PCS class ($f_{PCS}$), the fraction in NC class ($f_{NC}$), and z-score ($|z|$) of the B-Test, and for non dummy variables: the statistic $D$ of the KS-Test. All $p_v\ll 0.01$.}
\end{table*}

\begin{table*}[htb!]
\begin{tabular}{llllllllll}
\hline\hline
\multicolumn{5}{c}{\textbf{Dummy Variables}}                                                                                                                           &  &  & \multicolumn{3}{c}{\textbf{Non Dummy Variables}}                                                       \\
\hline
Type & Variable & $f_{PCS}$ & $f_{NC}$ & $|z|$ &  &  & Type & Variable       & $D$    \\
\hline
i)                           & S.NOM                            & 0.6349                            & 0.1417                           & 97.52                         &  &  & i)                           & BeginningWeek                          & 0.1661                         \\
     & PC.N     & 0.3640    & 0.7482   & 61.95 &  &  &      & Spending       & 0.1417 \\
                             & CT.ADQ                           & 0.8695                            & 0.5824                           & 41.12                         &  &  &                              & EBWeeks                                & 0.1166                         \\
     & PC.ITLC  & 0.4330    & 0.1720   & 48.32 &  &  & ii)  & T.Spending.Max    & 0.2045 \\
                             & CT.S                             & 0.1072                            & 0.3541                           & 36.48                         &  &  &                              & T.Cont.Max                                    & 0.1900                         \\
     & S.NA     & 0.0197    & 0.2569   & 38.45 &  &  &     &           &  \\
                             & S.PEQ                            & 0.0825                            & 0.2060                           & 21.60                         &  &  &                              &                                  &                        \\
     & PC.I     & 0.2028    & 0.0796   & 31.74 &  &  &      &     &  \\
\hline\hline                             
\end{tabular}
\caption{{\it B-Test and KS-Test results for the 2nd period (PCS vs NC):} Set of Dummy and Non Dummy Variables for which were significant differences between PCS and NC classes. We specified the variable type (Type), its name (Variable), for dummy variables: the fraction in PCS class ($f_{PCS}$), the fraction in NC class ($f_{NC}$), and z-score ($|z|$) of the B-Test, and for non dummy variables: the statistic $D$ of the KS-Test. All $p_v\ll 0.01$.}
\end{table*}

\begin{figure*}[htb!]
\begin{center}
\includegraphics*[width=0.275\textwidth]{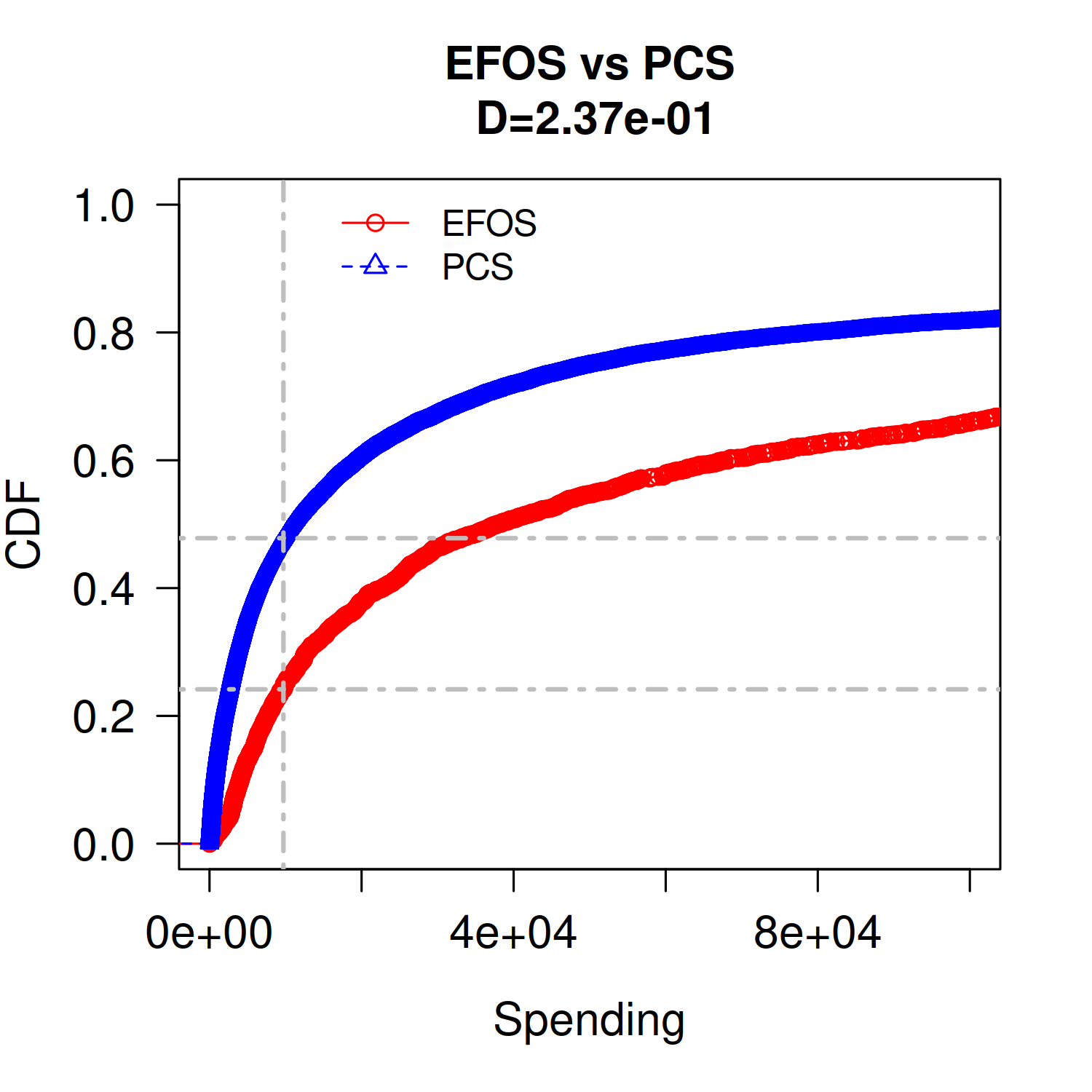}
\includegraphics*[width=0.275\textwidth]{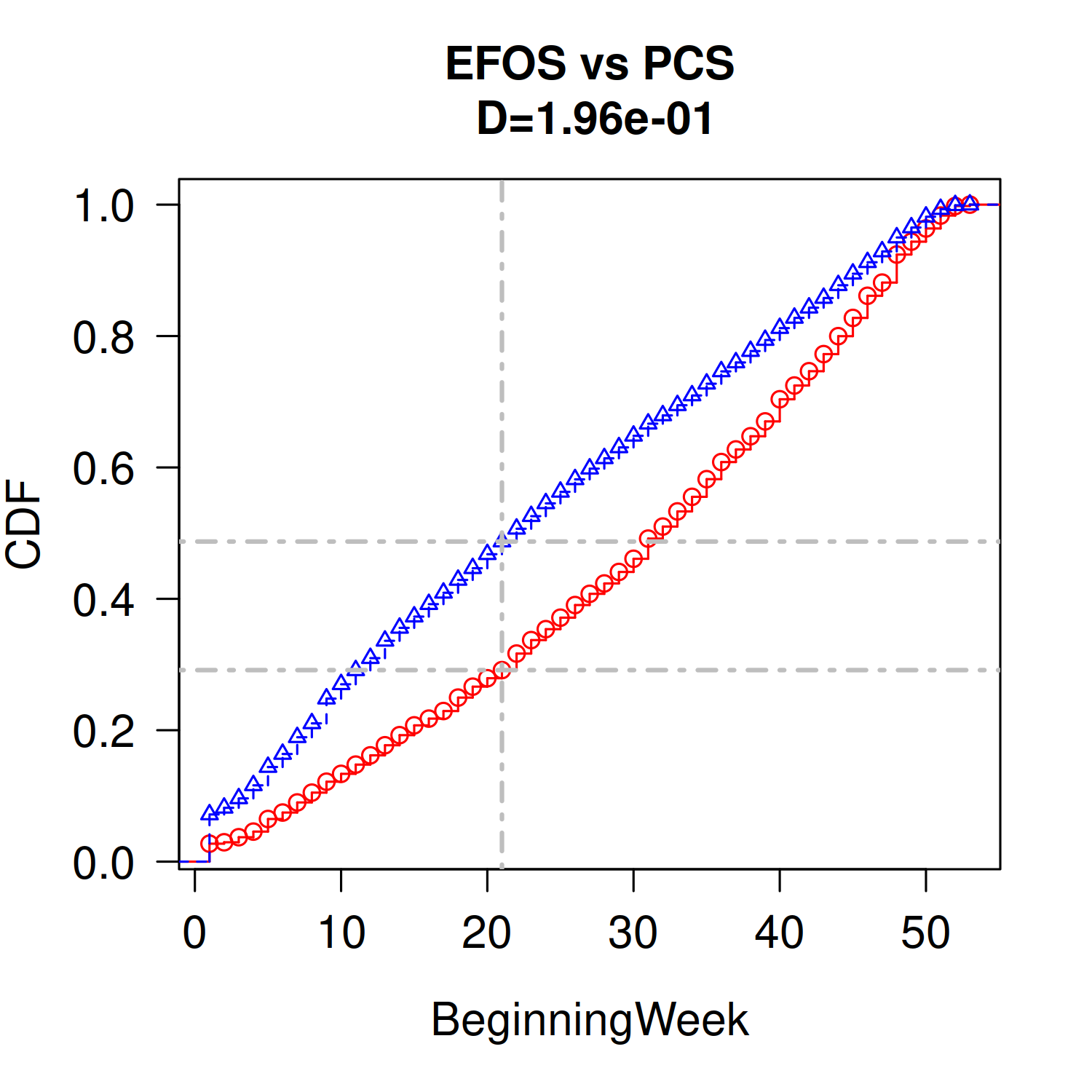}
\includegraphics*[width=0.275\textwidth]{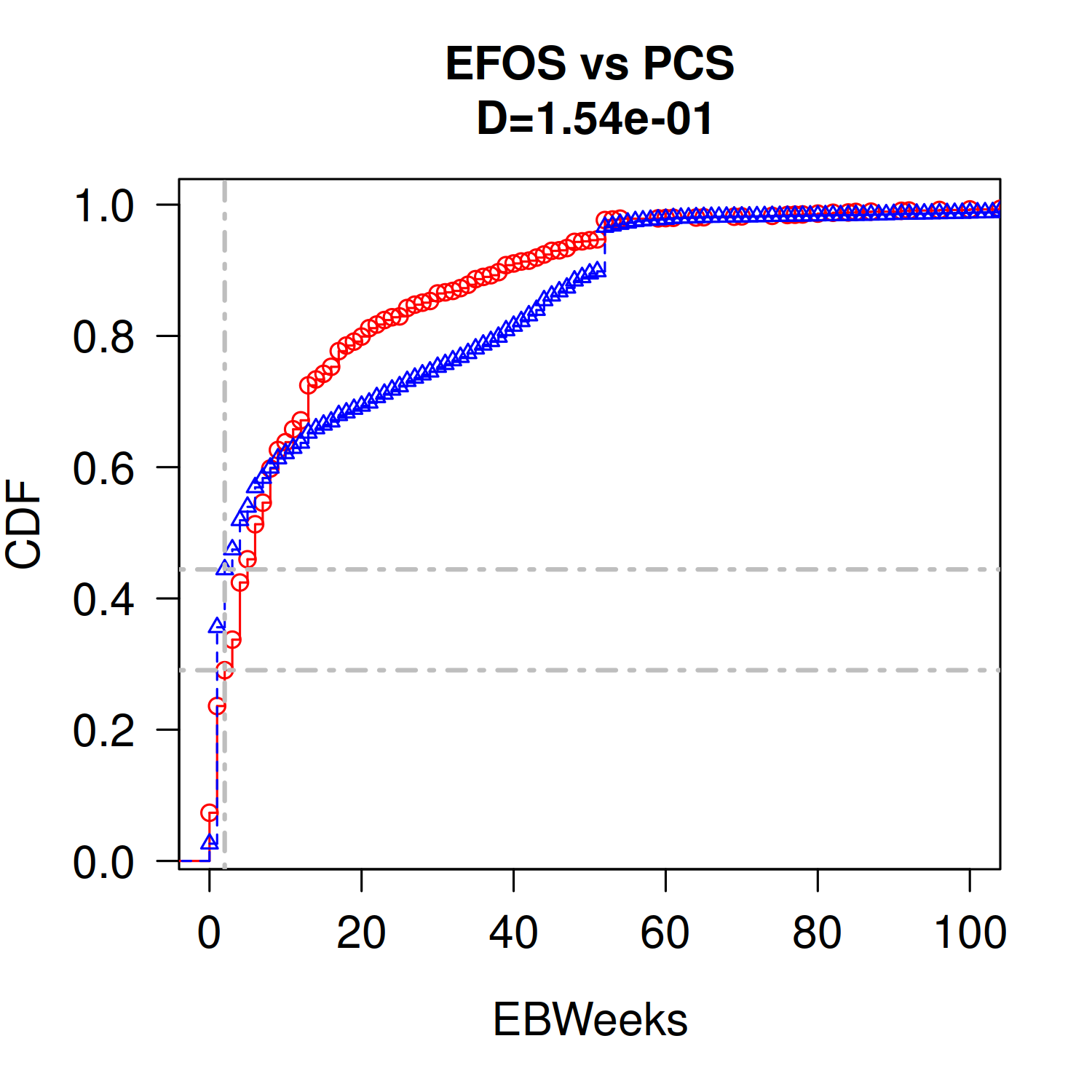}
\end{center}

    \caption{{\it KS-Test results for the 1st period (EFOS vs PCS):} The graph shows the cumulative distribution function (CDF) for each significant non-dummy variable in EFOS class (red circles solid line) and PCS class (blue triangles dashed line). The grey dashed line shows the value from which the CDF's had the maximum difference. In all graphs the scale of the $X$ axis corresponds to that for which the statistic $D$ is clearly visible.}
    \label{kst-1st-ep}
\end{figure*}

\begin{figure*}[htb!]

\begin{center}
\includegraphics*[width=0.3\textwidth]{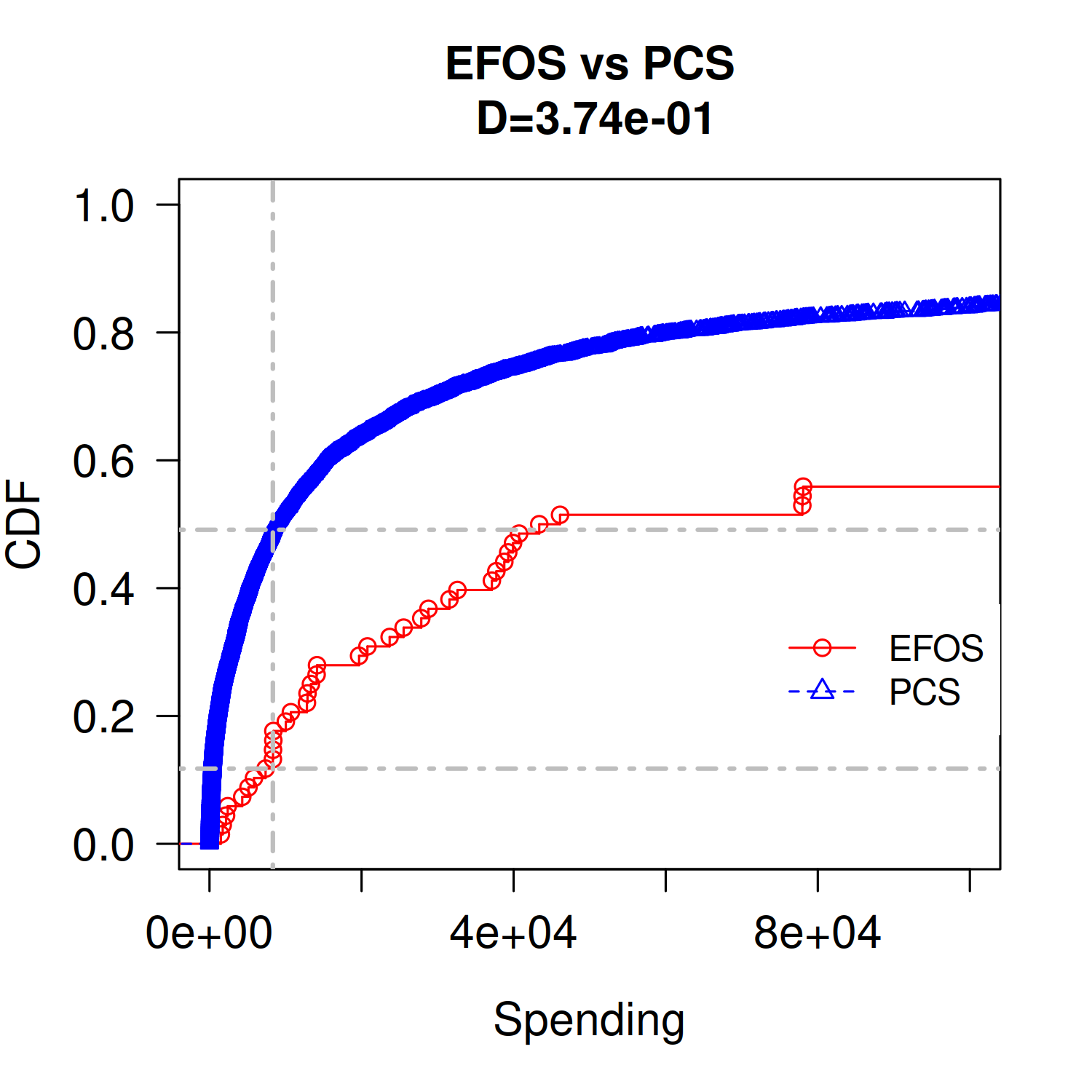}
\includegraphics*[width=0.3\textwidth]{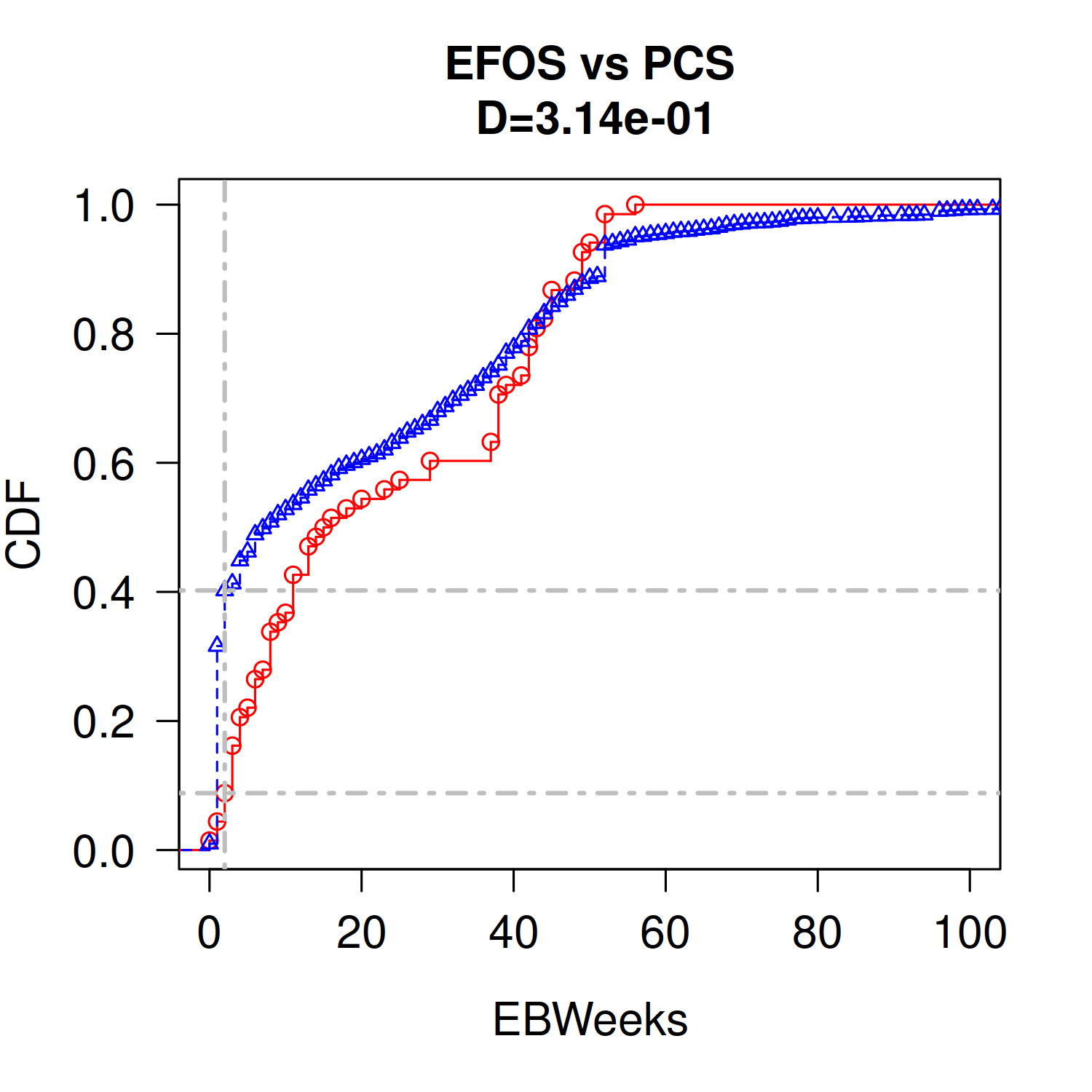}
\end{center}
    \caption{{\it KS-Test results for the 2nd period (EFOS vs PCS):} The graph shows the cumulative distribution function (CDF) for each significant non-dummy variable in EFOS class (red circles solid line) and PCS class (blue triangles dashed line). The grey dashed line shows the value from which the CDF's had the maximum difference. In all graphs the scale of the $X$ axis corresponds to that for which the statistic $D$ is clearly visible.}
    \label{kst-2nd-ep}
\end{figure*}

\begin{figure*}[htb!]
\begin{center}
\includegraphics*[width=0.225\textwidth]{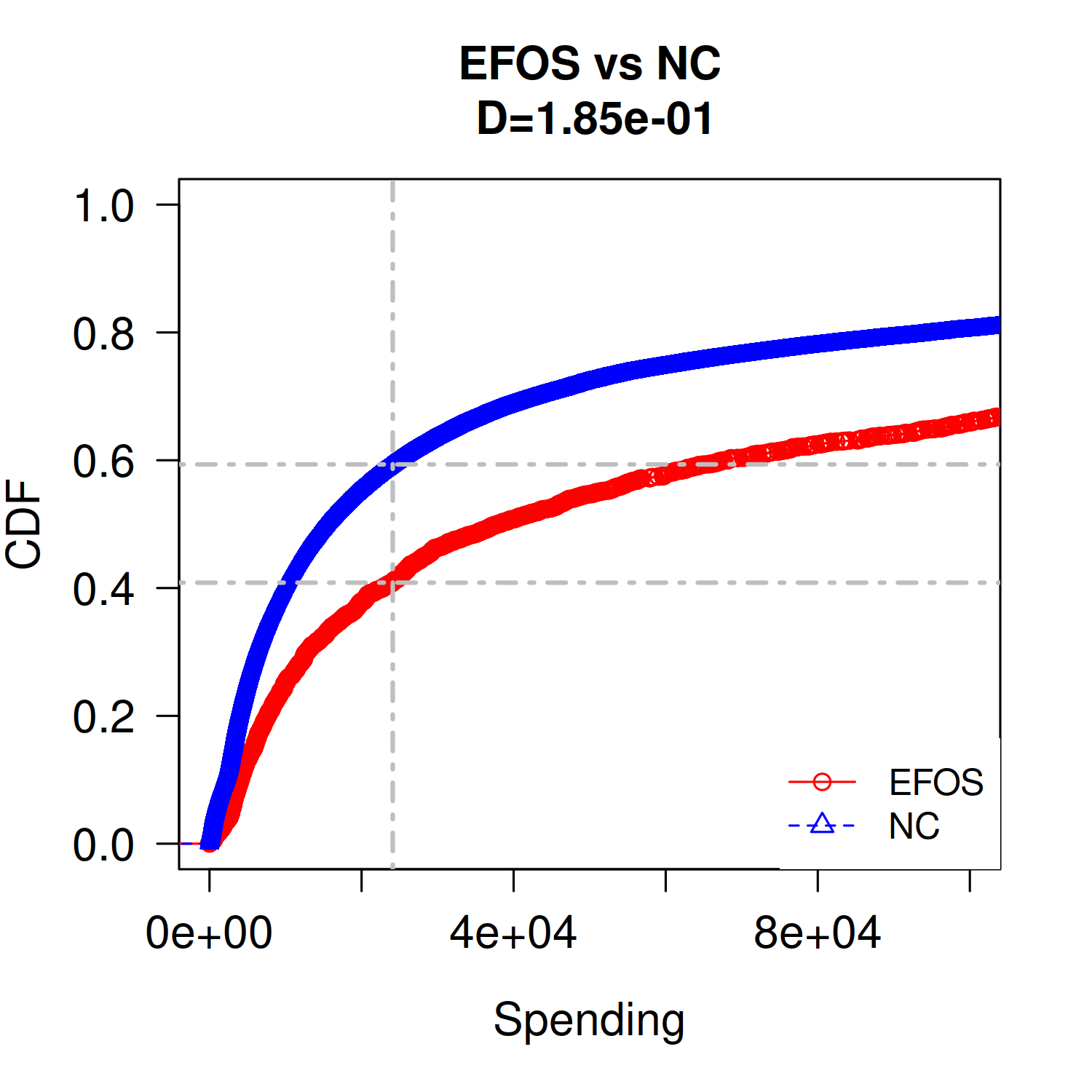}
\includegraphics*[width=0.225\textwidth]{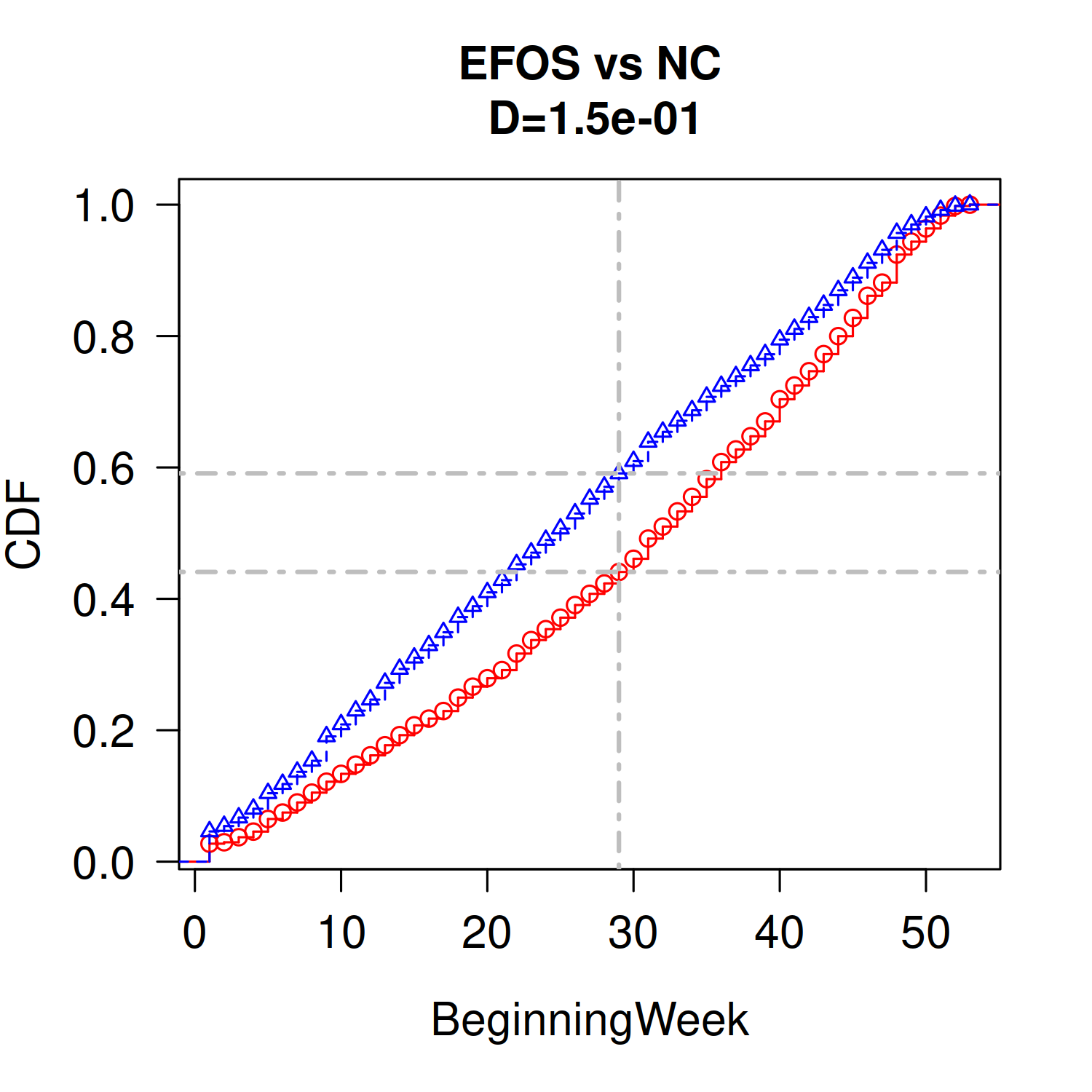}
\includegraphics*[width=0.225\textwidth]{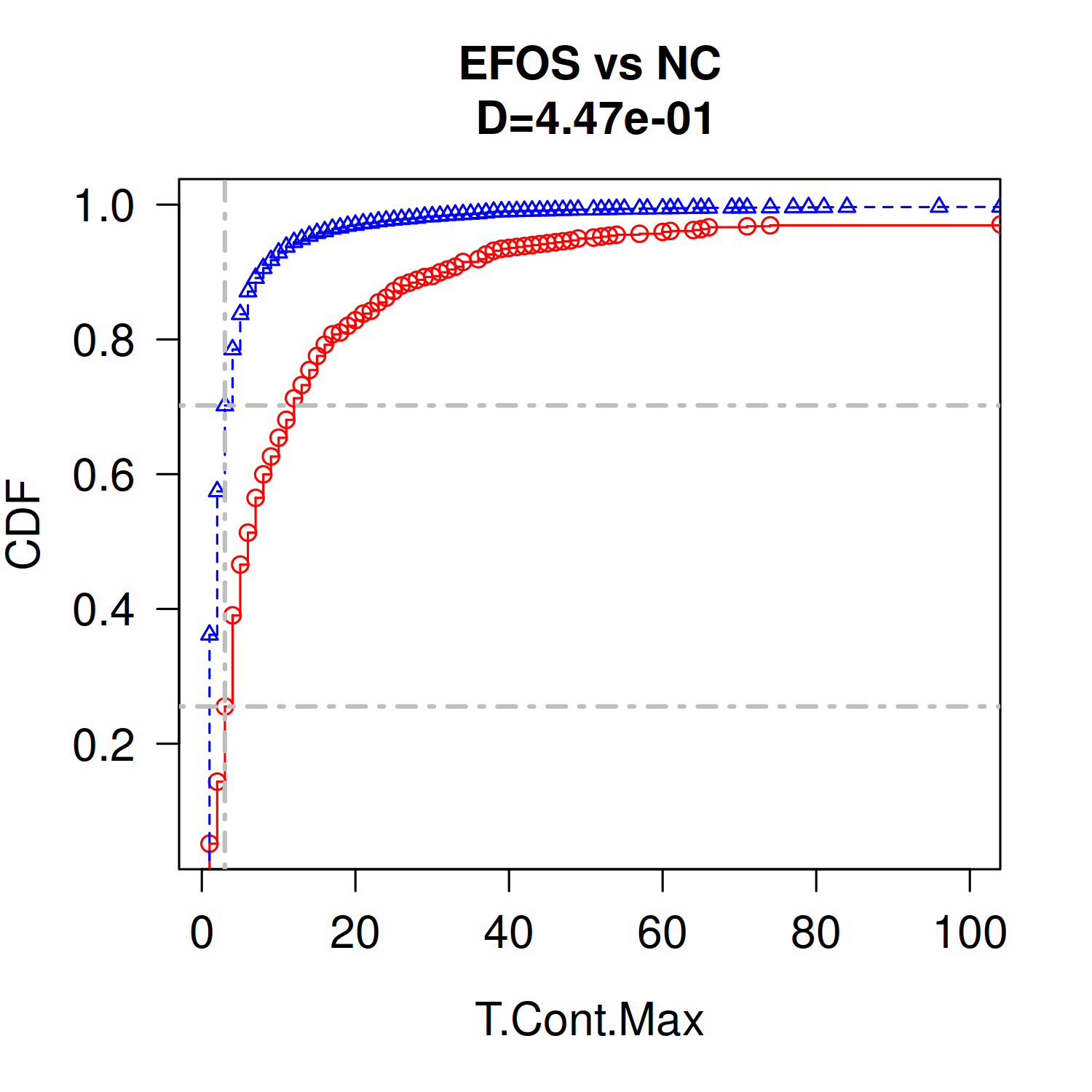}
\includegraphics*[width=0.225\textwidth]{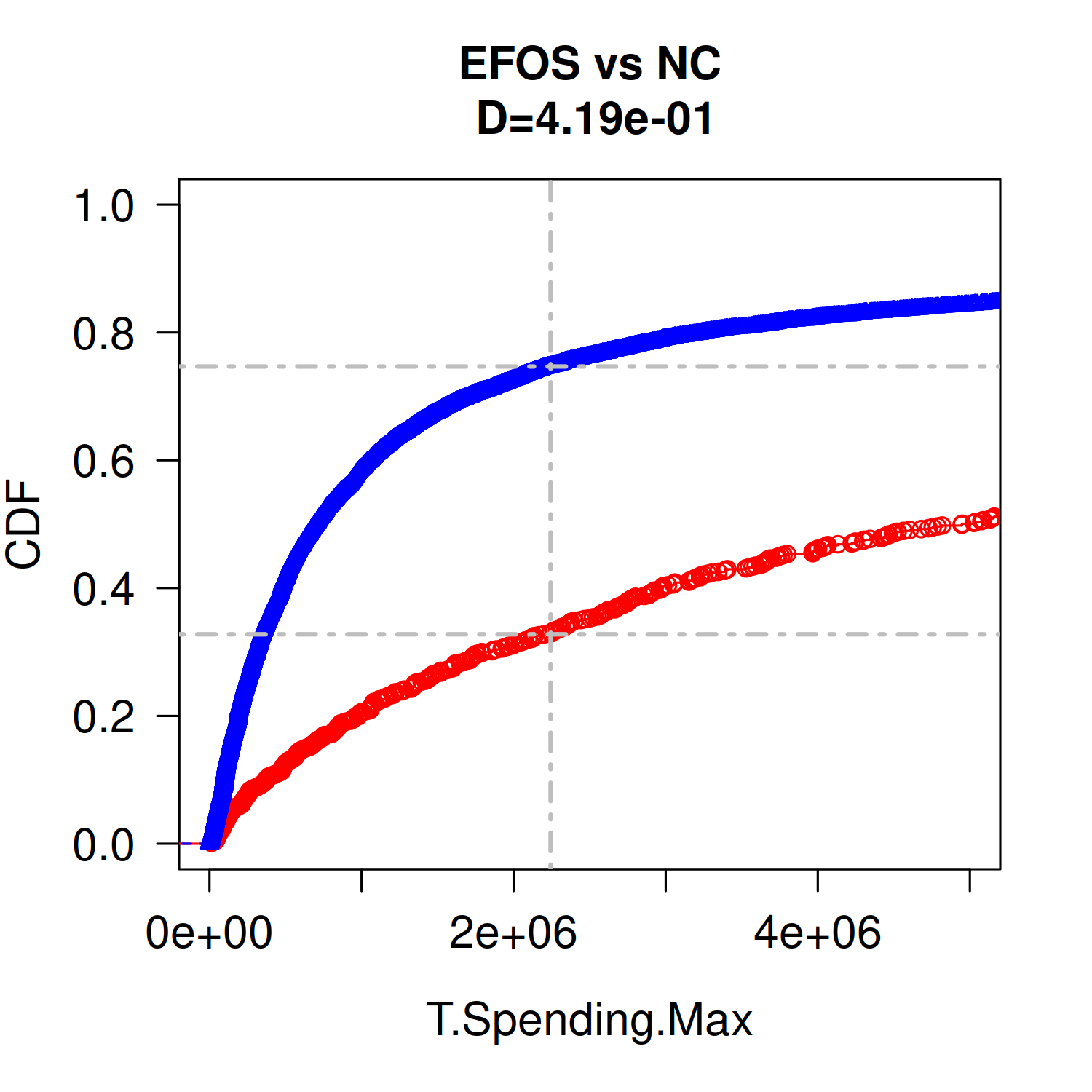}
\end{center}
    \caption{{\it KS-Test results for the 1st period (EFOS vs NC):}The graph shows the cumulative distribution function (CDF) for each significant non-dummy variable in EFOS class (red circles solid line) and NC class (blue triangles dashed line). The grey dashed line shows the value from which the CDF's had the maximum difference. In all graphs the scale of the $X$ axis corresponds to that for which the statistic $D$ is clearly visible.}
    \label{kst-1st-en}
\end{figure*}

\begin{figure*}[htb!]
\begin{center}
\includegraphics*[width=0.275\textwidth]{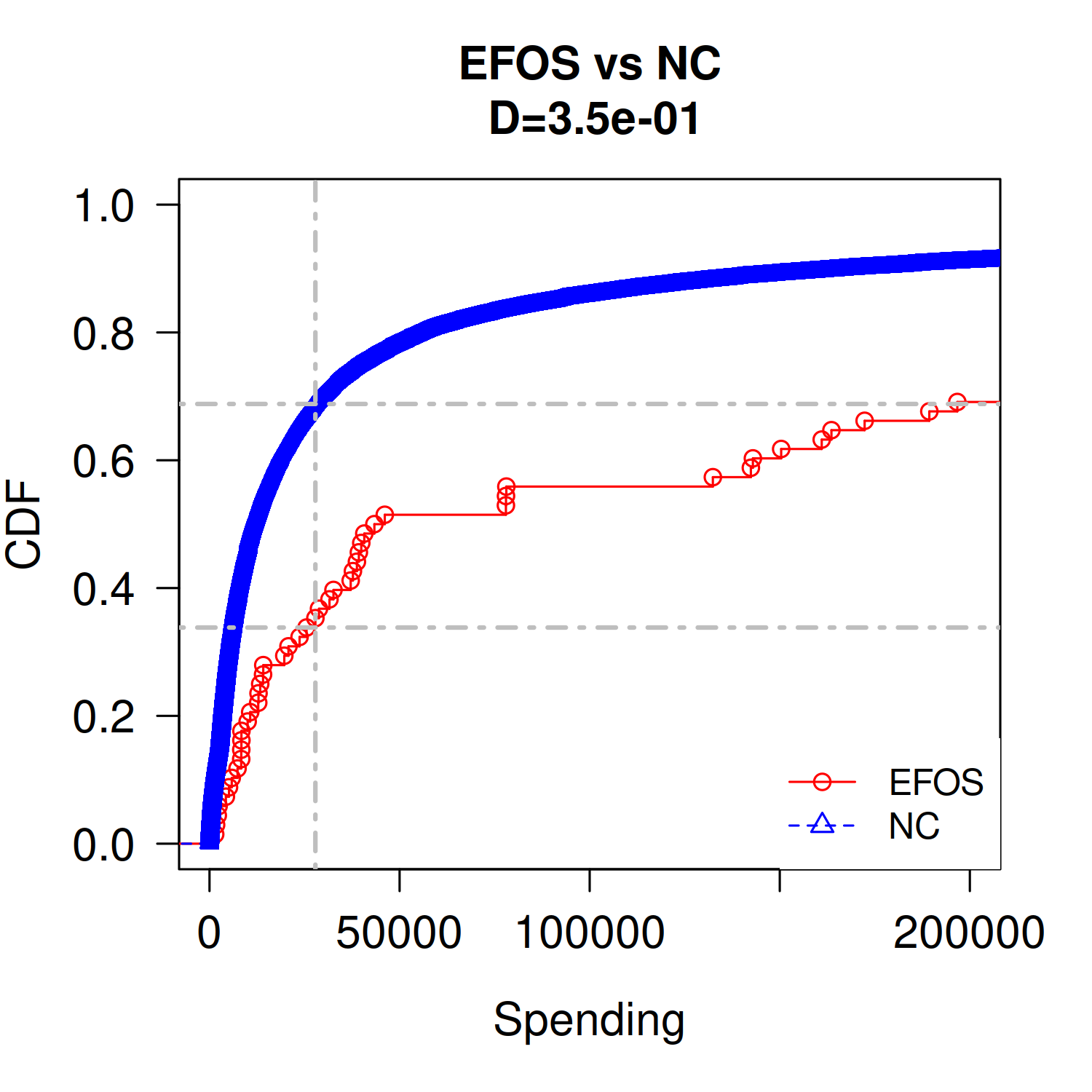}
\includegraphics*[width=0.275\textwidth]{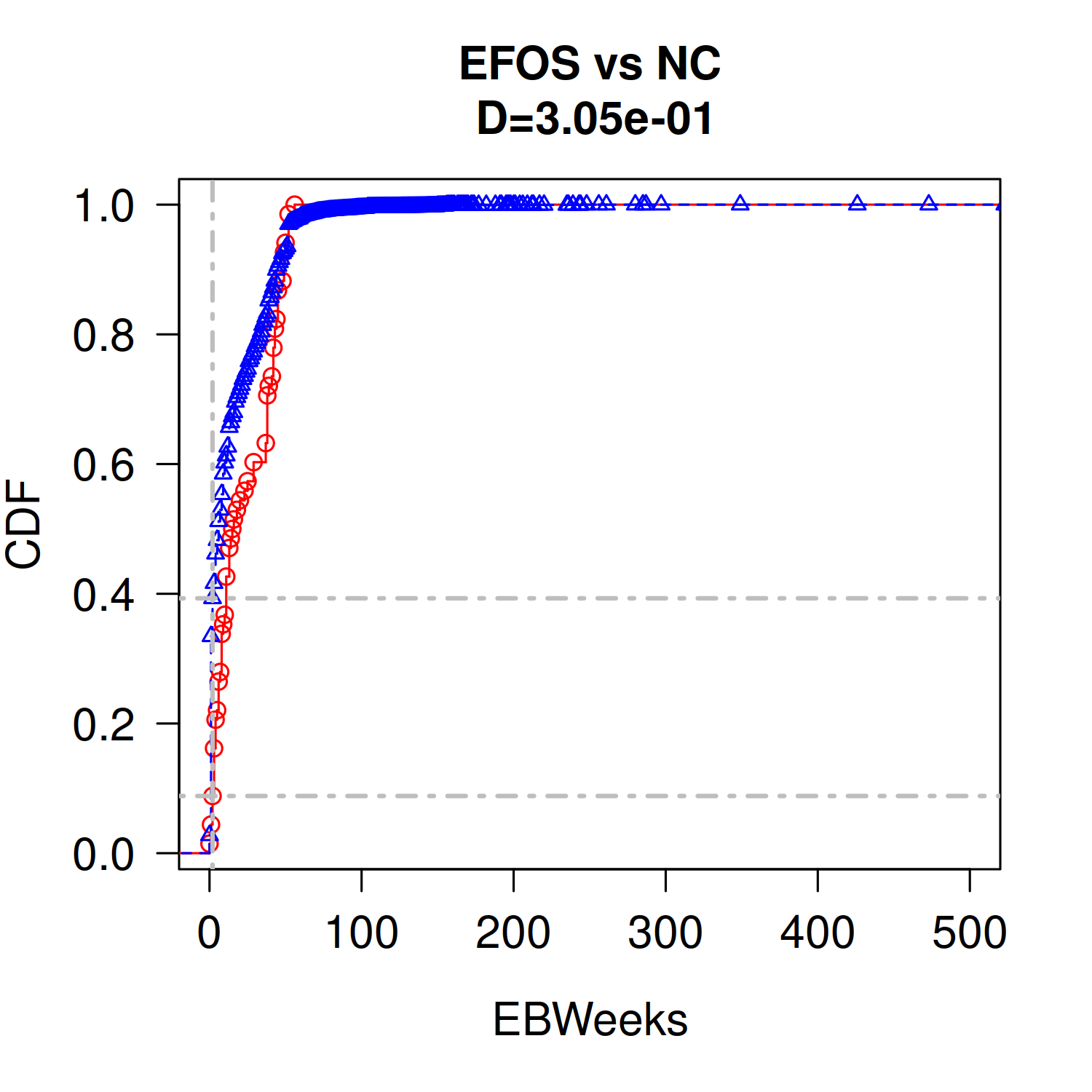}
\includegraphics*[width=0.275\textwidth]{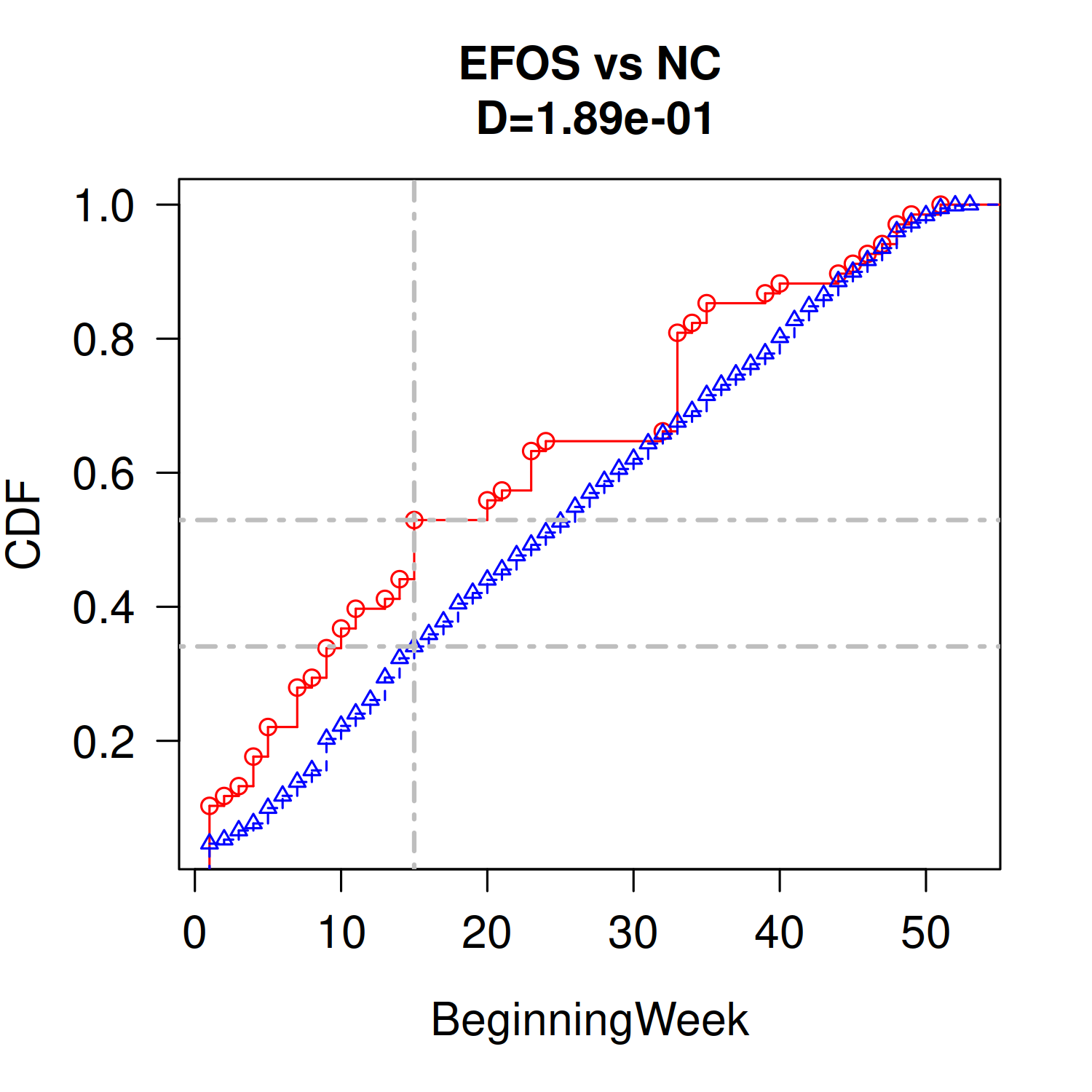}
\end{center}
\begin{center}
\includegraphics*[width=0.275\textwidth]{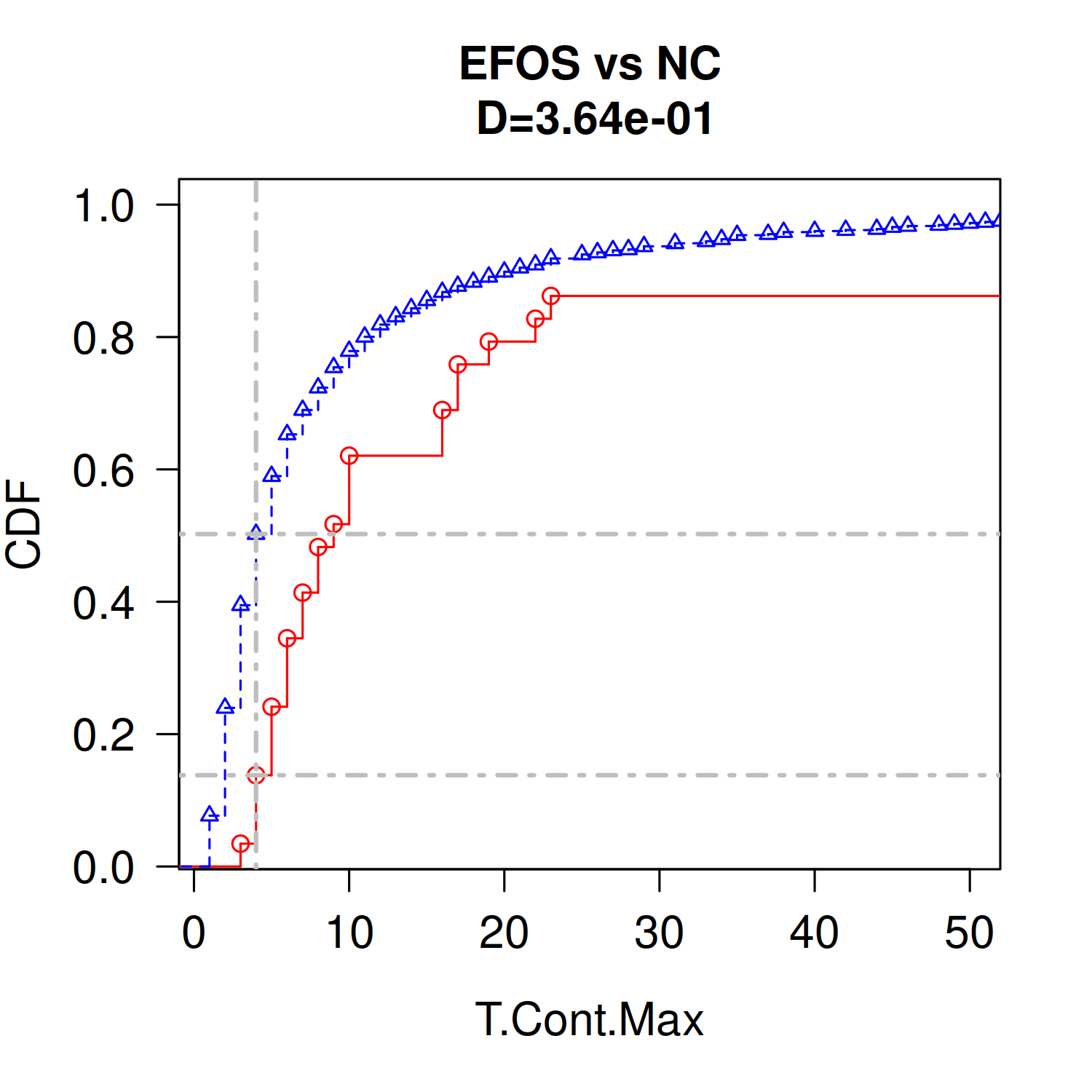}
\includegraphics*[width=0.275\textwidth]{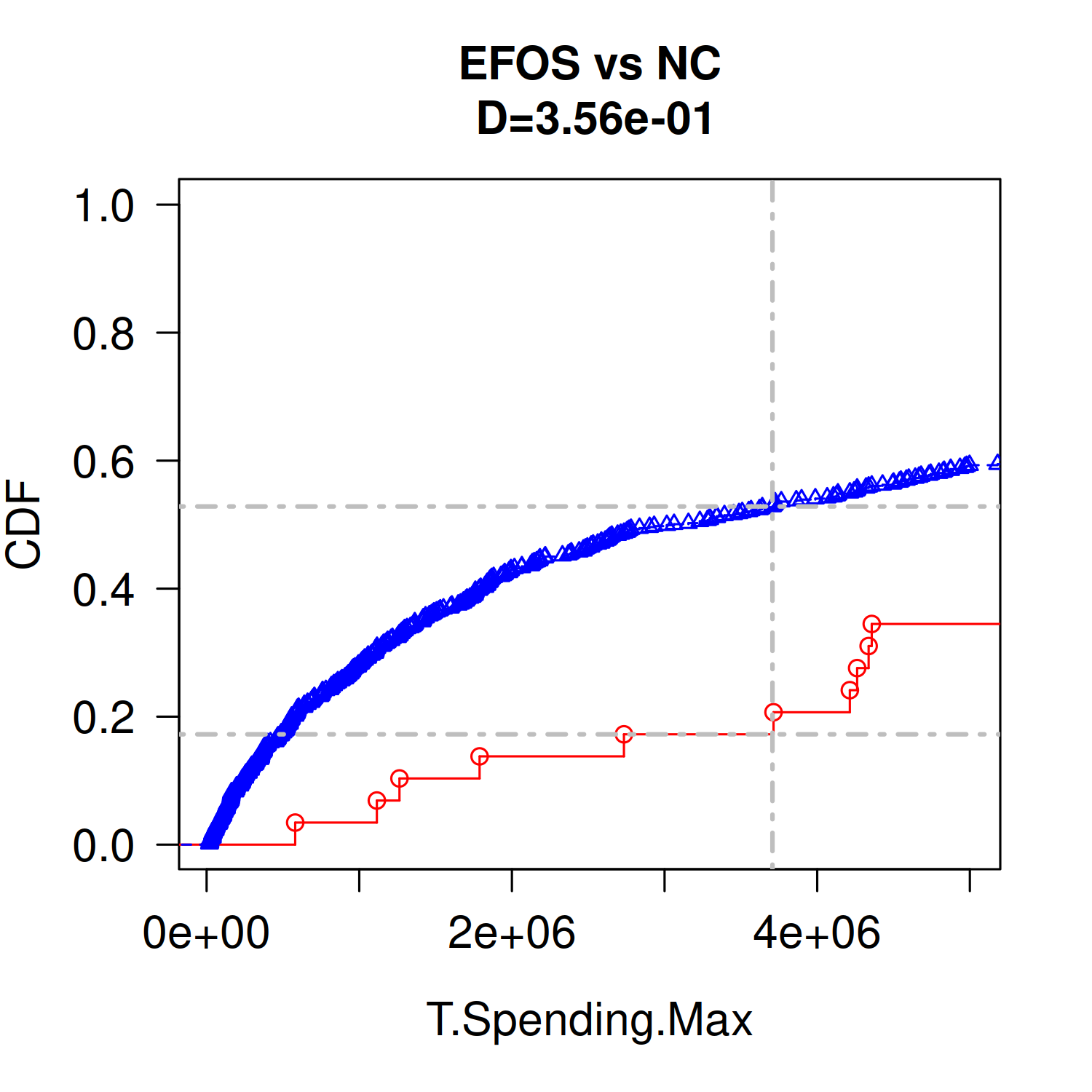}
\end{center}
    \caption{{\it KS-Test results for the 2nd period (EFOS vs NC):} The graph shows the cumulative distribution function (CDF) for each significant non-dummy variable in EFOS class (red circles solid line) and NC class (blue triangles dashed line). The grey dashed line shows the value from which the CDF's showed the maximum difference. In all graphs the scale of the $X$ axis corresponds to that for which the statistic $D$ is clearly visible.}
    \label{kst-2nd-en}
\end{figure*}

\begin{figure*}[htb!]
\begin{center}
\includegraphics*[width=0.275\textwidth]{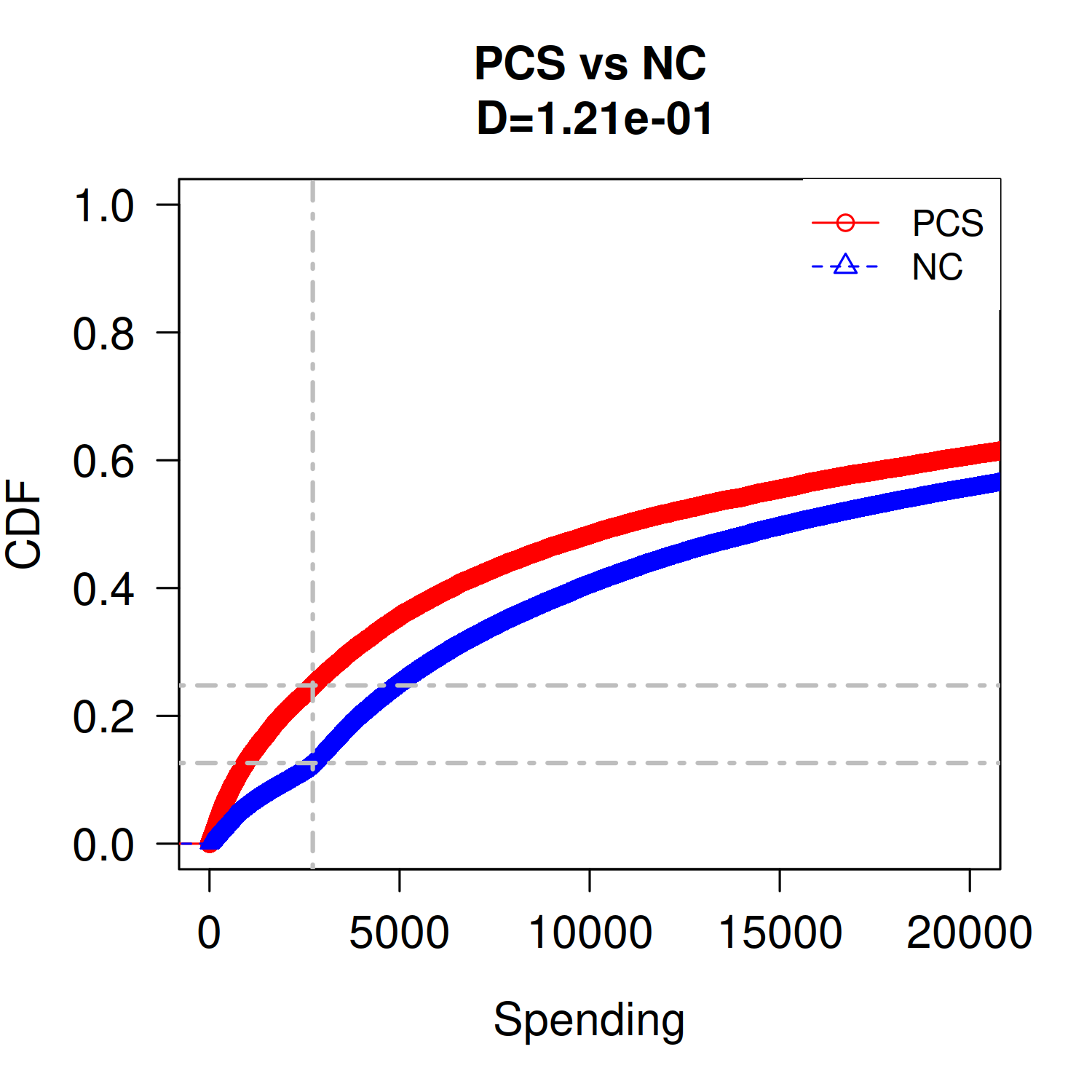}
\includegraphics*[width=0.275\textwidth]{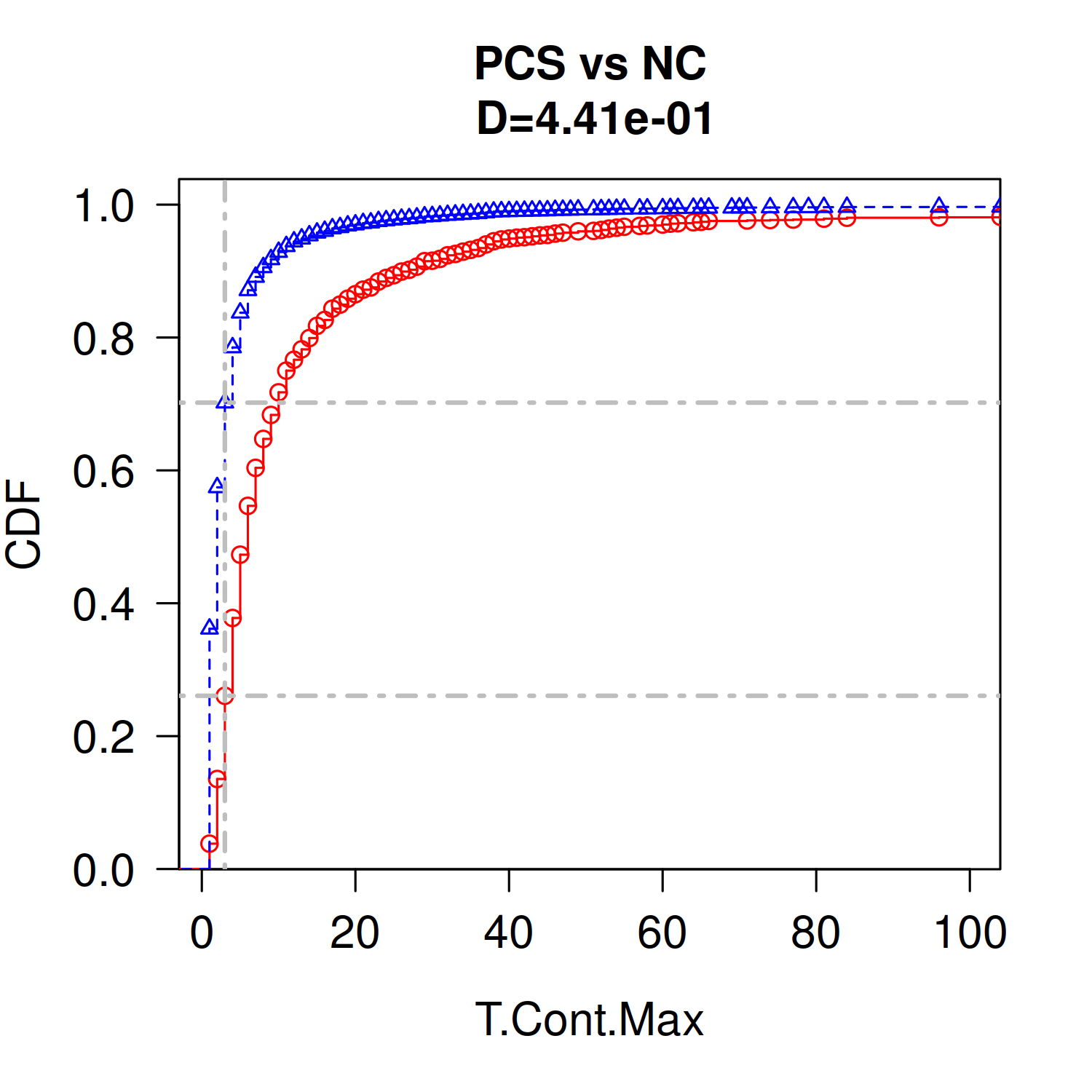}
\includegraphics*[width=0.275\textwidth]{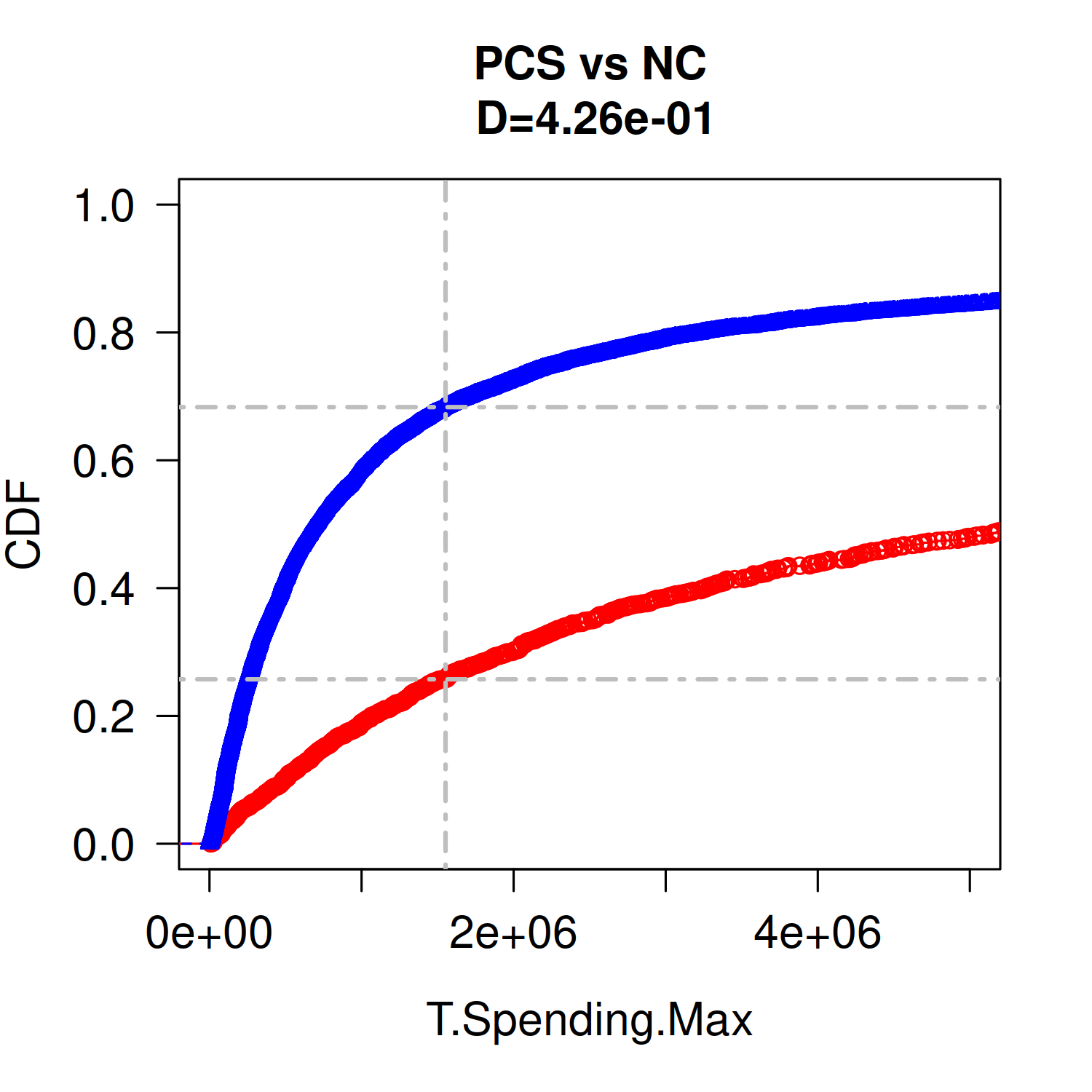}
\end{center}
    \caption{{\it KS-Test results for the 1st period (PCS vs NC):} The graph shows the cumulative distribution function (CDF) for each significant non-dummy variable in PCS class (red circles solid line) and NC class (blue triangles dashed line). The grey dashed line shows the value from which the CDF's had the maximum difference. In all graphs the scale of the $X$ axis corresponds to that for which the statistic $D$ is clearly visible.}
    \label{kst-1st-pn}
\end{figure*}

\begin{figure*}[htb!]
\begin{center}
\includegraphics*[width=0.275\textwidth]{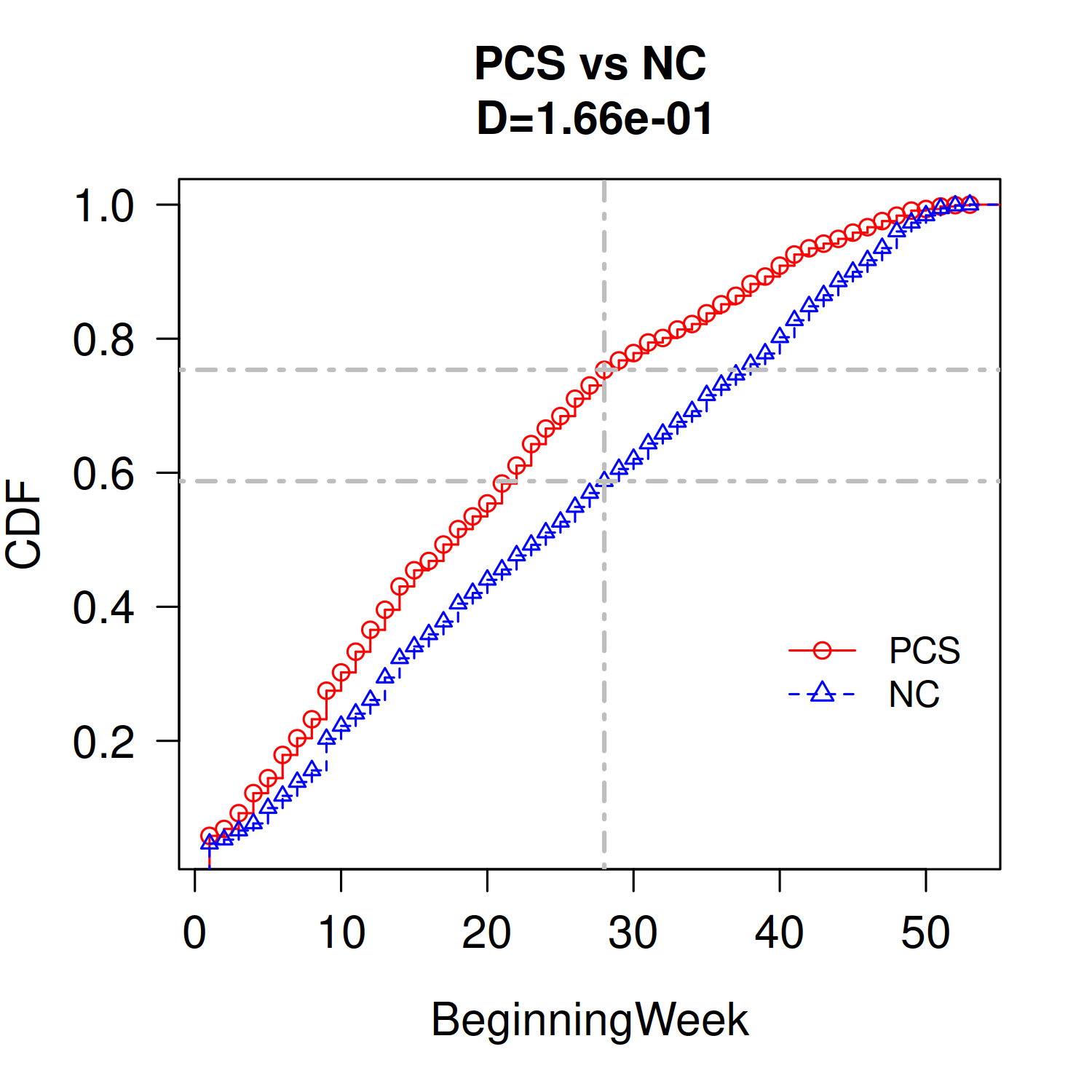}
\includegraphics*[width=0.275\textwidth]{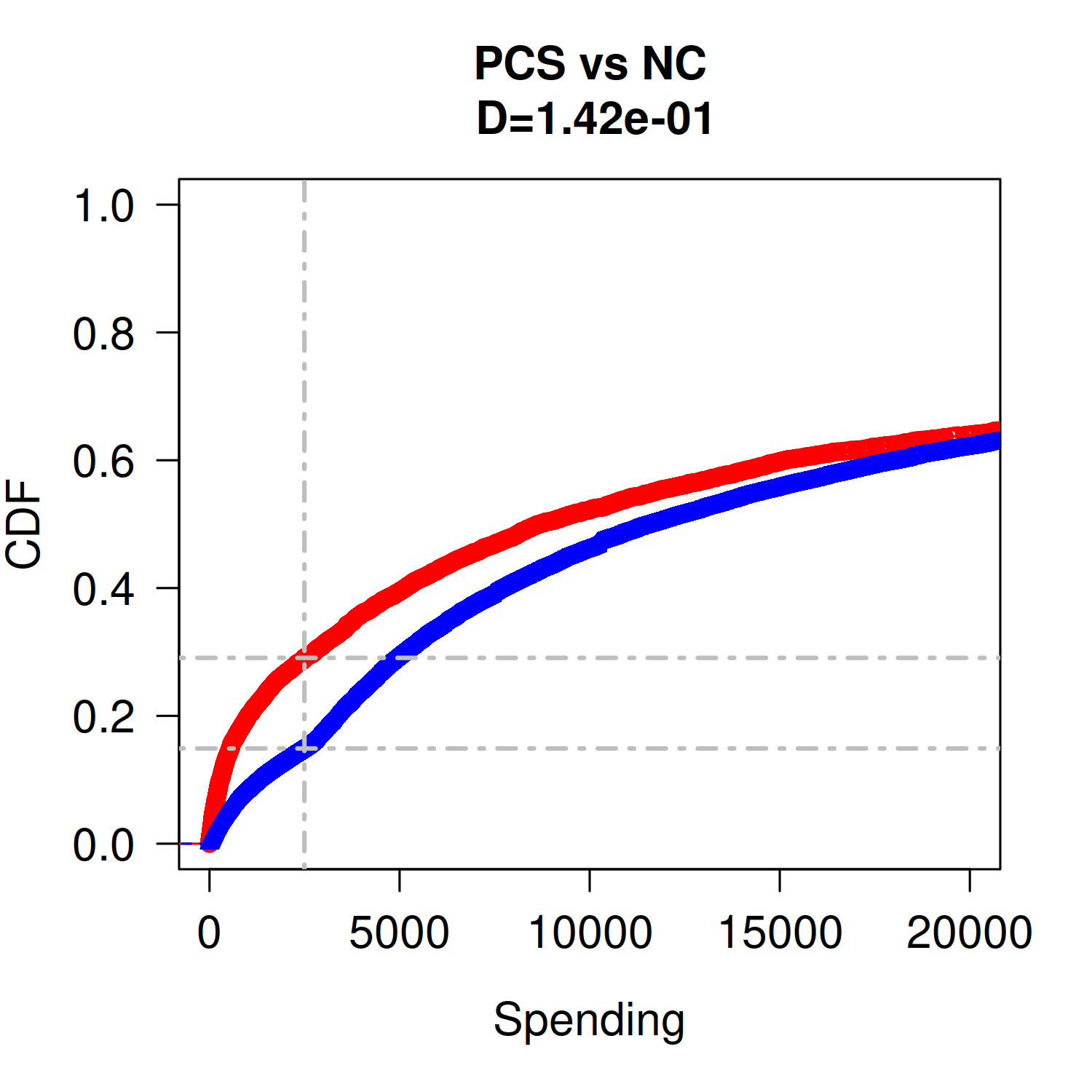}
\includegraphics*[width=0.275\textwidth]{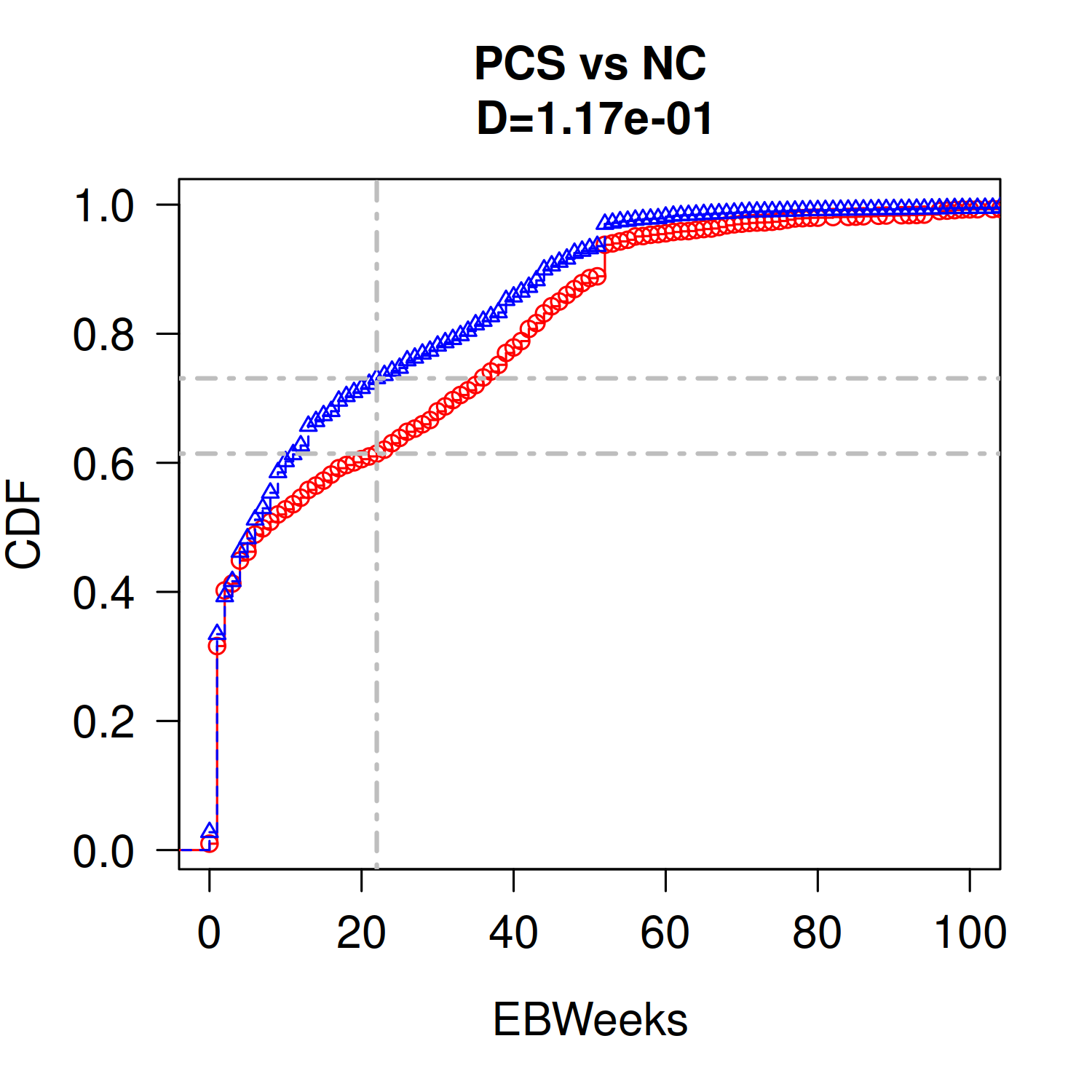}
\end{center}
\begin{center}
\includegraphics*[width=0.275\textwidth]{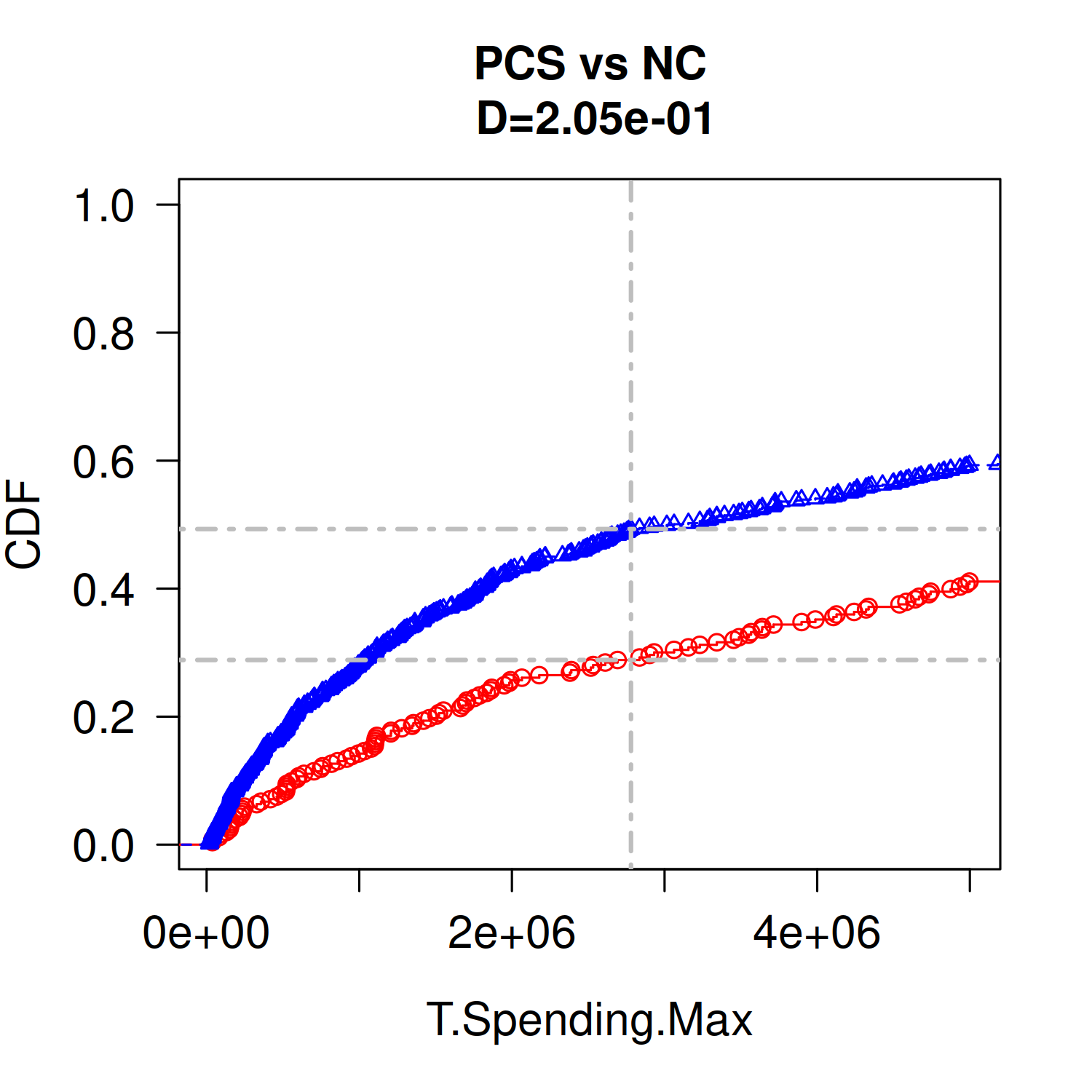}
\includegraphics*[width=0.275\textwidth]{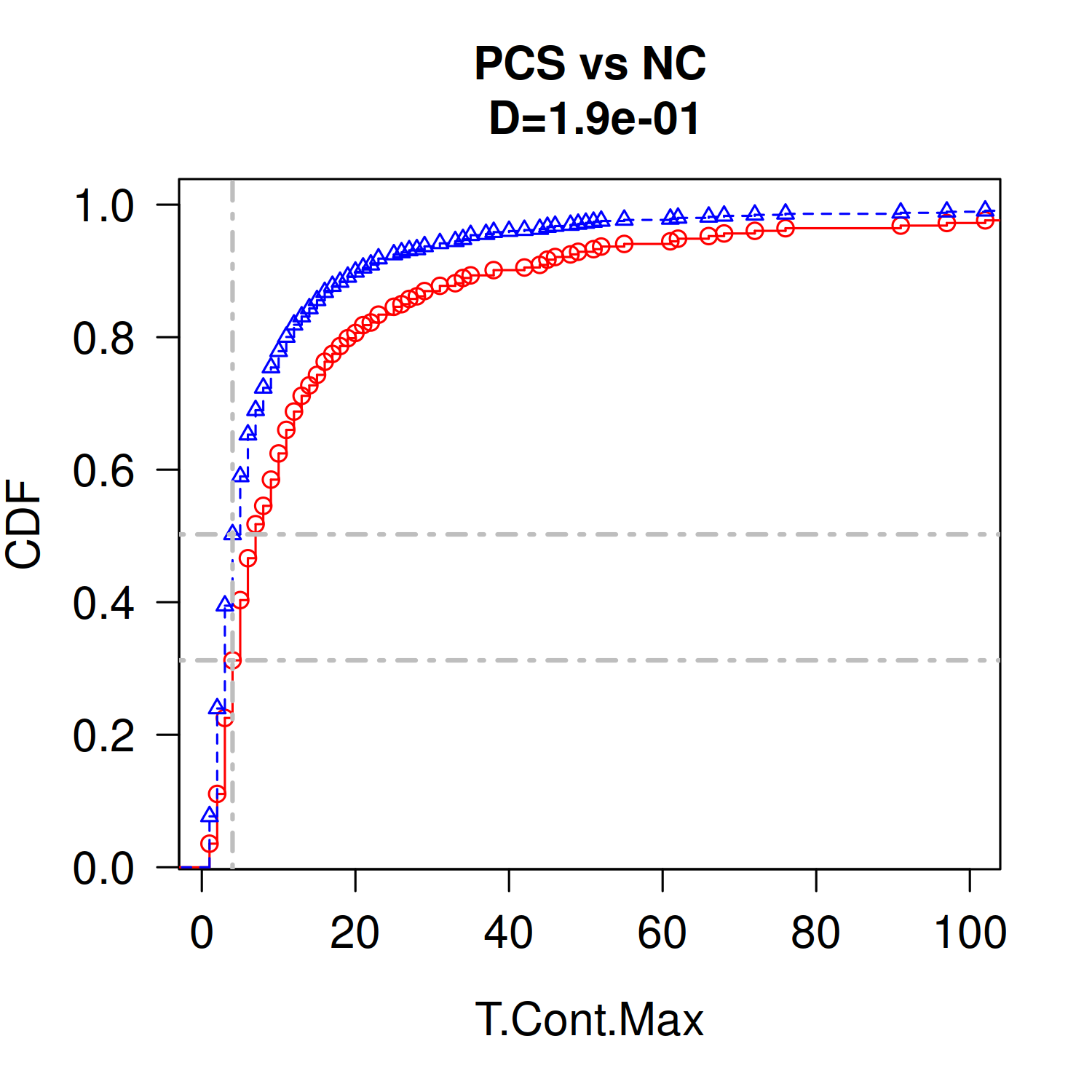}
\end{center}
    \caption{{\it KS-Test results for the 2nd period (PCS vs NC):} The graph shows the cumulative distribution function (CDF) for each significant non-dummy variable in PCS class (red circles solid line) and NC class (blue triangles dashed line). The grey dashed line shows the value from which the CDF's had the maximum difference. In all graphs the scale of the $X$ axis corresponds to that for which the statistic $D$ is clearly visible.}
    \label{kst-2nd-pn}
\end{figure*}

\begin{figure*}[htb!]
\begin{center}
\includegraphics*[width=0.225\textwidth]{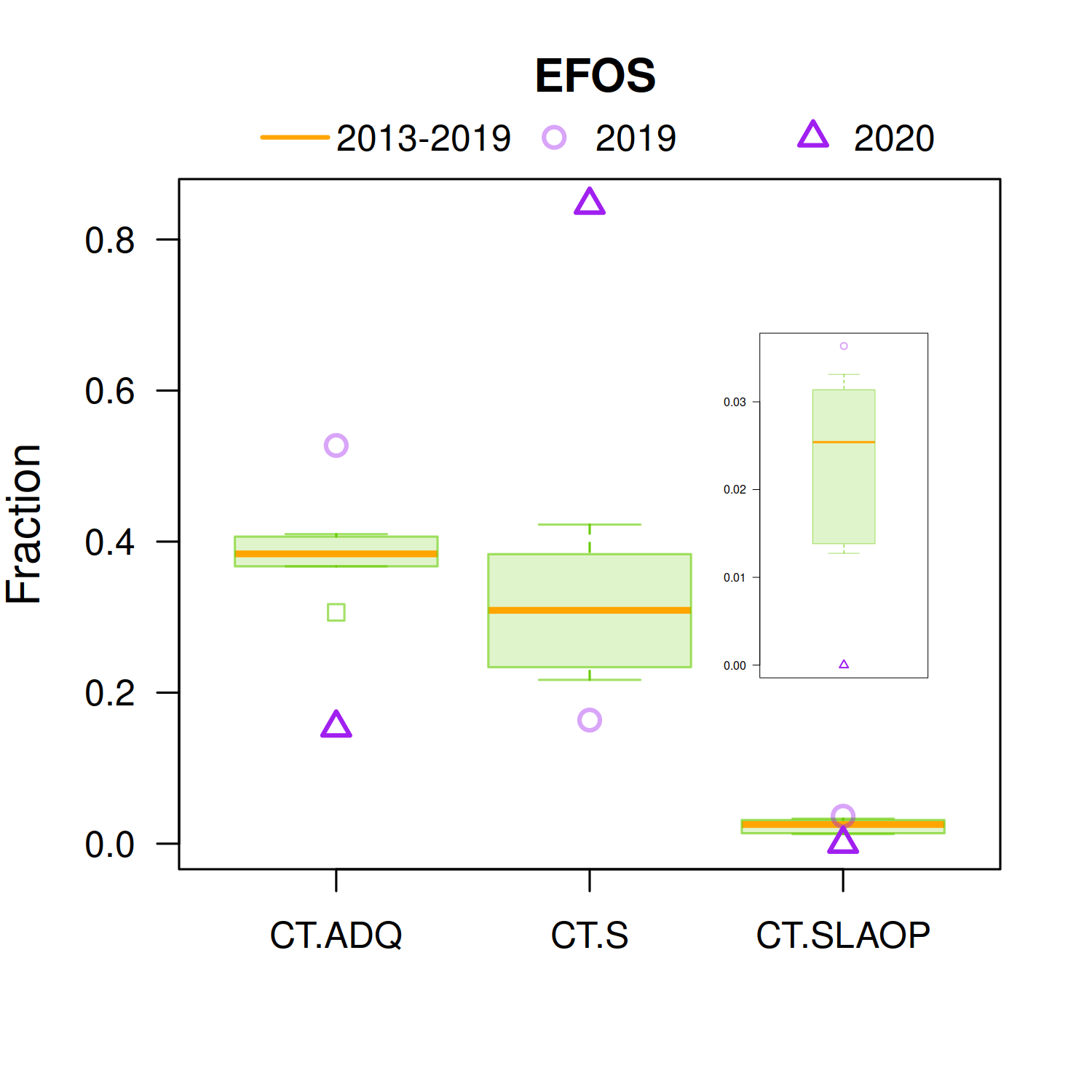}
\includegraphics*[width=0.225\textwidth]{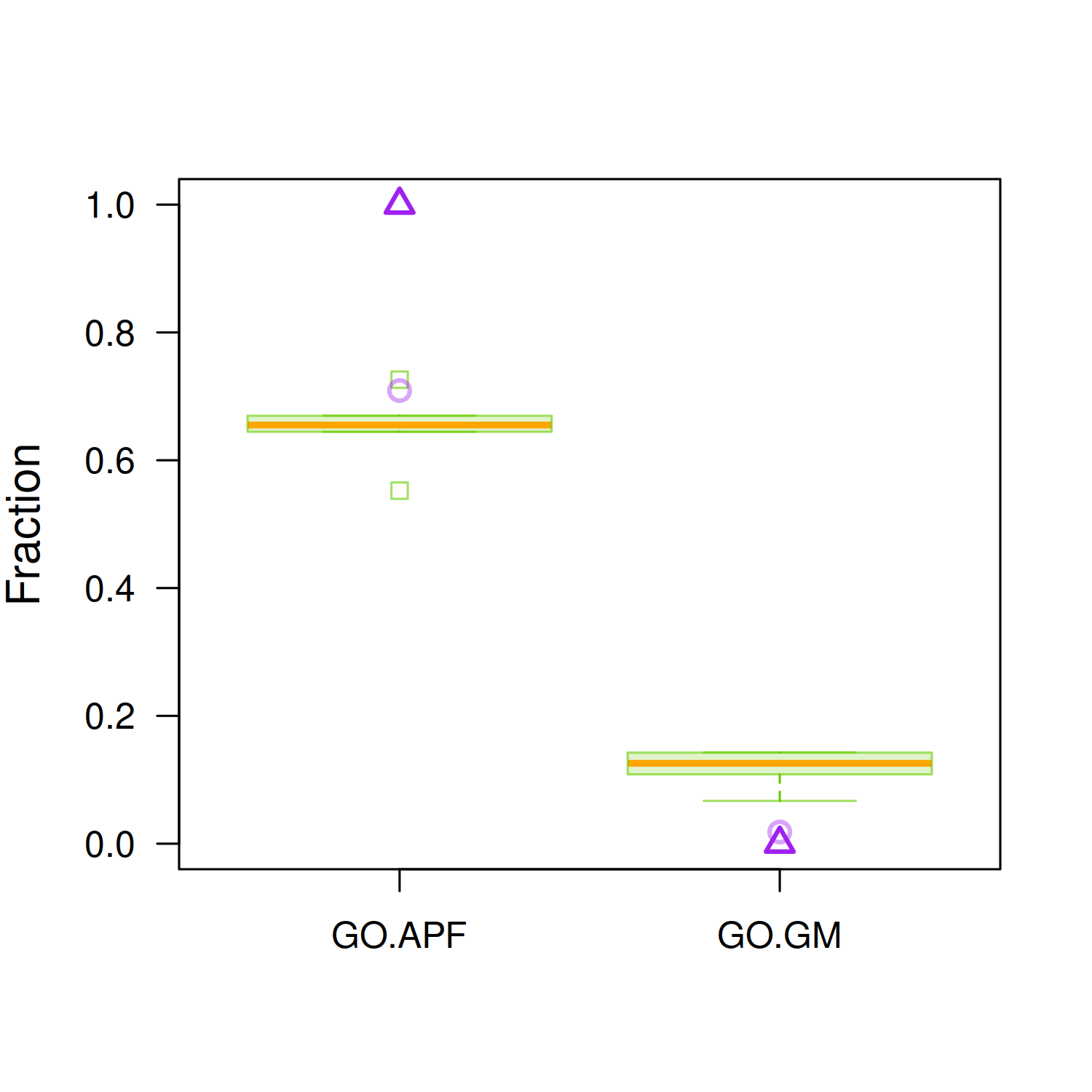}
\includegraphics*[width=0.225\textwidth]{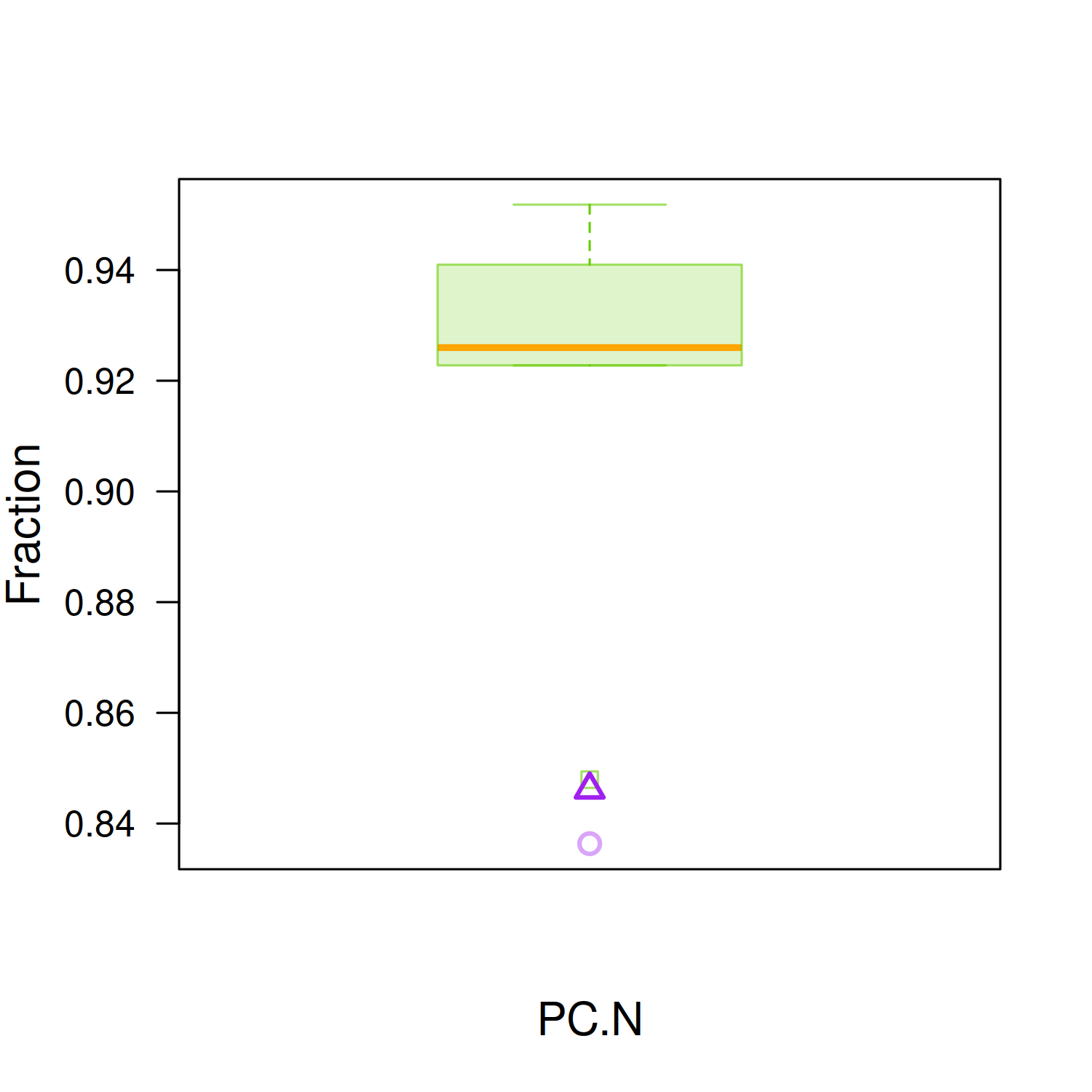}
\includegraphics*[width=0.225\textwidth]{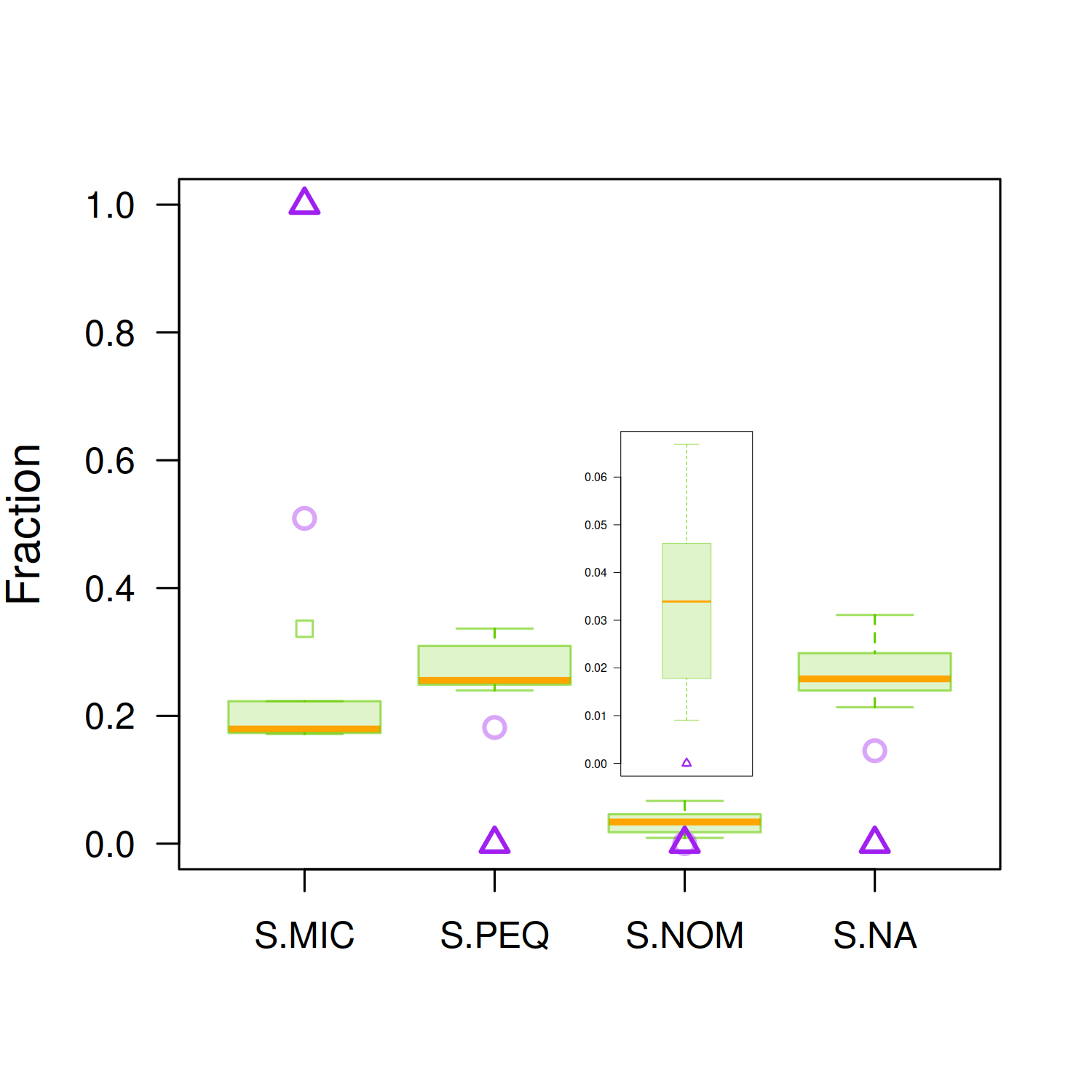}
\end{center}
\begin{center}
\includegraphics*[width=0.225\textwidth]{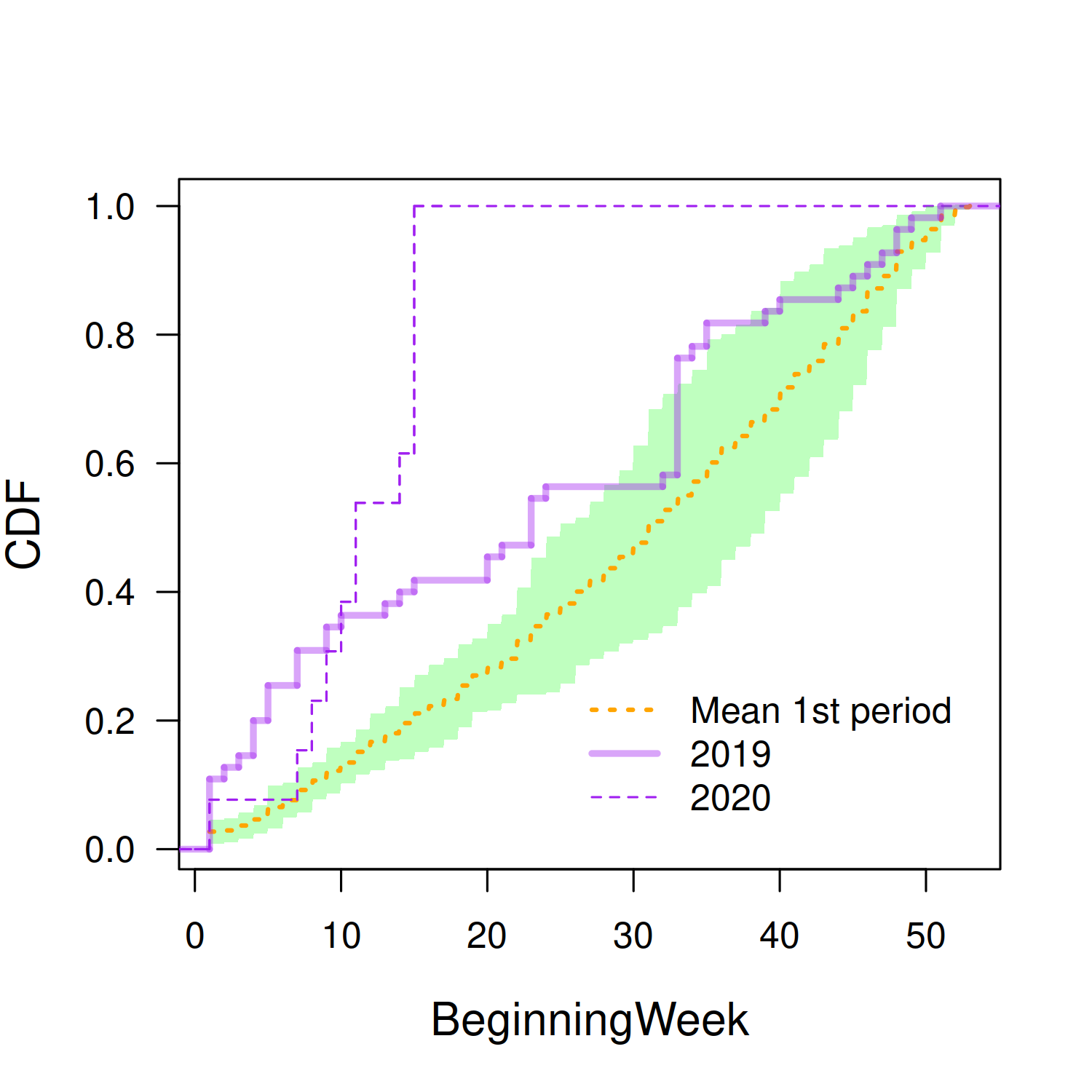}
\includegraphics*[width=0.225\textwidth]{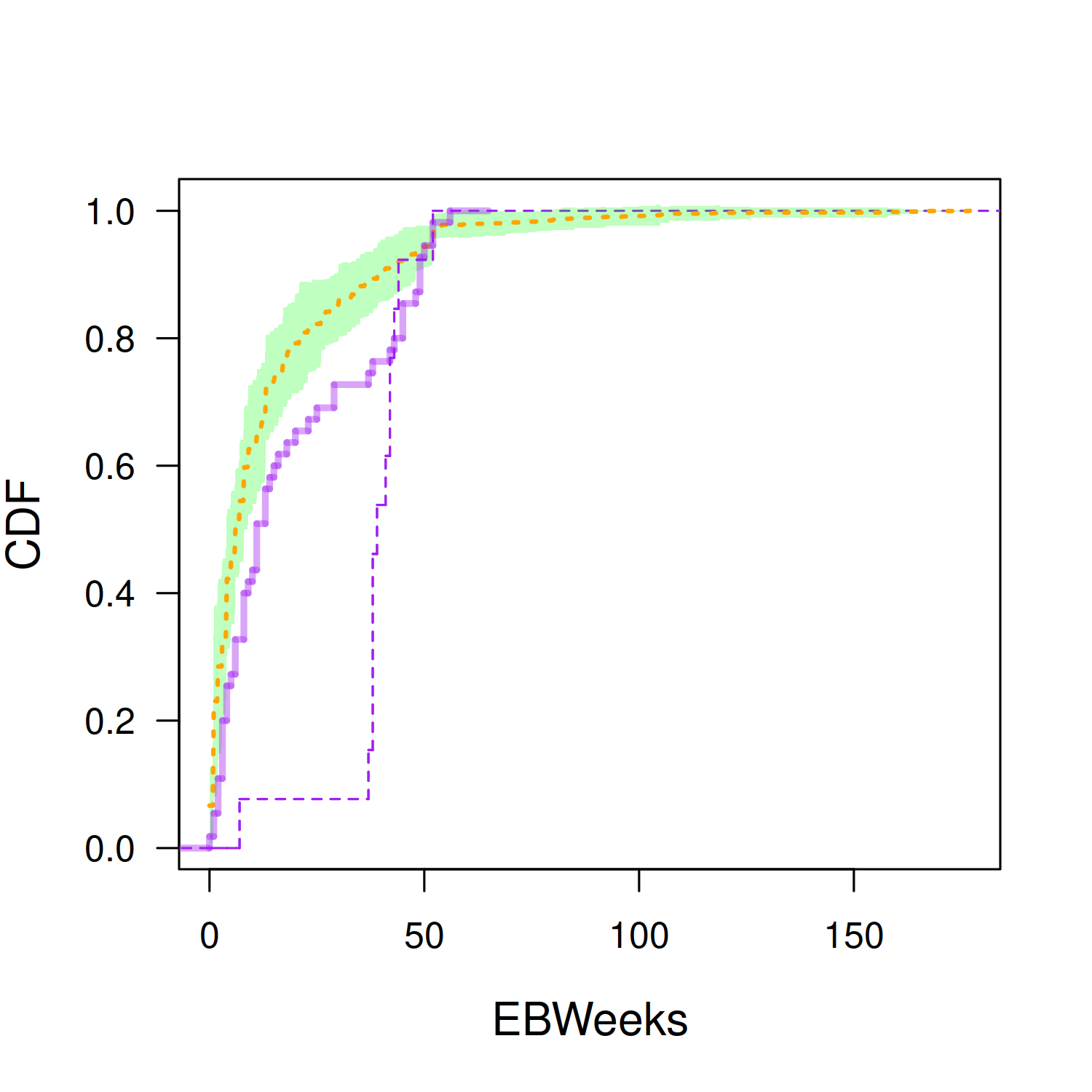}
\includegraphics*[width=0.225\textwidth]{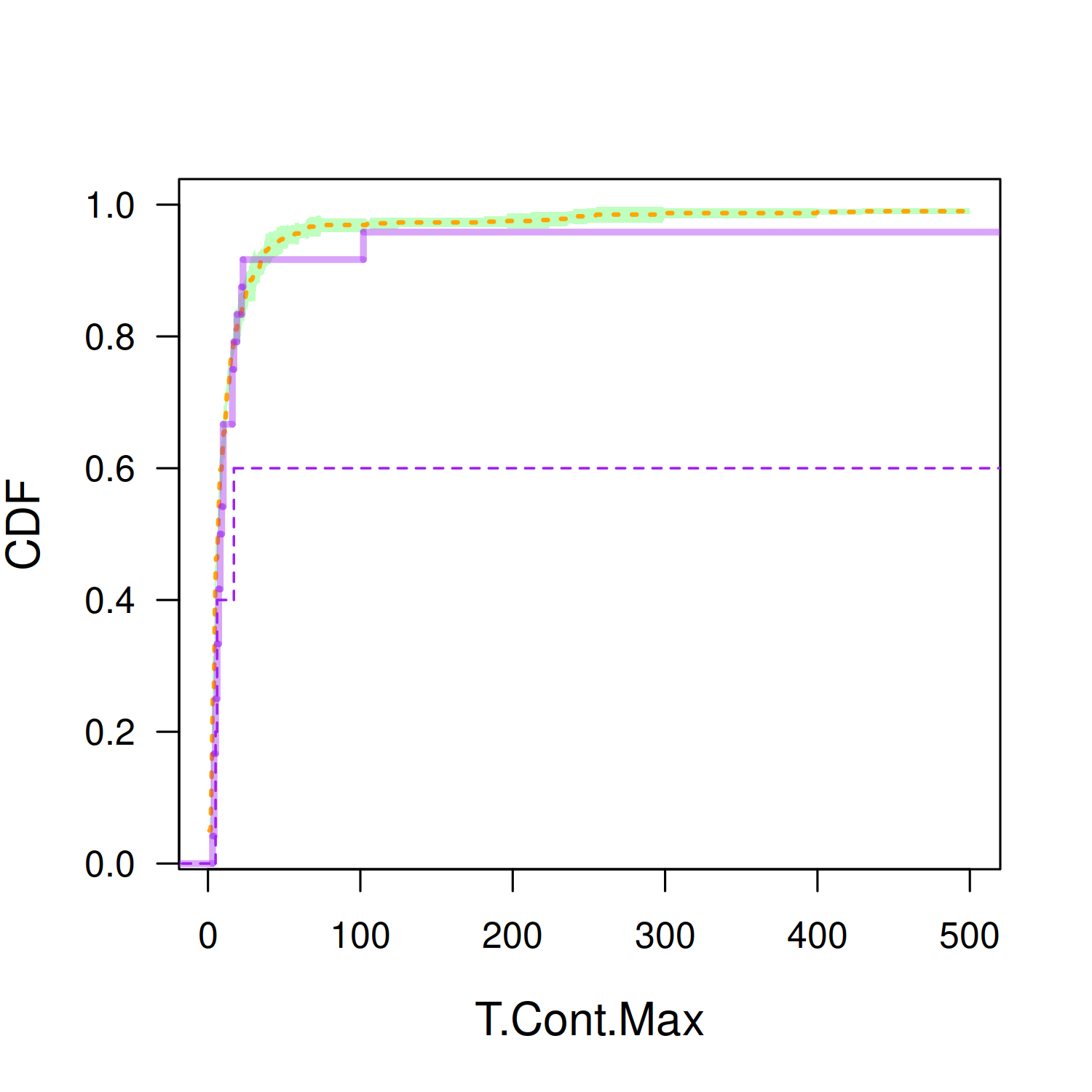}
\includegraphics*[width=0.225\textwidth]{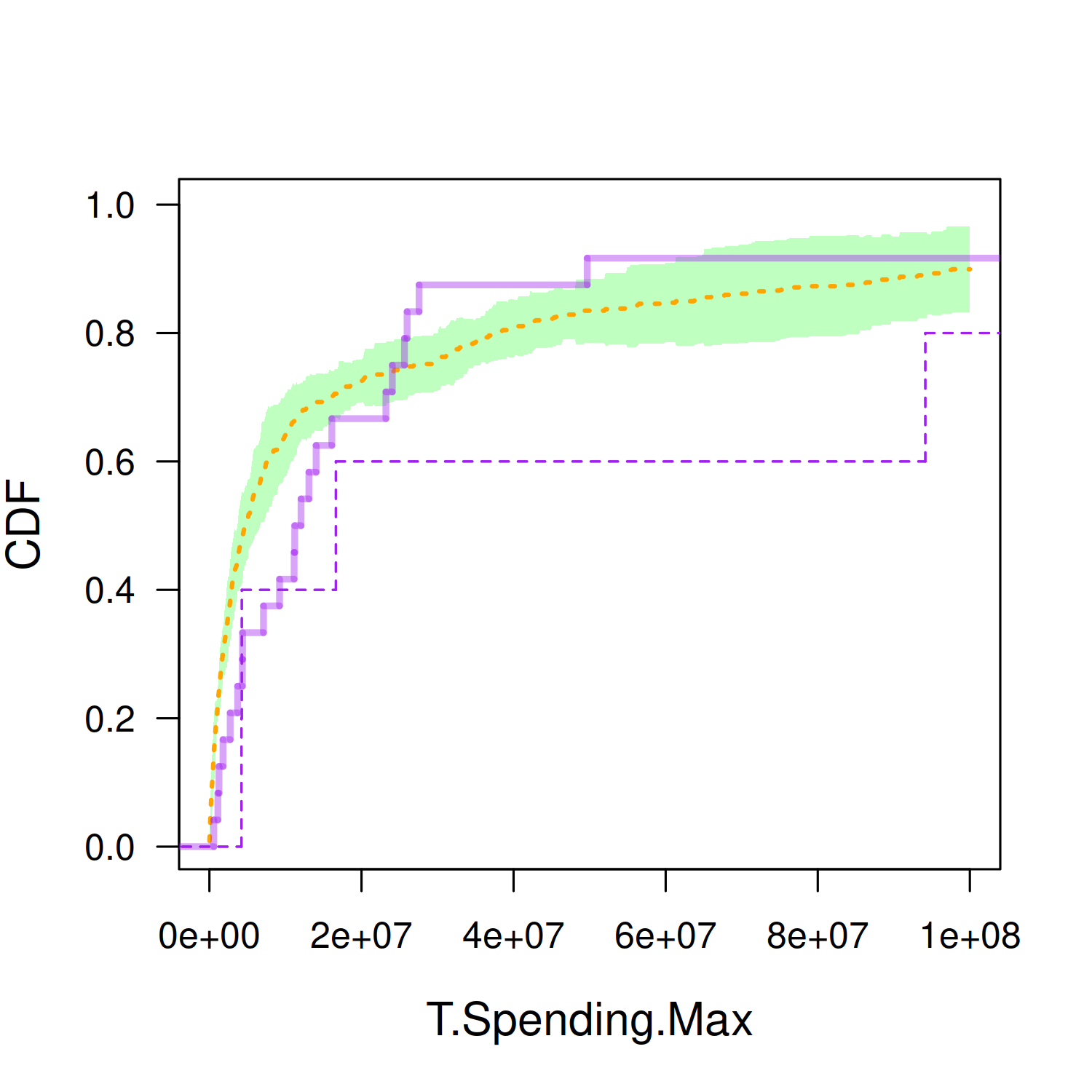}
\end{center}
    \caption{{\it Differences between both government periods for contracts in the EFOS class:} For the dummy variables ({\bf Top}): The boxplot (in green) groups the data of the 6 years comprising the 1st government period. For the remaining non-dummy variables ({\bf Bottom}) the graph shows the cumulative distribution function (CDF), and the green area represents the CI at 99\% generated by the data of the 1st period. In all graphs, the orange line corresponds to the mean of the 1st government period, and two purple colored curves (symbols) correspond to the first two years of the second government period. The insets appearing in some graphs are a close up to the behavior of the corresponding variable.}
    \label{ci-efos-dif}
\end{figure*}

\begin{figure*}[htb!]
\begin{center}
\includegraphics*[width=0.275\textwidth]{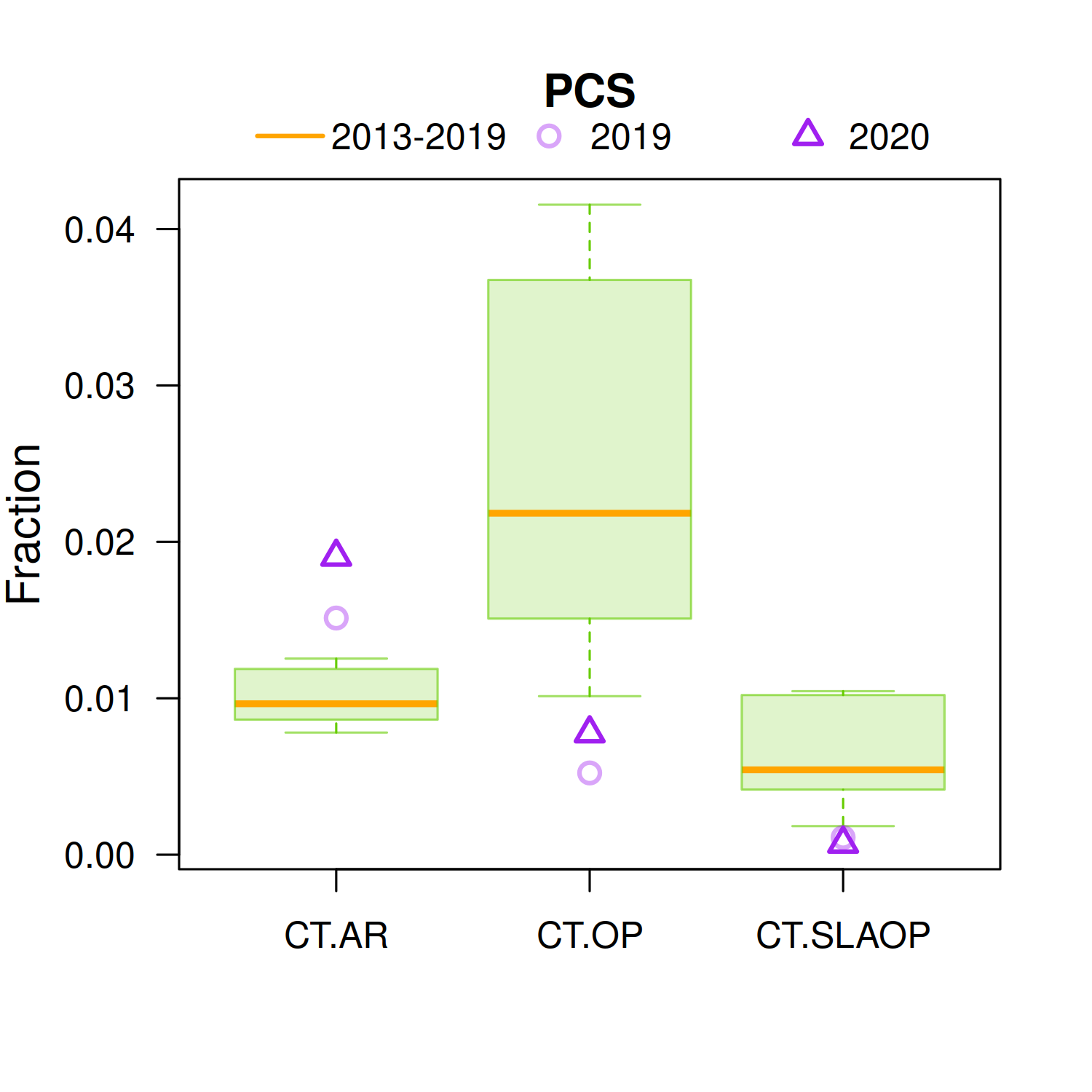}
\includegraphics*[width=0.275\textwidth]{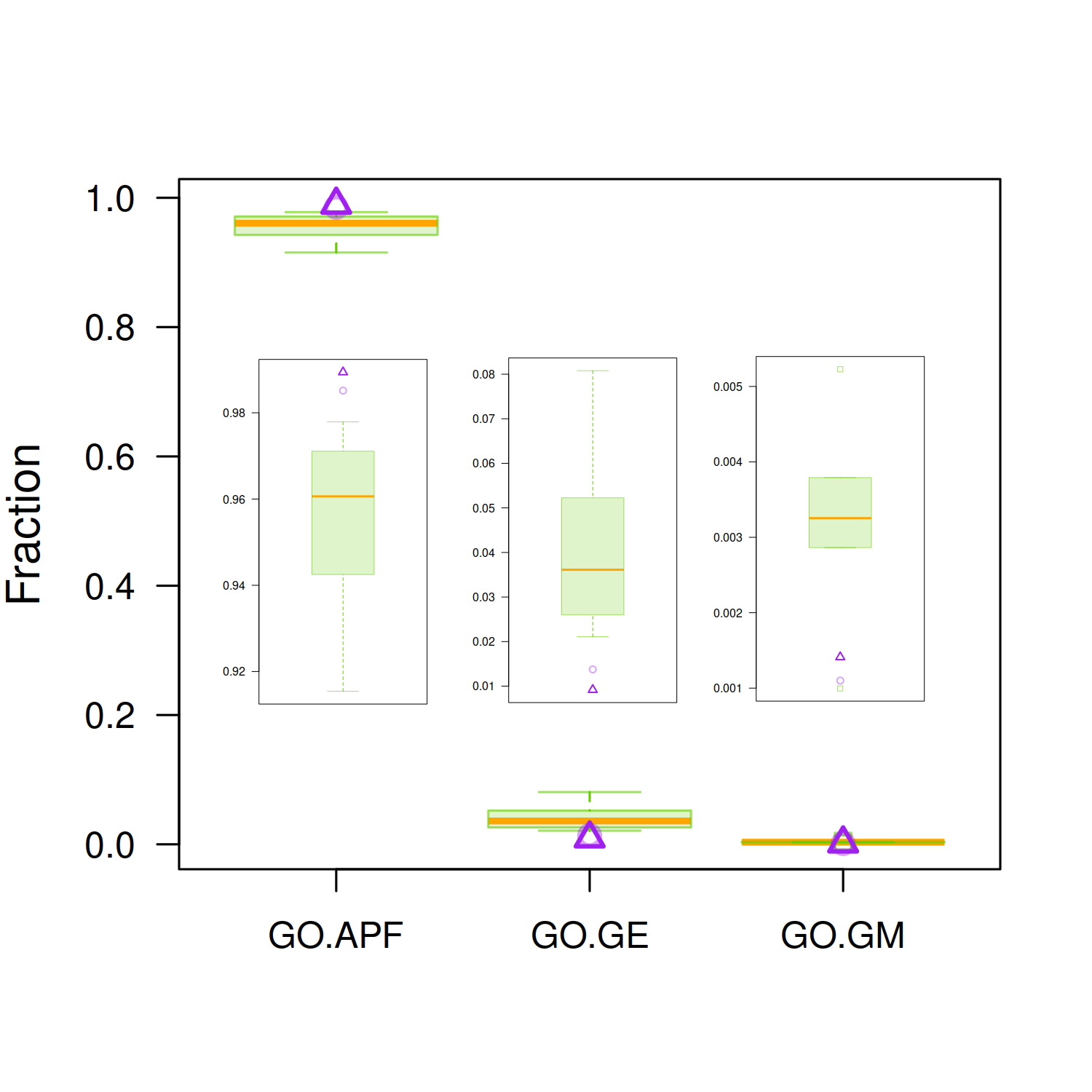}
\includegraphics*[width=0.275\textwidth]{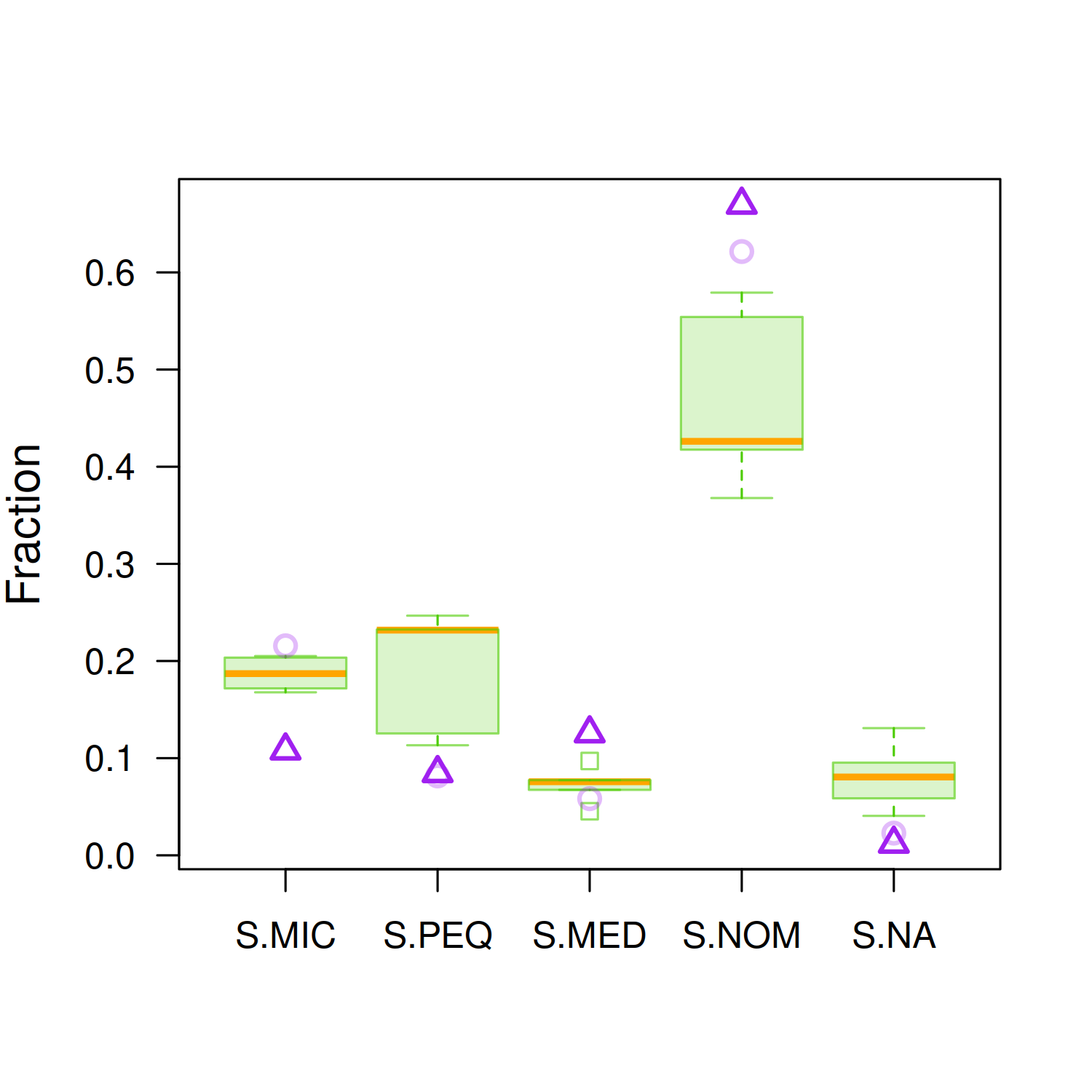}
\end{center}
\begin{center}
\includegraphics*[width=0.275\textwidth]{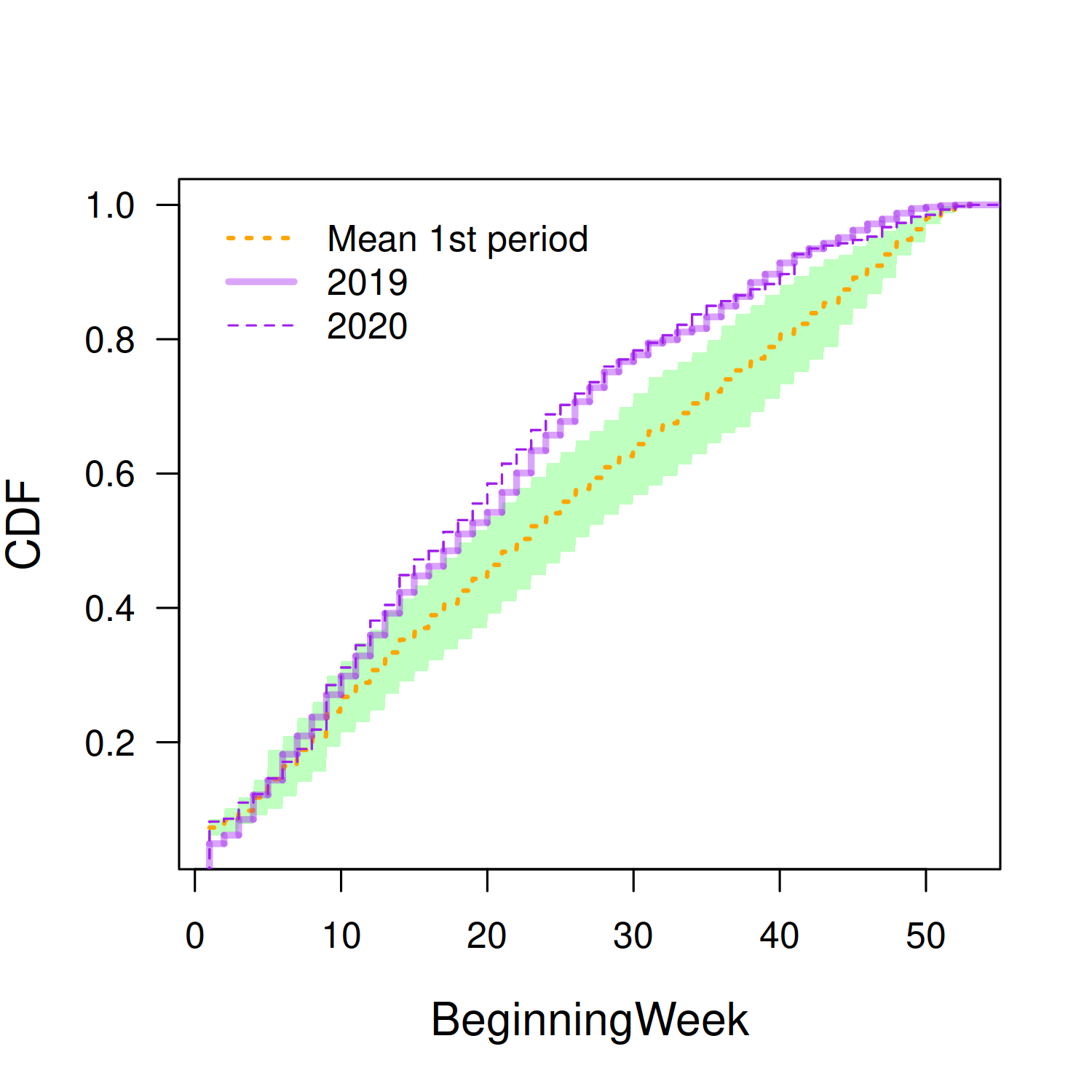}
\end{center}
    \caption{{\it Differences between both government periods for contracts in the PCS class:} For the dummy variables ({\bf Top}): The boxplot (in green) groups the data of the 6 years comprising the 1st government period. For the remaining non-dummy variables ({\bf Bottom}) the graph shows the cumulative distribution function (CDF), and the green area represents the CI at 99\% generated by the data of the 1st period. In all graphs, the orange line corresponds to the mean of the 1st government period, and two purple colored curves (symbols) correspond to the first two years of the second government period. The insets appearing in the top center graph are close ups to the behavior of the corresponding variables.}
    \label{ci-pcs-dif}
\end{figure*}

\begin{figure*}[htb!]
\begin{center}
\includegraphics*[width=0.25\textwidth]{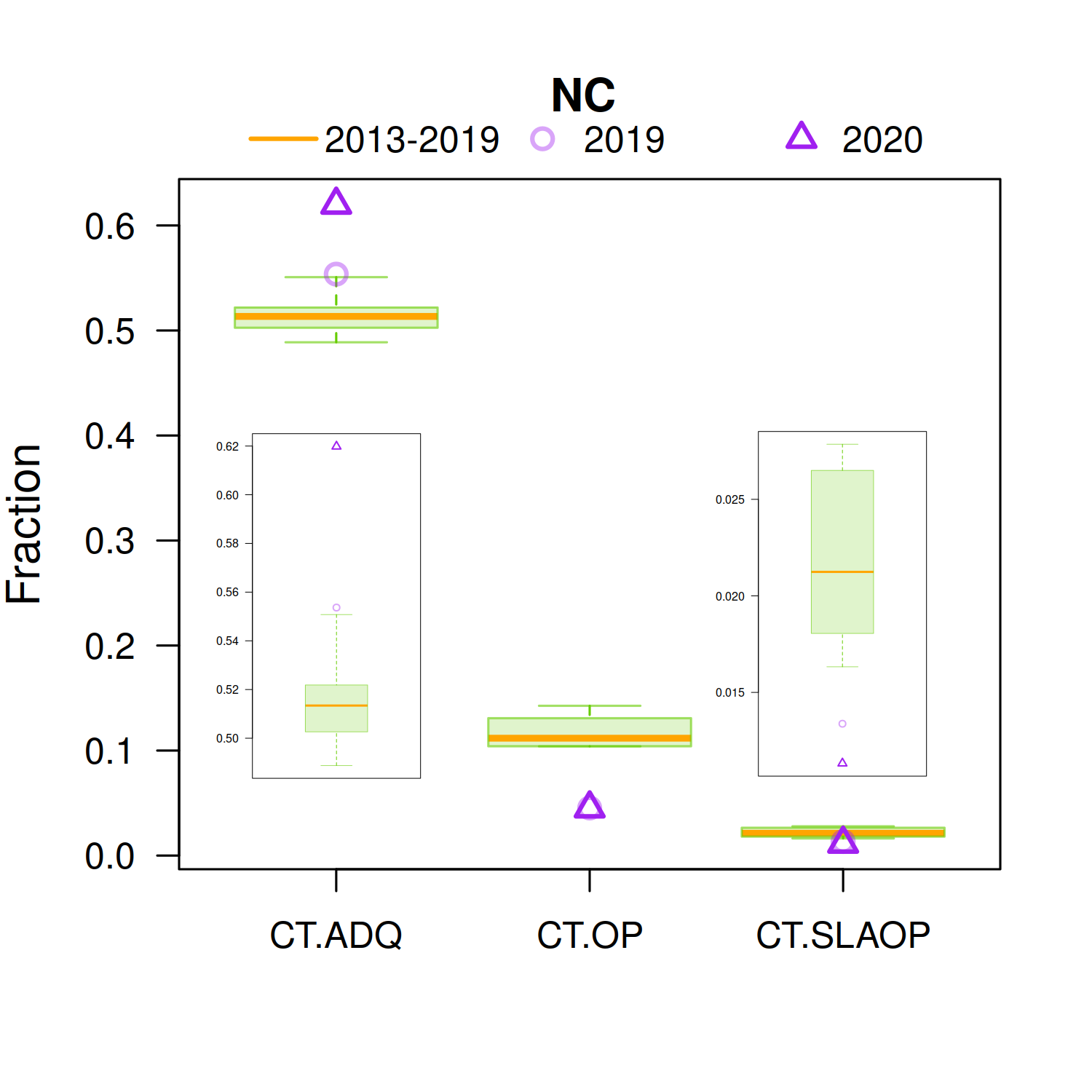}
\includegraphics*[width=0.25\textwidth]{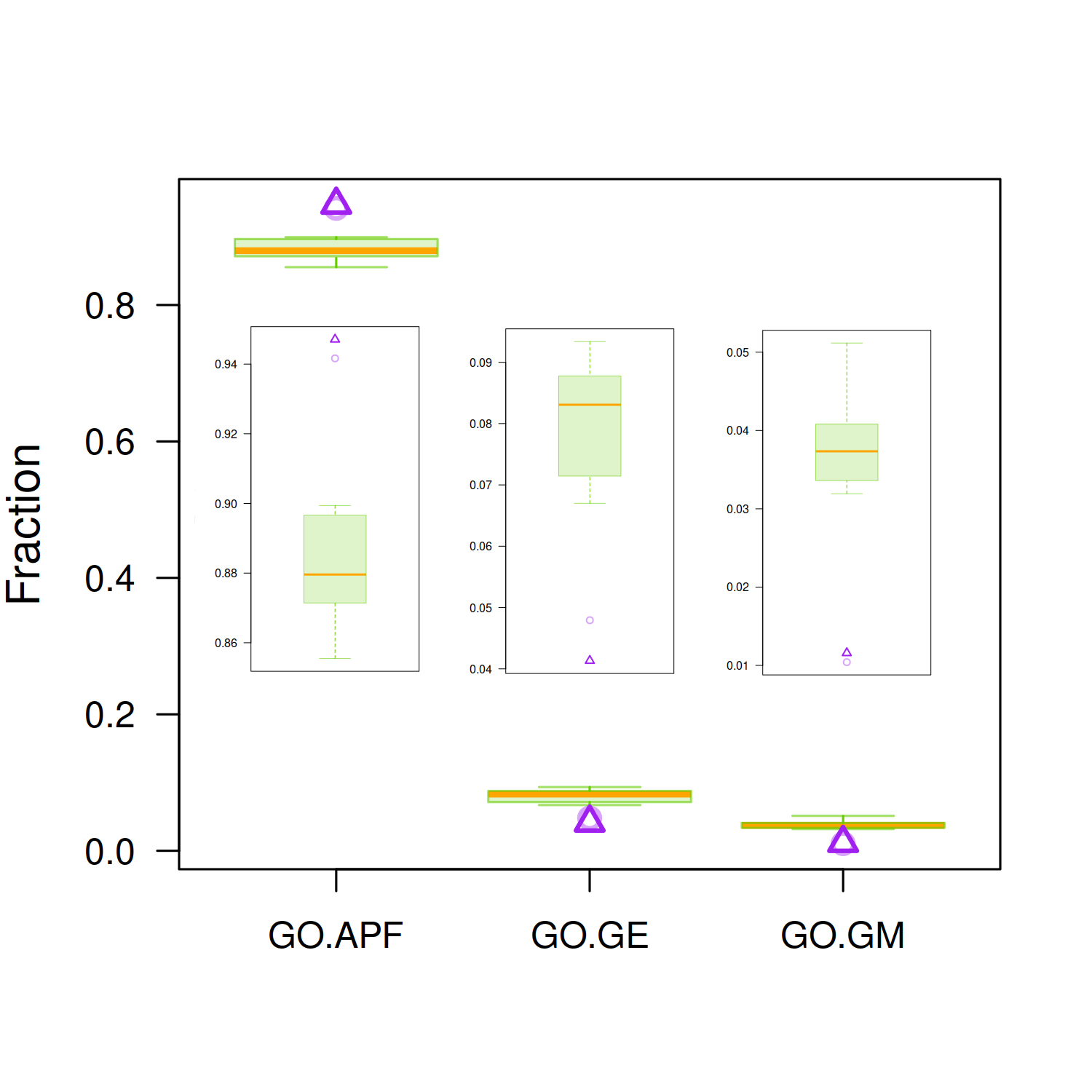}
\includegraphics*[width=0.25\textwidth]{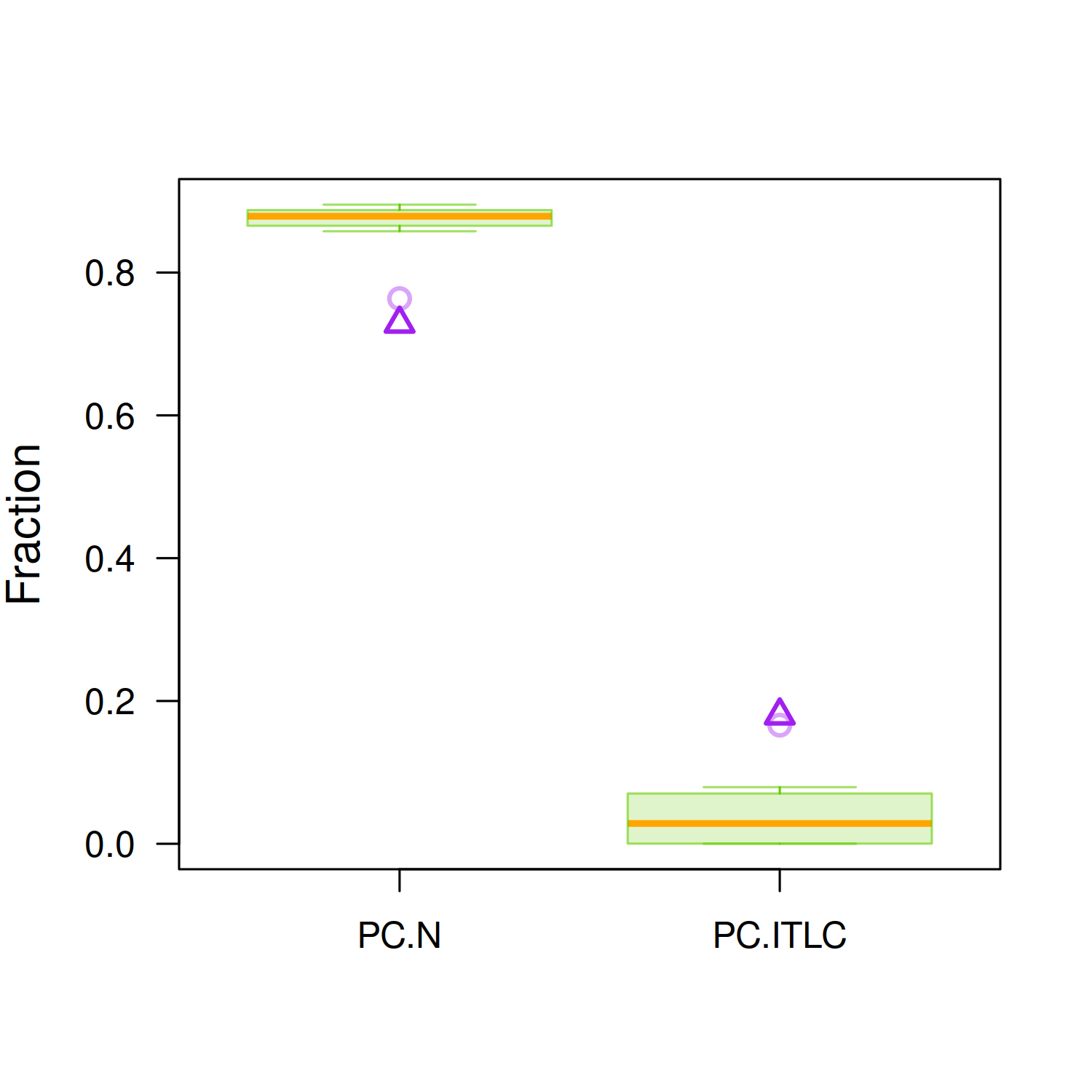}
\end{center}
\begin{center}
\includegraphics*[width=0.25\textwidth]{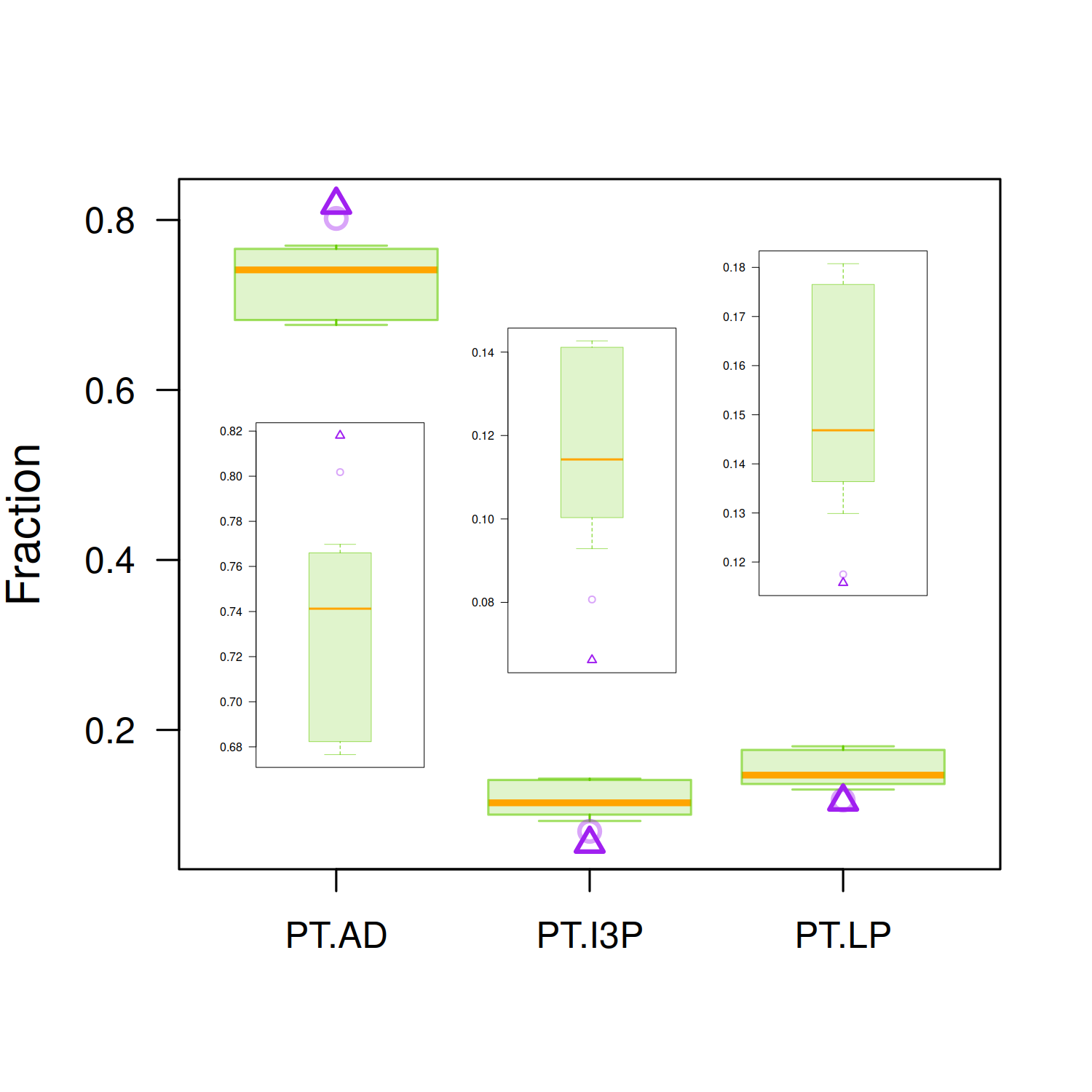}
\includegraphics*[width=0.25\textwidth]{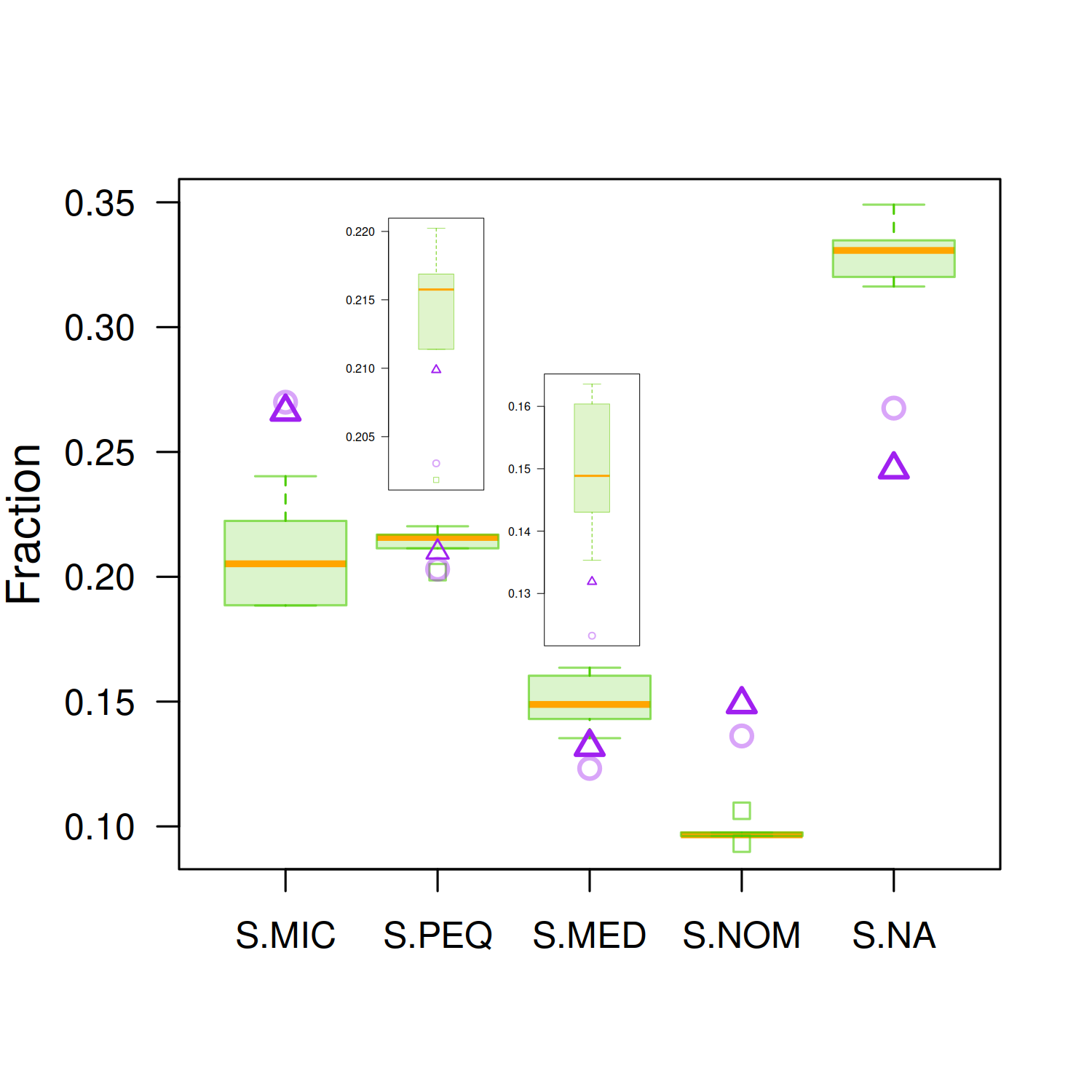}
\end{center}
\includegraphics*[width=0.25\textwidth]{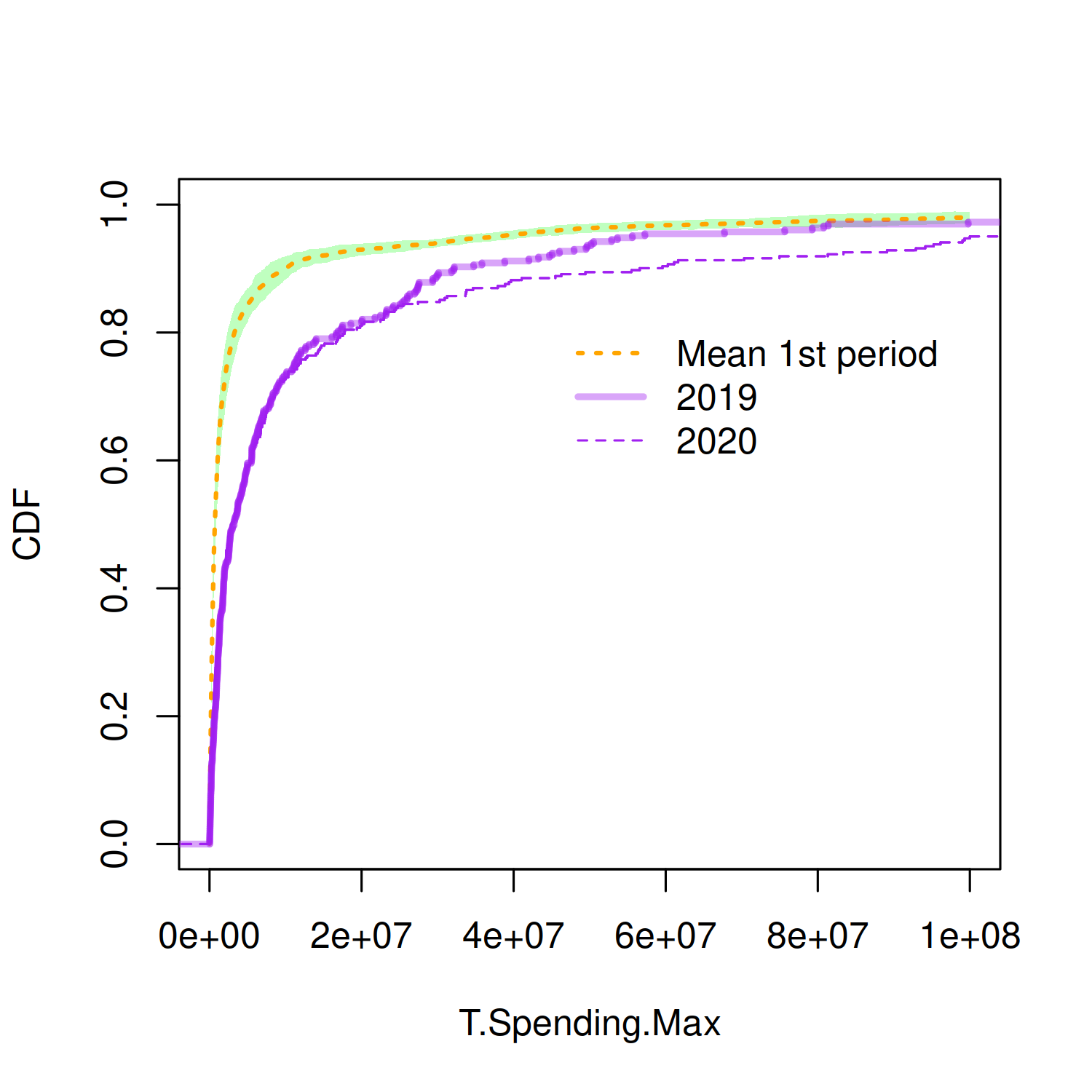}
\includegraphics*[width=0.25\textwidth]{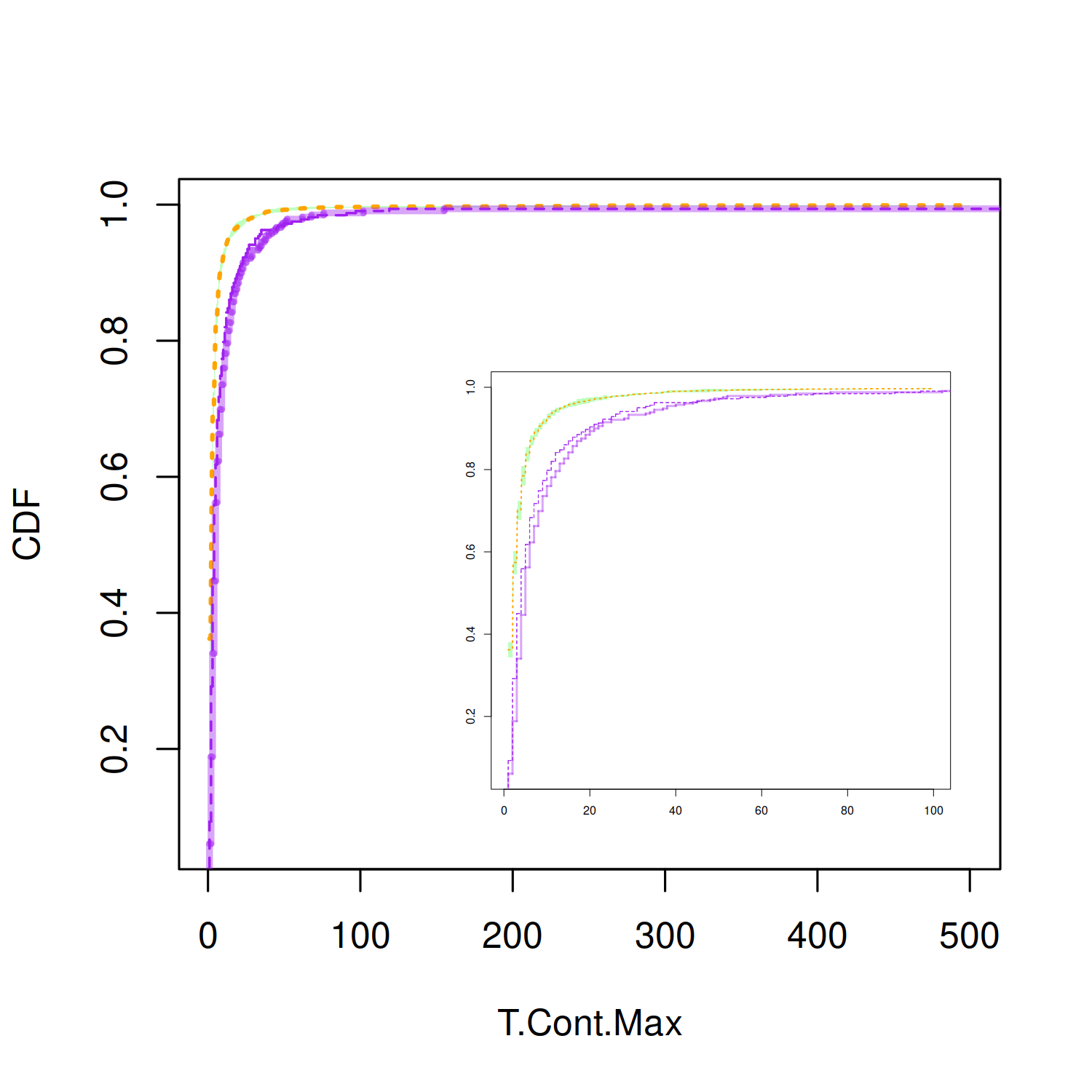}
\begin{center}
    
\end{center}

    \caption{{\it Differences between both government periods for contracts in the NC class:} For the dummy variables ({\bf Top and Center}): The boxplot (in green) groups the data of the 6 years comprising the 1st government period. For the remaining non-dummy variables ({\bf Bottom}) the graph shows the cumulative distribution function (CDF), and the green area represents the CI at 99\% generated by the data of the 1st period. In all graphs, the orange line corresponds to the mean of the 1st government period, and two purple colored curves (symbols) correspond to the first two years of the second government period. The insets appearing in some graphs are a close up to the behavior of the corresponding variable.}
    \label{ci-nc-dif}
\end{figure*}

\begin{figure*}[htb!]
\begin{center}
\includegraphics*[width=0.3\textwidth]{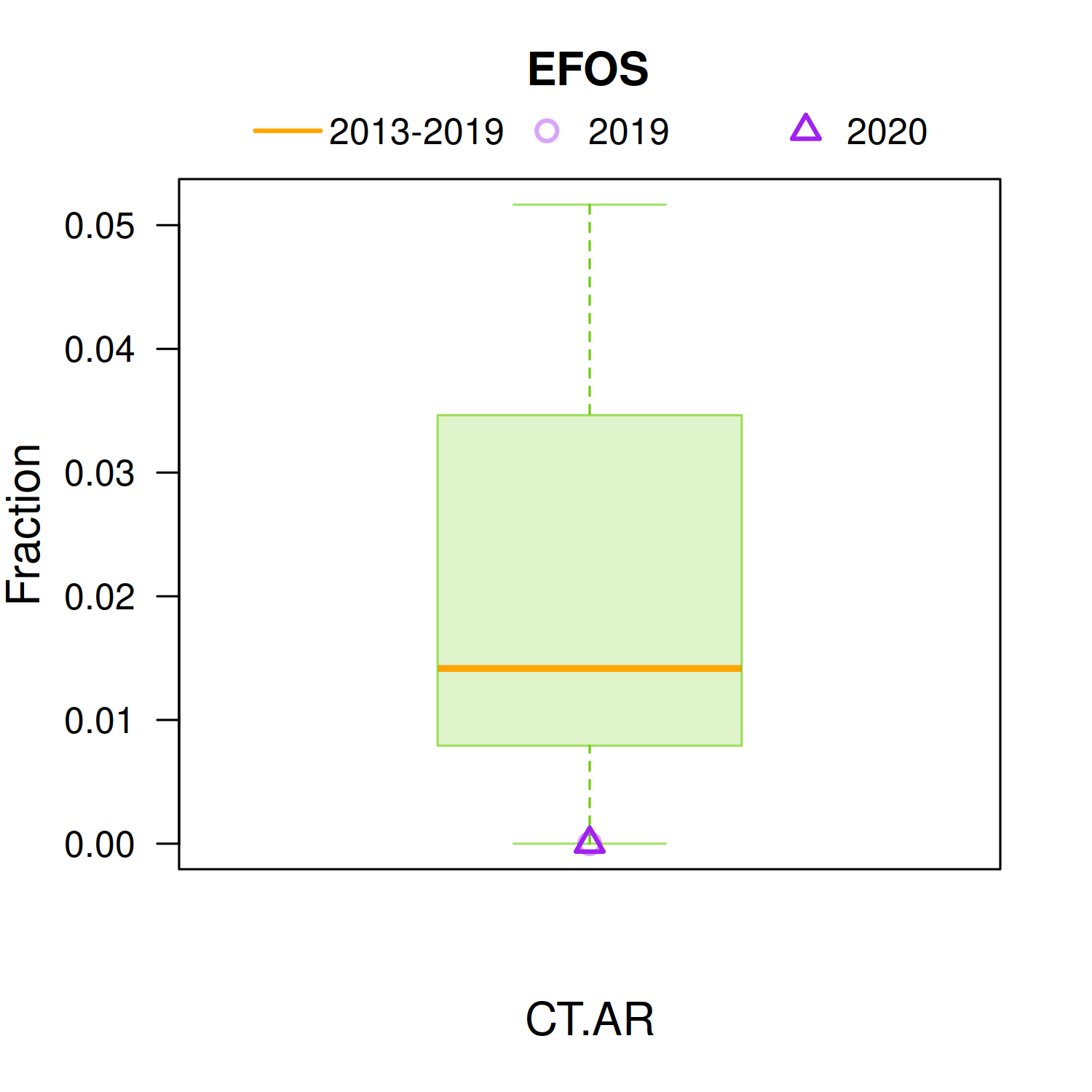}
\includegraphics*[width=0.3\textwidth]{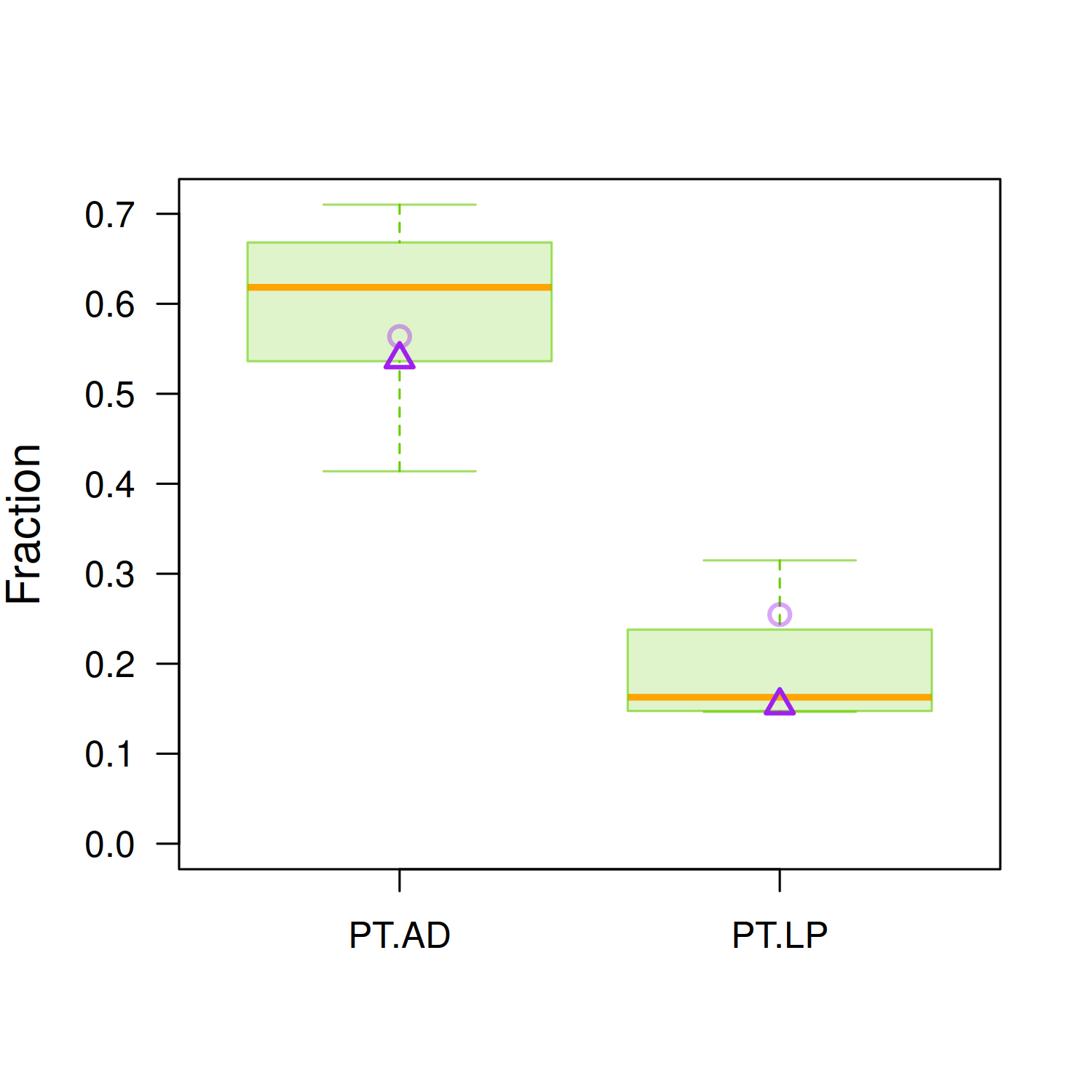}
\end{center}
    \caption{{\it Similarities between both government periods for EFOS class:} For the dummy variables: The boxplot (in green) groups the data of the 6 years comprising the 1st government period, the orange line corresponds to the mean of the 1st government period, and two purple symbols correspond to the first two years of the second government period.}
    \label{ci-efos-sim}
\end{figure*}

\begin{figure*}[htb!]
\begin{center}
\includegraphics*[width=0.35\textwidth]{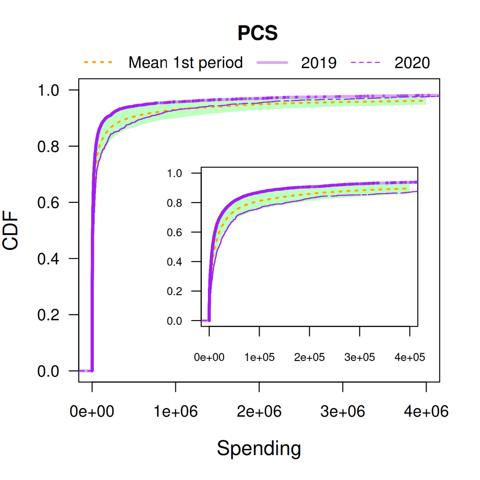}
\end{center}
    \caption{{\it Similarities between both government periods for PCS class:} For the non-dummy variable the graph shows the cumulative distribution function (CDF), and the green area represents the CI at 99\% generated by the data of the 1st period. The orange line corresponds to the mean of the 1st government period, and two purple colored curves correspond to the first two years of the second government period. The inset is a close up to the behavior of the variable for small values of the X axis.}
    \label{ci-pcs-sim}
\end{figure*}

\begin{figure*}[htb!]
\begin{center}
\includegraphics*[width=0.3\textwidth]{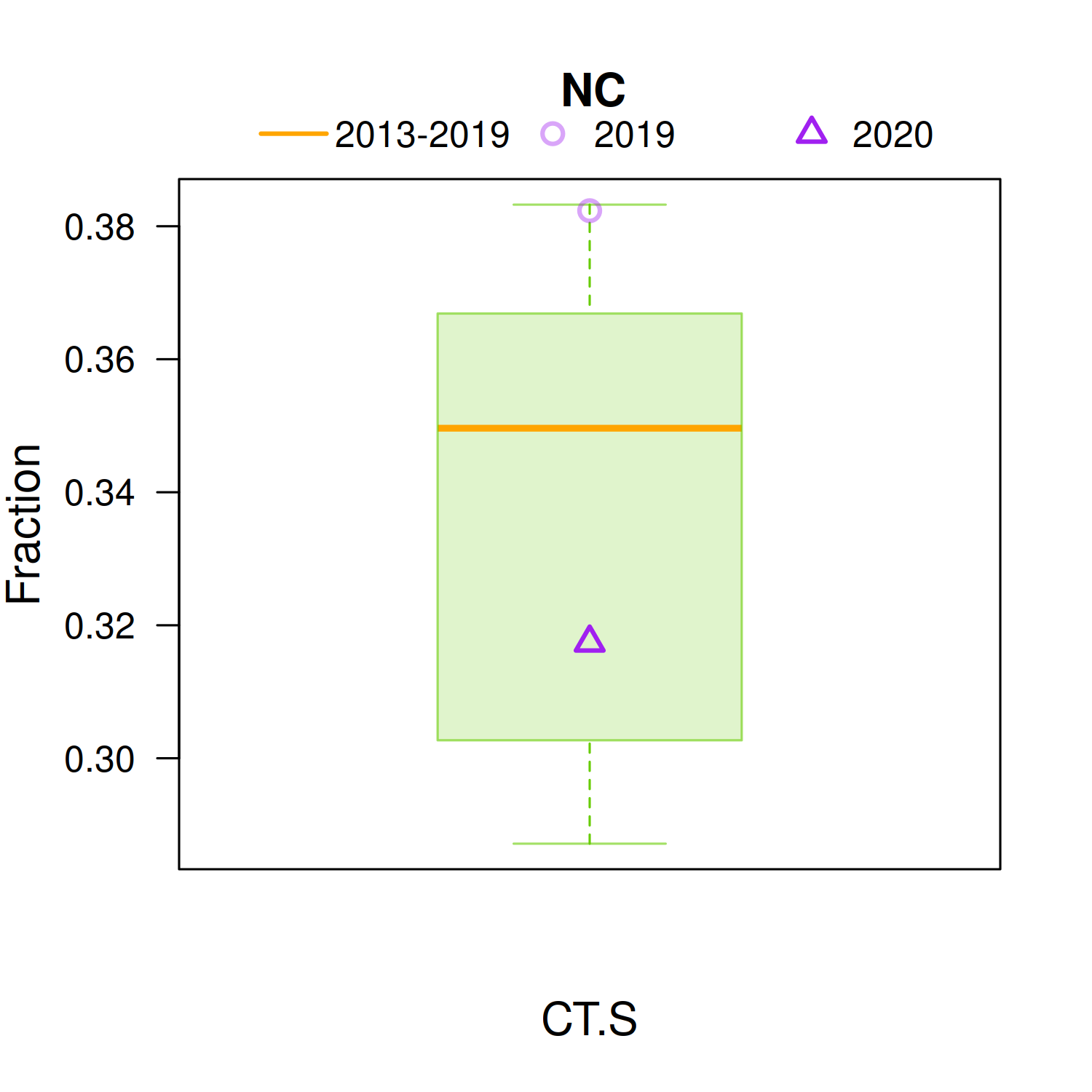}
\includegraphics*[width=0.3\textwidth]{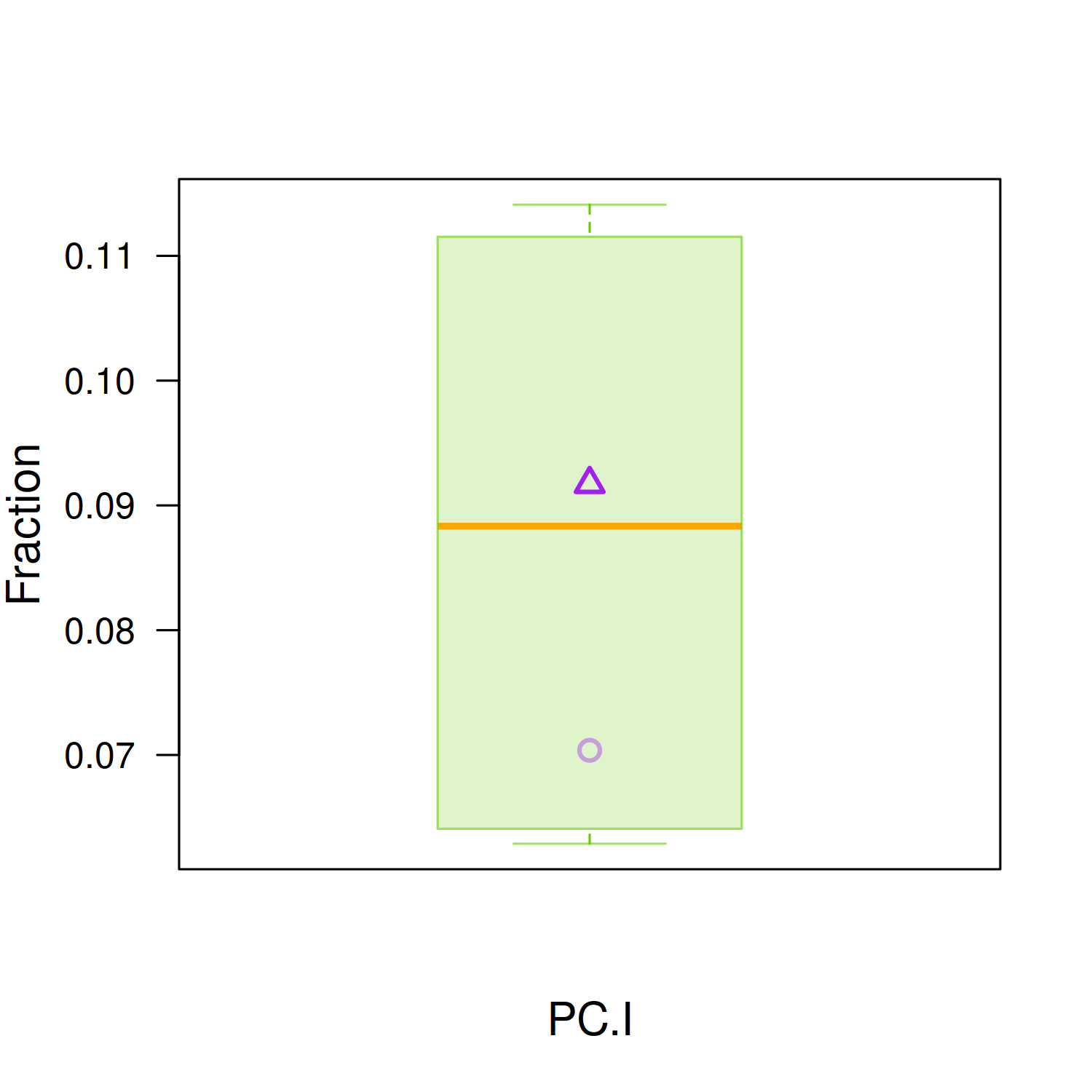}
\end{center}
\begin{center}
\includegraphics*[width=0.3\textwidth]{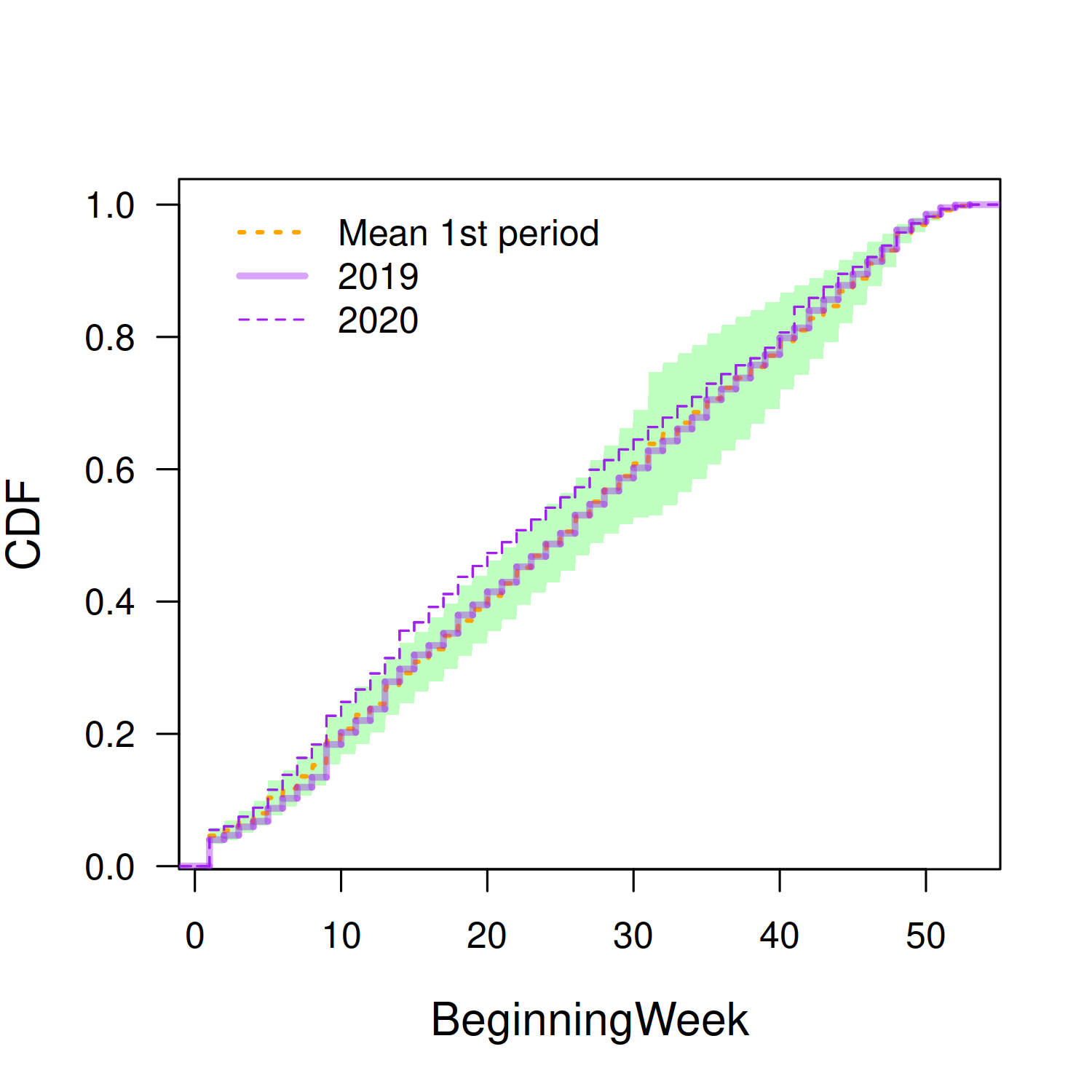}
\includegraphics*[width=0.3\textwidth]{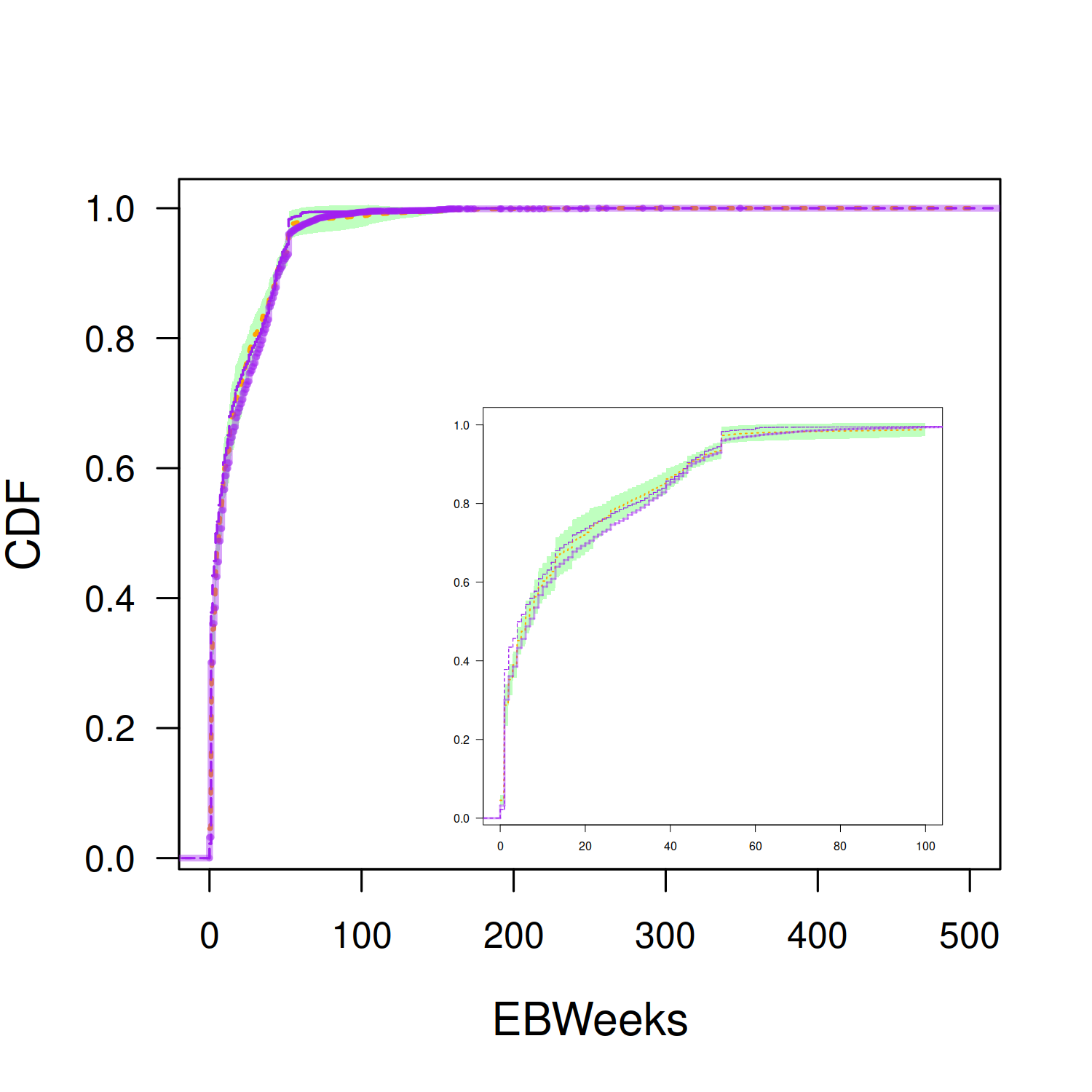}
\end{center}

    \caption{{\it Similarities between both government periods for NC class:} For the dummy variables ({\bf Top}): The boxplot (in green) groups the data of the 6 years comprising the 1st government period. For the remaining non-dummy variables ({\bf Bottom}) the graph shows the cumulative distribution function (CDF), and the green area represents the CI at 99\% generated by the data of the 1st period. In all graphs, the orange line corresponds to the mean of the 1st government period, and two purple colored curves (symbols) correspond to the first two years of the second government period. The inset in the last graph is a close up to the behavior of the variable for small values of the X axis. }
    \label{ci-nc-sim}
\end{figure*}

\begin{figure*}[htb!]
\begin{center}
\includegraphics*[width=0.3\textwidth]{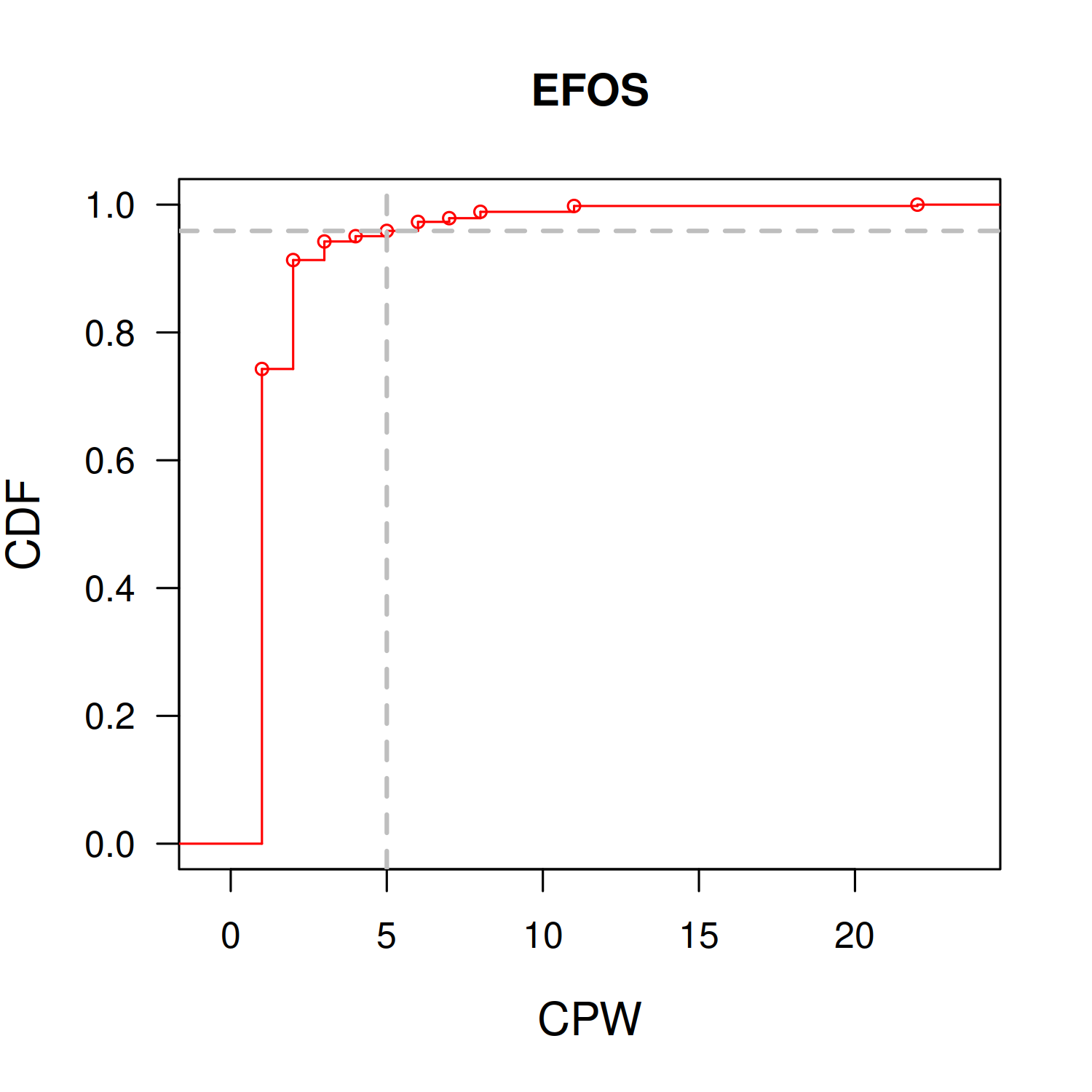}
\includegraphics*[width=0.3\textwidth]{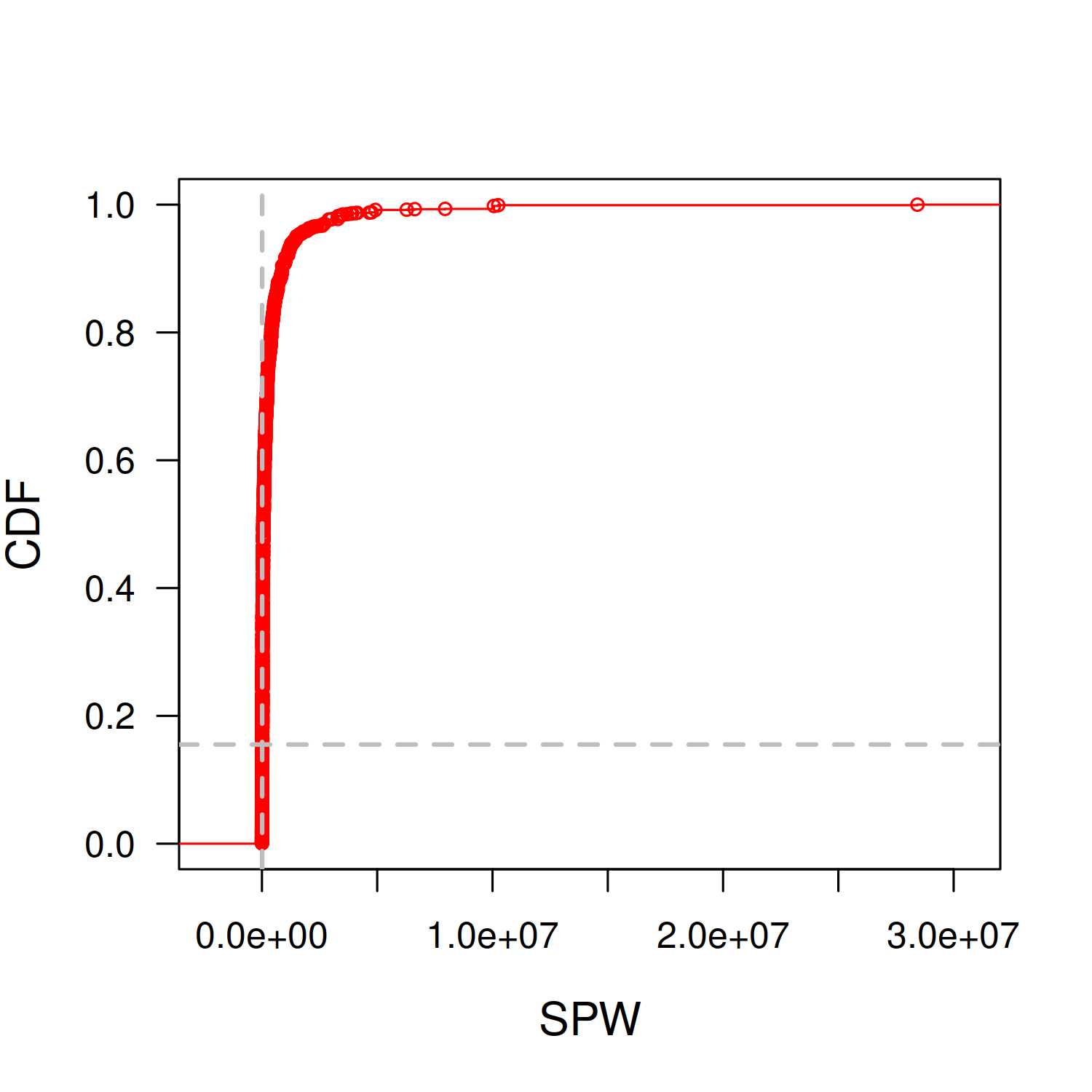}
\end{center}
\begin{center}
\includegraphics*[width=0.3\textwidth]{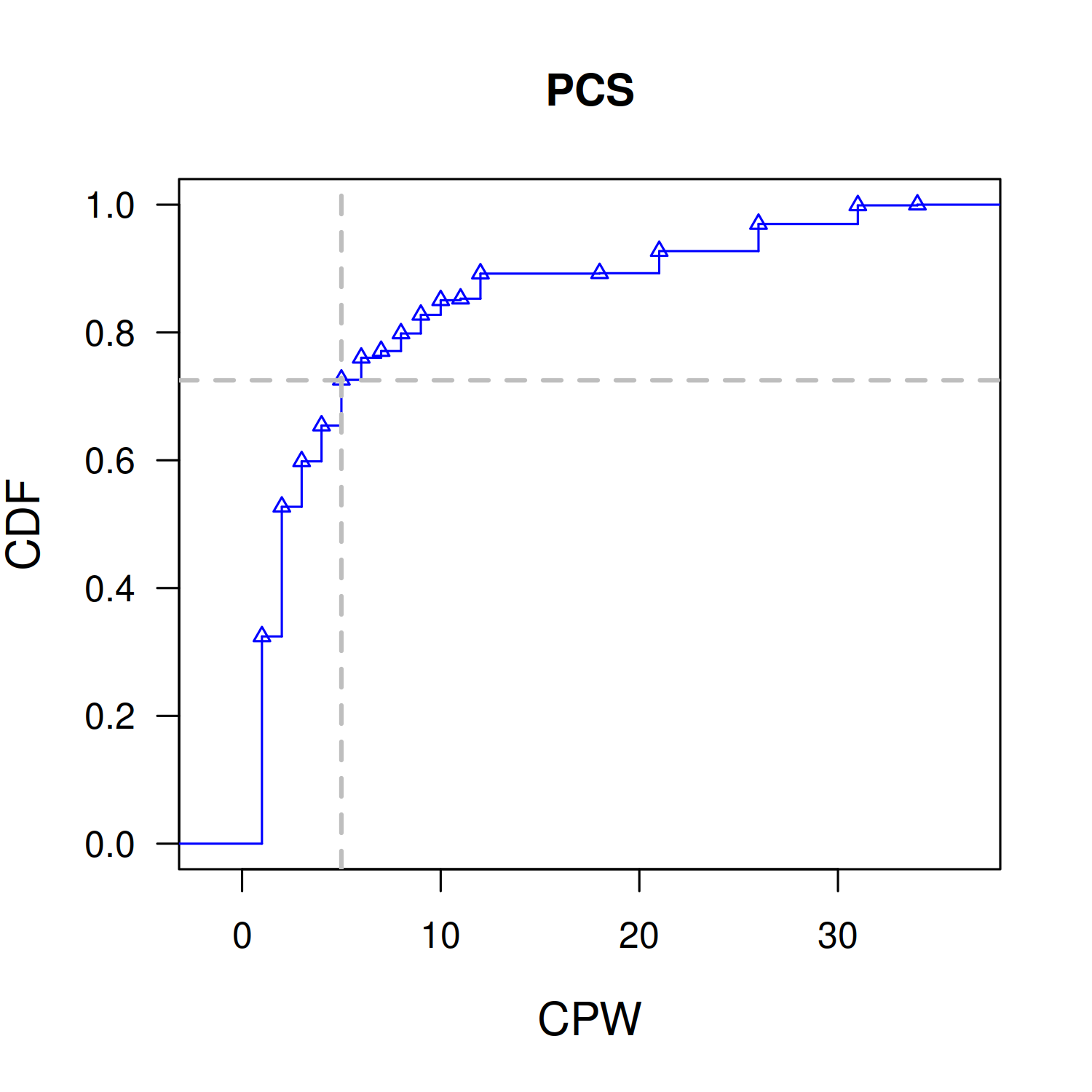}
\includegraphics*[width=0.3\textwidth]{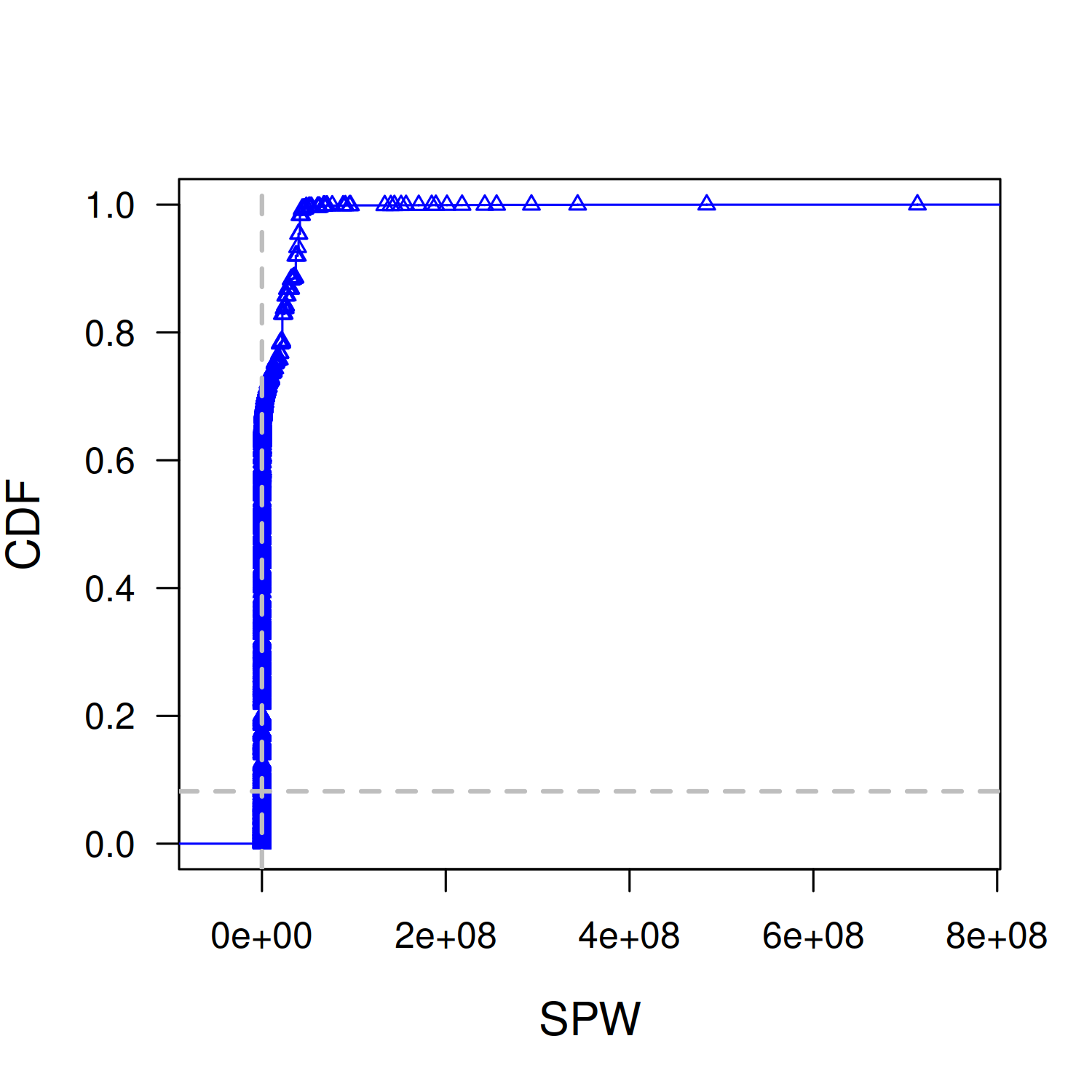}
\end{center}

\begin{center}
\includegraphics*[width=0.3\textwidth]{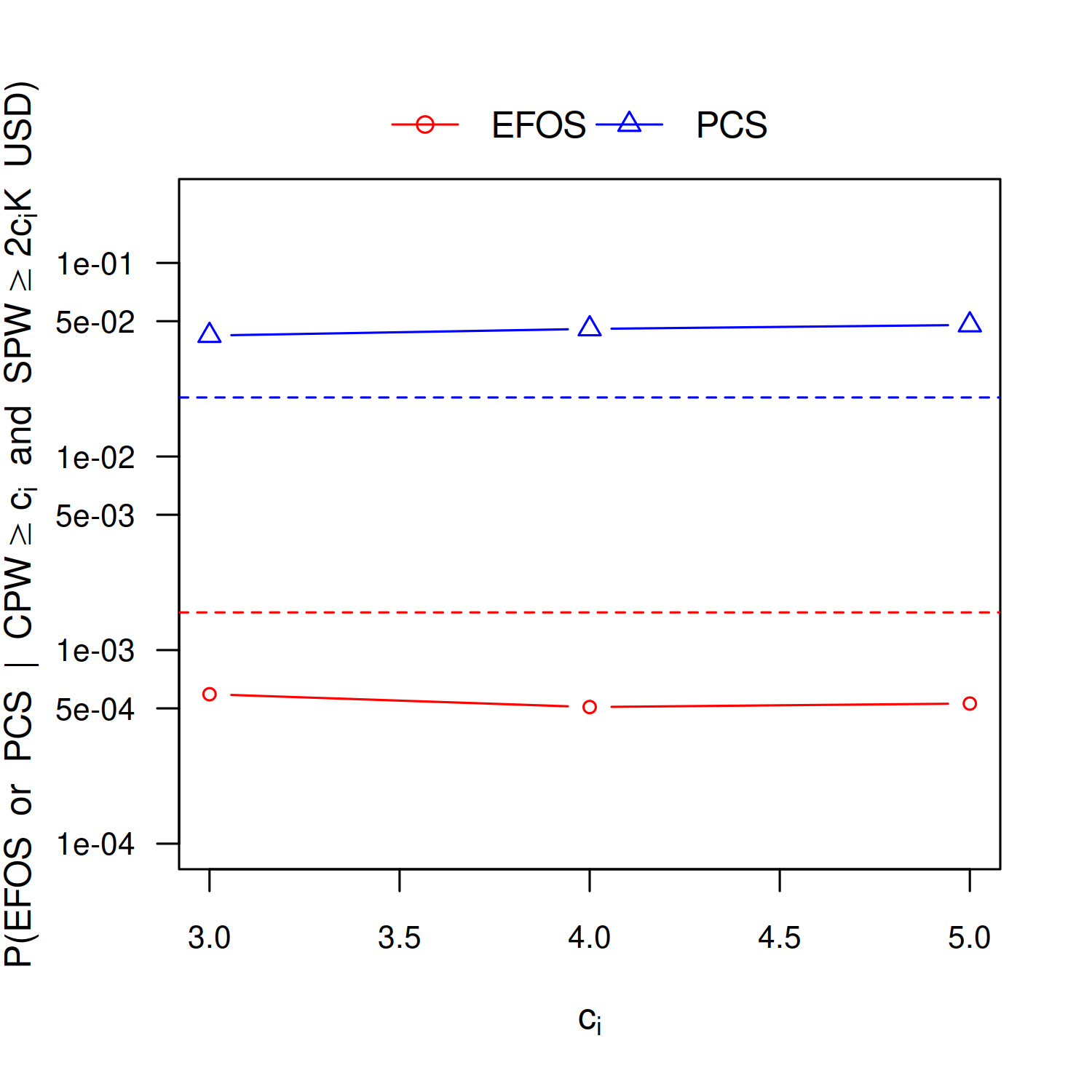}
\end{center}

    \caption{First and second experiments for {\bf CPW} and {\bf SPW} risk factors. For the first experiment the graphs show the CDF for {\bf CPW} and {\bf SPW} variables ({\bf Left} and {\bf Right}, respectively) for the two different classes EFOS ({\bf Up}) and PCS ({\bf Center}). The dashed grey lines indicate the threshold for each risk factor and the value of the CDF at which it was crossed. This value in the CDF indicates the probability to find a contract with the risk factor above the marked threshold given that the contract belongs to the class. For the second experiment ({\bf Bottom}) the graph shows the probability for a contract to be corrupt (EFOS or PCS, solid lines) given it has an {\bf CPW}$\geq c_i$ and a {\bf SPW} $\geq 2c_i$K USD, for several values of the threshold, $c_i=\{3, 4, 5\}$. The dashed lines correspond to the probability to find an EFOS (red) or PCS (blue) contract in the source list.}
    \label{testriskfactor}
\end{figure*}

\begin{figure*}[htb!]
\begin{center}
\includegraphics*[width=0.3\textwidth]{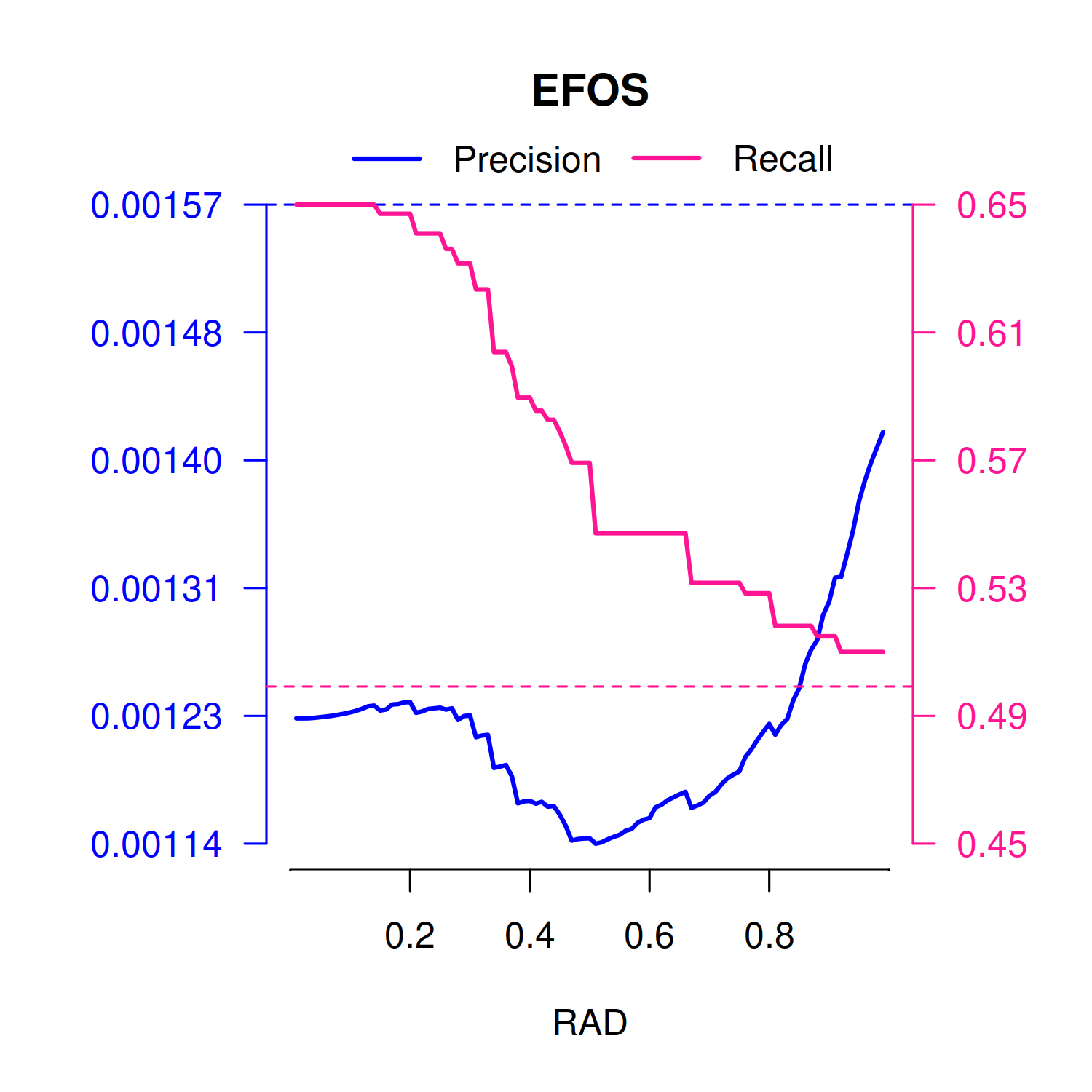}
\includegraphics*[width=0.3\textwidth]{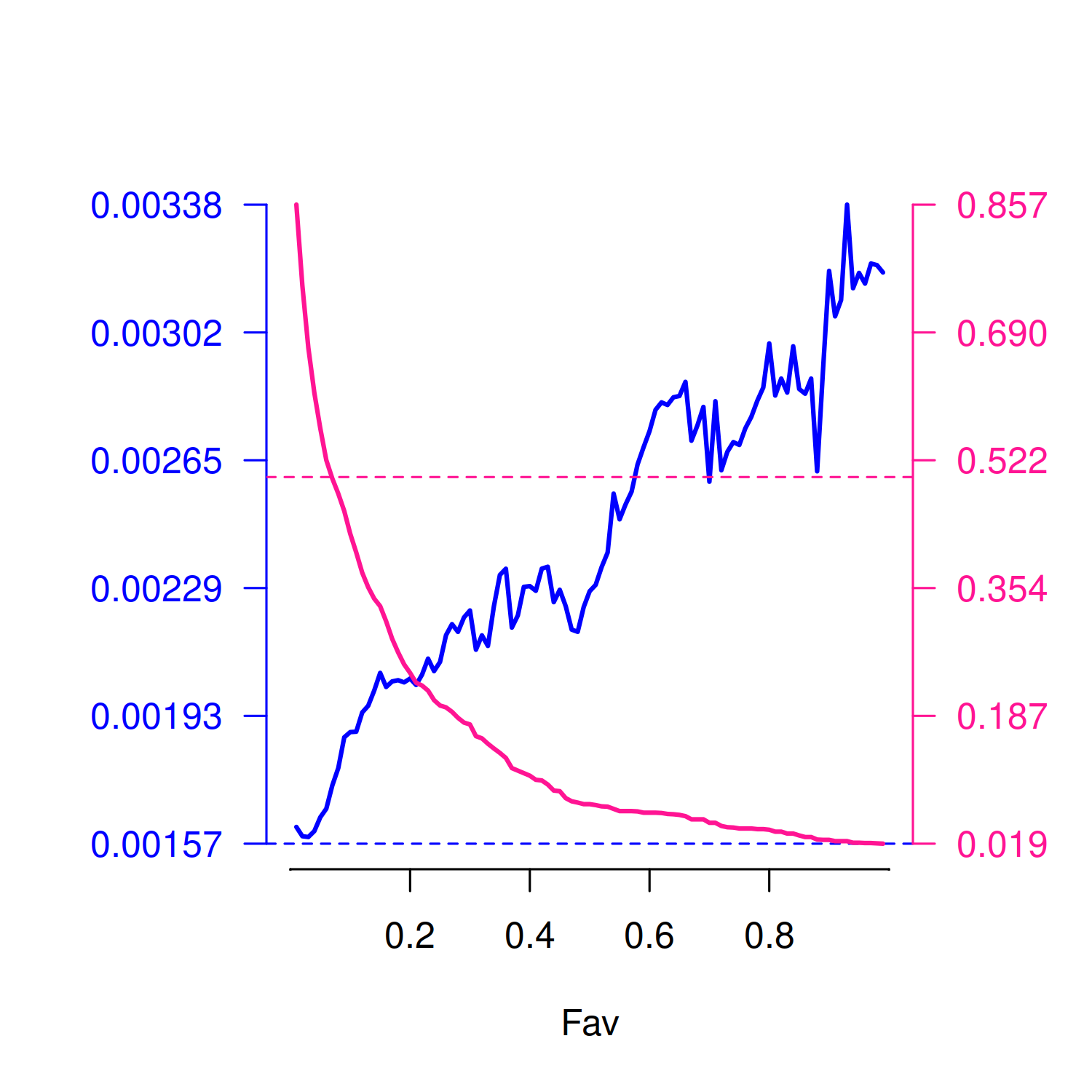}
\includegraphics*[width=0.3\textwidth]{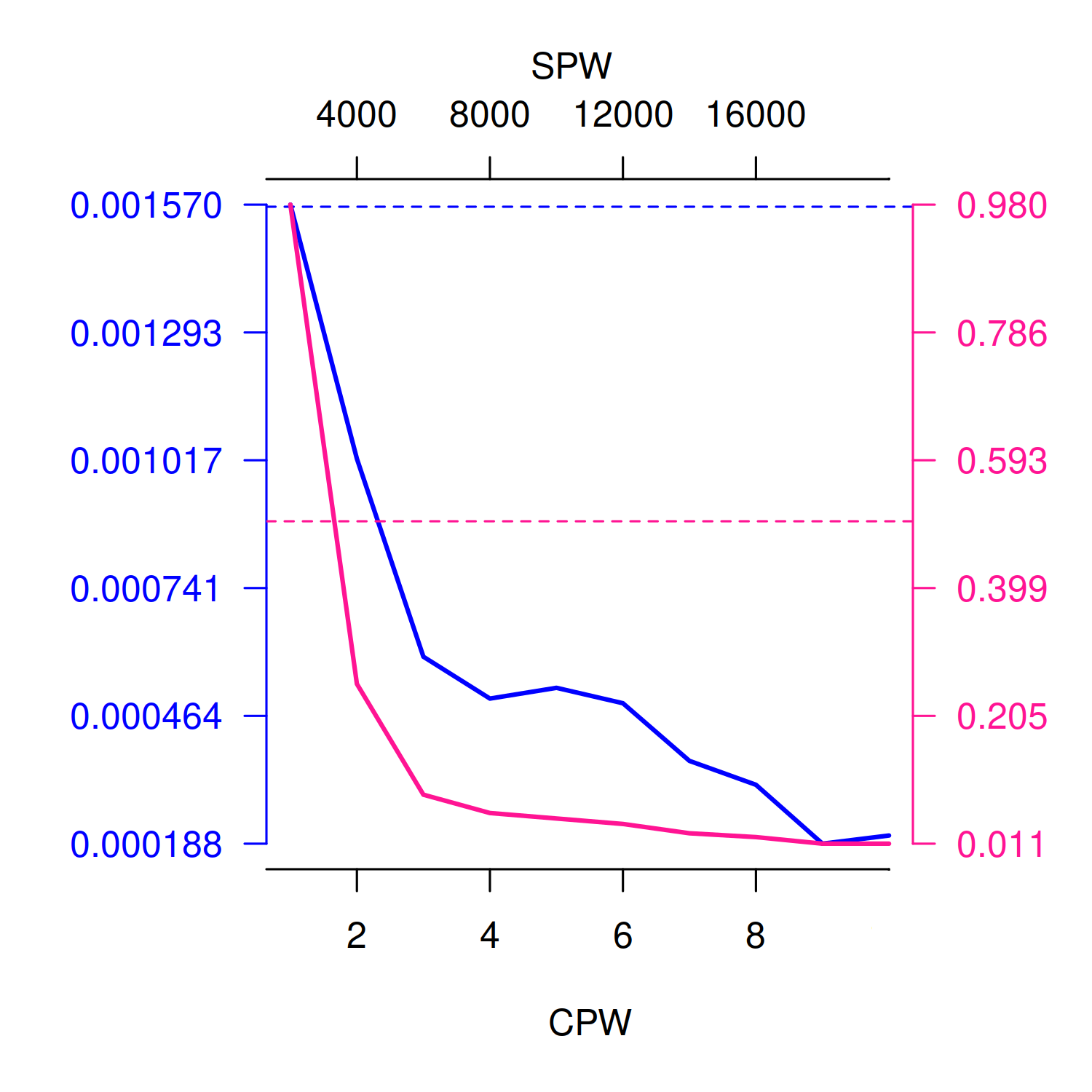}
\end{center}
\begin{center}
\includegraphics*[width=0.3\textwidth]{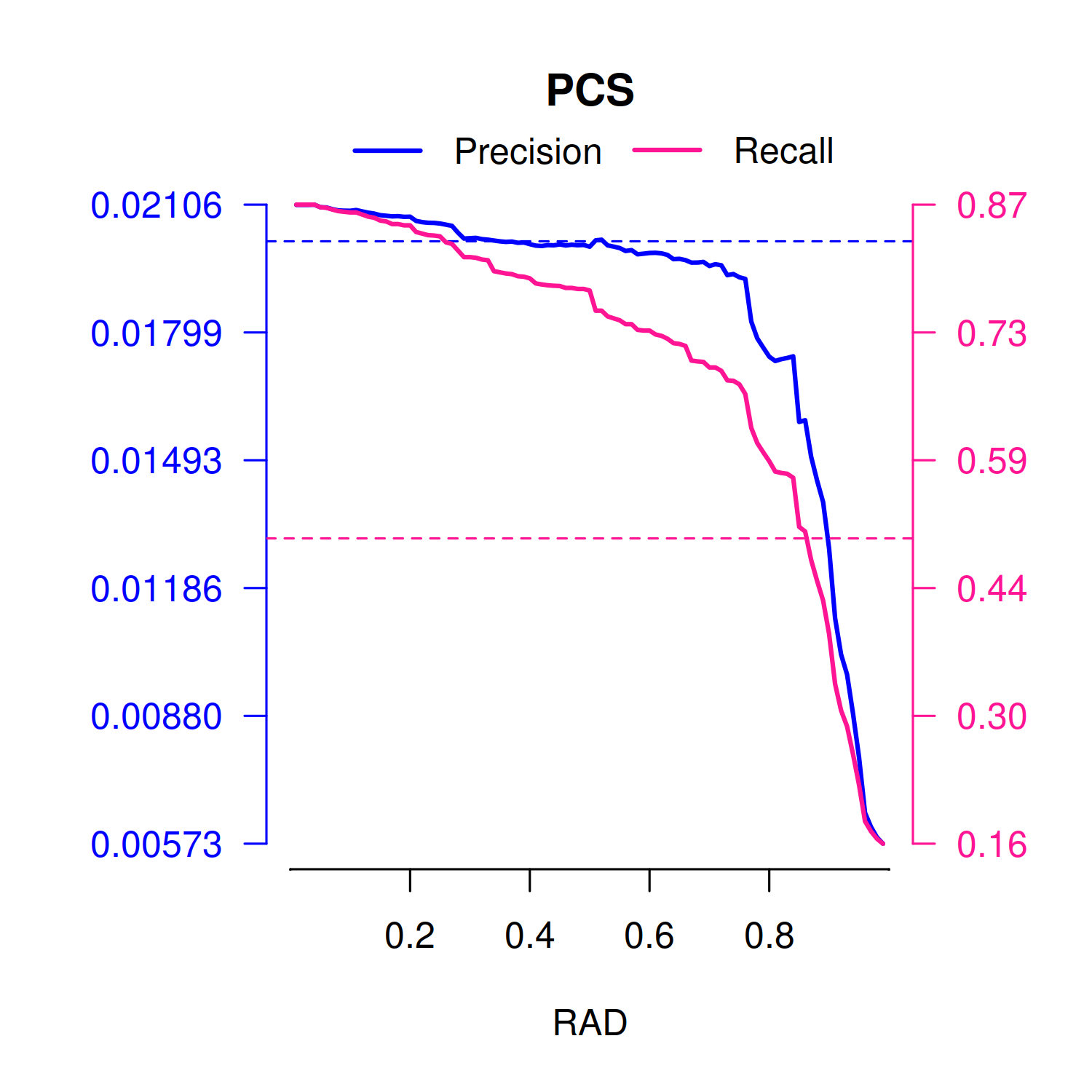}
\includegraphics*[width=0.3\textwidth]{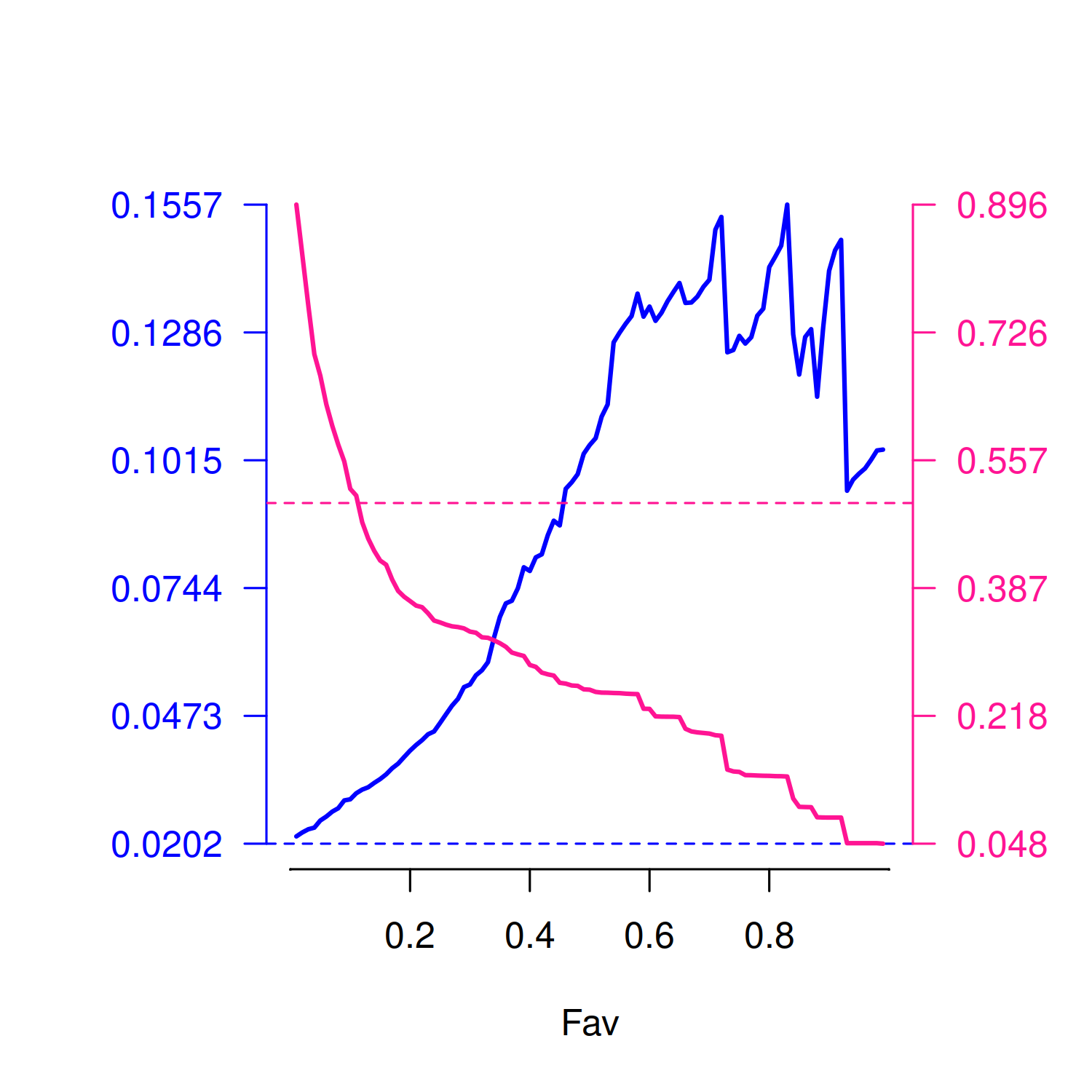}
\includegraphics*[width=0.3\textwidth]{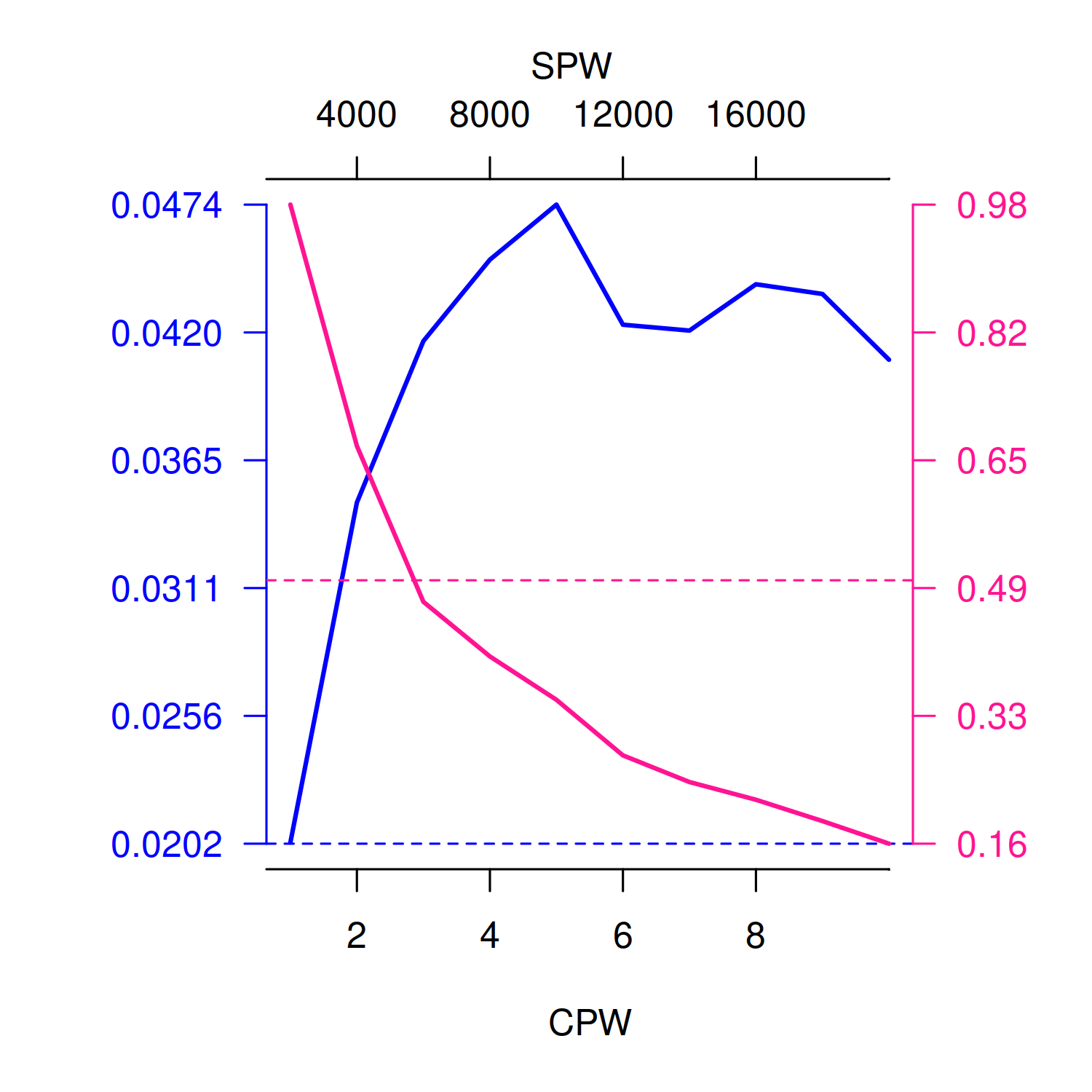}
\end{center}
    \caption{Precision/Recall curves for each risk factor. {\bf Top} the EFOS class. {\bf Bottom} the PCS class. The blue and pink straight lines indicate respectively the precision and recall values calculated over the range of possible thresholds for each risk factor ({\bf RAD}, {\bf Fav}, and {\bf SPW-CPW}). {\it Precision}, defined as the ratio between the actual positive instances and all the cases predicted as positive by the classifier, is directly related to the conditional probability that a contract belongs to an EFOS or PCS class, given that it exceeds a certain threshold of each risk factor (experiment 2). On the other side, {\it recall}, defined as the ratio between the correctly detected instances by the classifier and all positive instances, is equivalent to the conditional probability that a contract exceeds the risk factor threshold, given that it belongs to a corrupt class (experiment 1). The dashed lines are shown the success threshold for each curve (experiment): If the precision curve is over the probability to randomly find a corrupt contract in the source list (blue dashed line), then the second experiment was successful; If the recall curve is over 0.5 (pink dashed line), {\it i.e.} the risk factor describes more than 50\% of the contracts in each corrupt class, then the first experiment was successful.}
    \label{PR}
\end{figure*}

\begin{figure*}[htb!]
\begin{center}
\includegraphics*[width=0.35\textwidth]{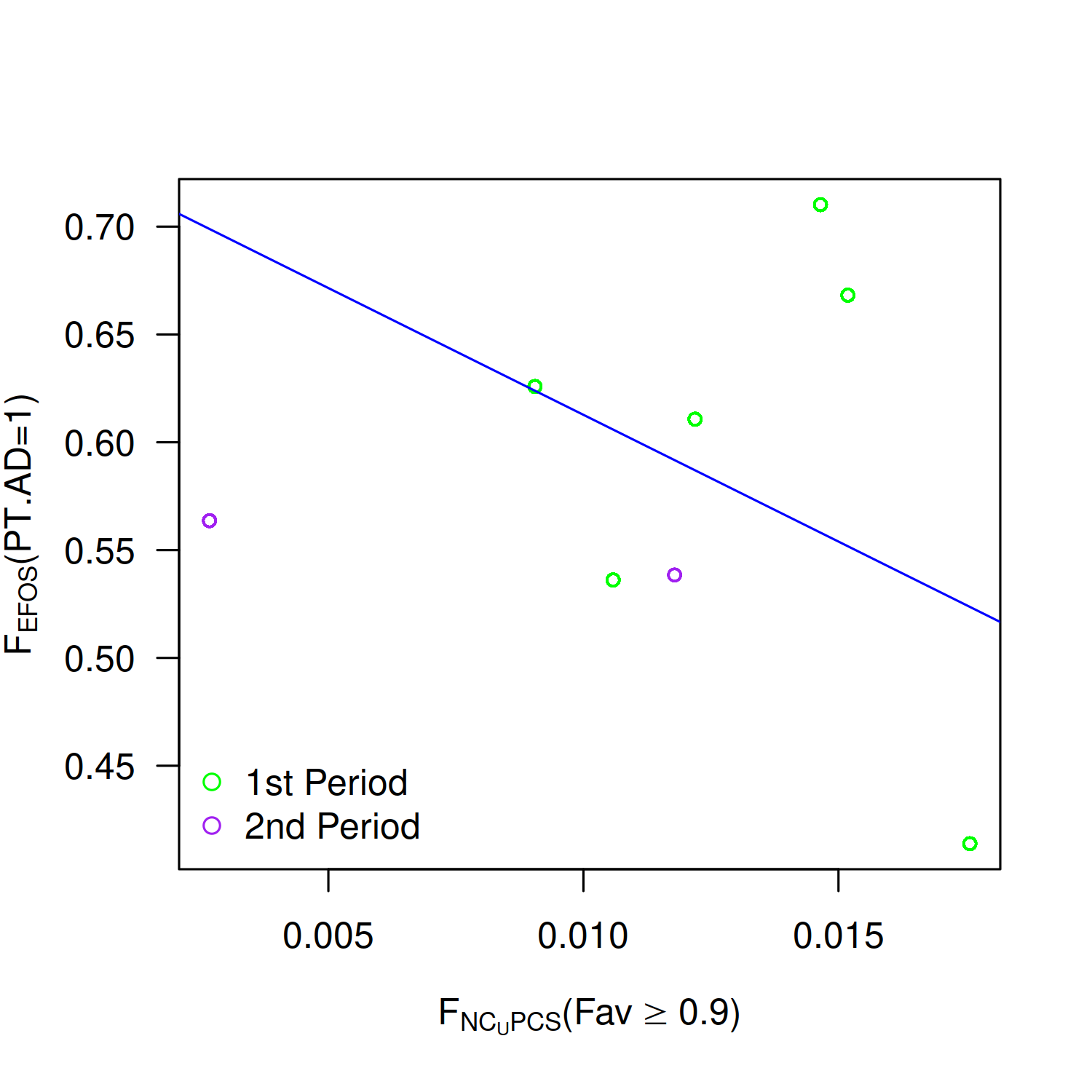}
\includegraphics*[width=0.35\textwidth]{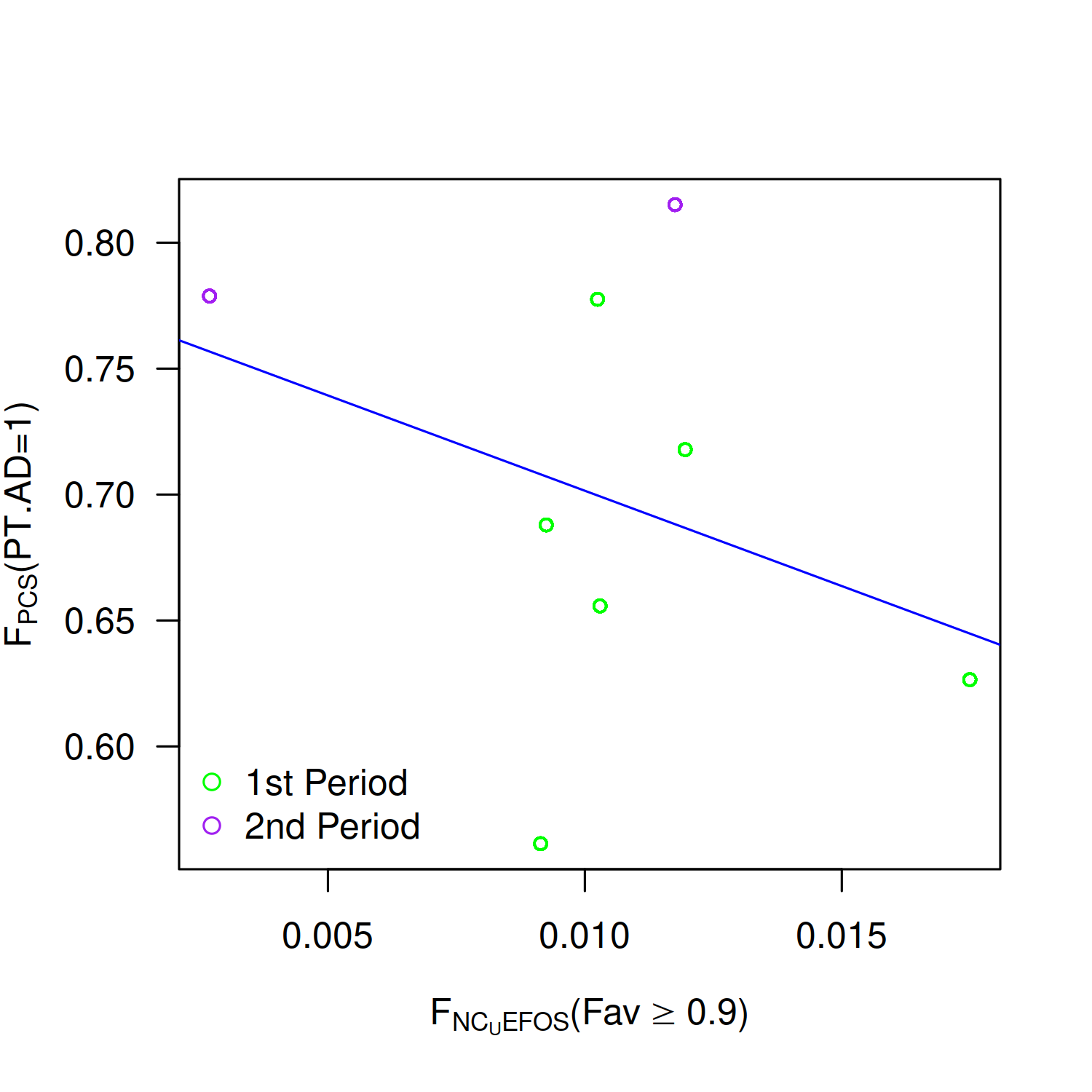}
\end{center}
\caption{Linear model to correlate the fraction of contracts with {\bf Fav}$\geq$0.9 per year (as independent variable) with the fraction of contracts by single-bidder per year ({\it i.e.} contracts for which {\bf PT.AD}=1), as dependent variable for EFOS and PCS class ({\bf Up} and {\bf Bottom}, respectively). The 1st period years are marked in green, and in purple the years corresponding to the 2nd period. Each data is weighted by the number of contracts detected per year in each class. For the first model (EFOS class) we obtained a $R^2 = 0.16$, an intercept of 0.73, and a slope of -11.75. For the second model (PCS class) we obtained a $R^2 = 0.14$, an intercept of 0.77, and a slope of -7.56. All  the parameters with a $p_v\ll 0.01$.}
\label{favvsptad}
\end{figure*}

\begin{table*}[htb!]
\begin{tabular}{lccccccccccc}
\hline\hline
\textbf{Variable} & \textbf{1}                                             & \textbf{2}                                             & \textbf{3}                                             & \textbf{4}                                             & \textbf{5}                                            & \textbf{6}                                              & \textbf{7}                                              & \textbf{8}                                             & \textbf{9}                                            & \textbf{10}                                            & \textbf{11}                                            \\
\hline
RAD               & \begin{tabular}[c]{@{}c@{}}1.35\\ (0.008)\end{tabular} & \begin{tabular}[c]{@{}c@{}}0.72\\ (0.01)\end{tabular}  & \begin{tabular}[c]{@{}c@{}}1.27\\ (0.01)\end{tabular}  & \begin{tabular}[c]{@{}c@{}}0.80\\ (0.01)\end{tabular}  & \begin{tabular}[c]{@{}c@{}}0.60\\ (0.01)\end{tabular} & \begin{tabular}[c]{@{}c@{}}1.12\\ (0.009)\end{tabular}  & \begin{tabular}[c]{@{}c@{}}0.83\\ (0.01)\end{tabular}   & \begin{tabular}[c]{@{}c@{}}1.30\\ (0.01)\end{tabular}  & \begin{tabular}[c]{@{}c@{}}1.07\\ (0.01)\end{tabular} & \begin{tabular}[c]{@{}c@{}}1.24\\ (0.01)\end{tabular}  & \begin{tabular}[c]{@{}c@{}}0.84\\ (0.01)\end{tabular}  \\
PC.N              & \begin{tabular}[c]{@{}c@{}}0.66\\ (0.008)\end{tabular} &                                                        &                                                        &                                                        &                                                       &                                                         &                                                         &                                                        &                                                       &                                                        &                                                        \\
PC.I              &                                                        & \begin{tabular}[c]{@{}c@{}}-0.16\\ (0.01)\end{tabular} &                                                        &                                                        &                                                       &                                                         &                                                         &                                                        &                                                       &                                                        &                                                        \\
PC.ITLC           &                                                        &                                                        & \begin{tabular}[c]{@{}c@{}}-0.49\\ (0.01)\end{tabular} &                                                        &                                                       &                                                         &                                                         &                                                        &                                                       &                                                        &                                                        \\
CT.ADQ            &                                                        &                                                        &                                                        & \begin{tabular}[c]{@{}c@{}}-0.11\\ (0.01)\end{tabular} &                                                       &                                                         &                                                         &                                                        &                                                       &                                                        &                                                        \\
CT.AR             &                                                        &                                                        &                                                        &                                                        & \begin{tabular}[c]{@{}c@{}}0.32\\ (0.01)\end{tabular} &                                                         &                                                         &                                                        &                                                       &                                                        &                                                        \\
S.NOM             &                                                        &                                                        &                                                        &                                                        &                                                       & \begin{tabular}[c]{@{}c@{}}-0.43\\ (0.009)\end{tabular} &                                                         &                                                        &                                                       &                                                        &                                                        \\
S.PEQ             &                                                        &                                                        &                                                        &                                                        &                                                       &                                                         & \begin{tabular}[c]{@{}c@{}}-0.032\\ (0.01)\end{tabular} &                                                        &                                                       &                                                        &                                                        \\
S.MIC             &                                                        &                                                        &                                                        &                                                        &                                                       &                                                         &                                                         & \begin{tabular}[c]{@{}c@{}}-0.51\\ (0.01)\end{tabular} &                                                       &                                                        &                                                        \\
S.NA              &                                                        &                                                        &                                                        &                                                        &                                                       &                                                         &                                                         &                                                        & \begin{tabular}[c]{@{}c@{}}0.38\\ (0.01)\end{tabular} &                                                        &                                                        \\
CPW - SPW         &                                                        &                                                        &                                                        &                                                        &                                                       &                                                         &                                                         &                                                        &                                                       & \begin{tabular}[c]{@{}c@{}}-0.45\\ (0.01)\end{tabular} &                                                        \\
Fav               &                                                        &                                                        &                                                        &                                                        &                                                       &                                                         &                                                         &                                                        &                                                       &                                                        & \begin{tabular}[c]{@{}c@{}}-0.02\\ (0.01)\end{tabular} \\
\hline
$R^2$             & 0.92                                                   & 0.73                                                   & 0.79                                                   & 0.73                                                   & 0.77                                                  & 0.85                                                    & 0.72                                                    & 0.78                                                   & 0.82                                                  & 0.78                                                   & 0.72                                                   \\
VIF               & 2.27                                                   & 2.62                                                   & 3.62                                                   & 1.20                                                   & 2.33                                                  & 1.54                                                    & 1.97                                                    & 4.39                                                   & 1.50                                                  & 3.72                                                   & 1.25 \\
\hline\hline
\end{tabular}
\caption{{\it Multivariate regression in the EFOS class:}  For all models the dependent variable is the fraction of contracts by single-bidder in the EFOS class, and one of the independent variables is always the fraction of contracts in which the relation buyer-supplier has an {\bf RAD $\geq$ 0.5}. The second independent variable is a control variable related to procedure character ({\bf PC}), contract type ({\bf CT}), supplier's size ({\bf S}), or one of the remaining risk factors: contracts and spending per week ({\bf CPW - SPW}), or favoritism ({\bf Fav}). Observations are weighted by the number of EFOS contracts per year. The table shows the standardized coefficients, their standard error (in parentheses), the $R^2$, and the mean variance inflation factor (VIF). Only models with VIF $<$ 5 are shown. All the reported parameters are highly significant with $p_v\ll 0.01$. {\it Note:} A \lq\lq complete'' model with the most significant independent variables in each model was performed, but the multicollinearity between the independent variables was very high. We do not report it.}
\end{table*}

\begin{table*}[htb!]
\begin{tabular}{lccccccccc}
\hline \hline
\textbf{Variable} & \textbf{1}                                             & \textbf{2}                                              & \textbf{3}                                             & \textbf{4}                                              & \textbf{5}                                             & \textbf{6}                                              & \textbf{7}                                             & \textbf{8}                                              & \textbf{9}                                             \\
\hline
RAD               & \begin{tabular}[c]{@{}c@{}}1.01\\ (0.004)\end{tabular} & \begin{tabular}[c]{@{}c@{}}0.77\\ (0.003)\end{tabular}  & \begin{tabular}[c]{@{}c@{}}0.79\\ (0.003)\end{tabular} & \begin{tabular}[c]{@{}c@{}}0.93\\ (0.002)\end{tabular}  & \begin{tabular}[c]{@{}c@{}}0.81\\ (0.002)\end{tabular} & \begin{tabular}[c]{@{}c@{}}1.03\\ (0.003)\end{tabular}  & \begin{tabular}[c]{@{}c@{}}1.10\\ (0.002)\end{tabular} & \begin{tabular}[c]{@{}c@{}}0.94\\ (0.003)\end{tabular}  & \begin{tabular}[c]{@{}c@{}}1.14\\ (0.002)\end{tabular} \\
PC.N              & \begin{tabular}[c]{@{}c@{}}0.13\\ (0.004)\end{tabular} &                                                         &                                                        &                                                         &                                                        &                                                         &                                                        &                                                         &                                                        \\
PC.I              &                                                        & \begin{tabular}[c]{@{}c@{}}-0.19\\ (0.003)\end{tabular} &                                                        &                                                         &                                                        &                                                         &                                                        &                                                         &                                                        \\
CT.S              & \multicolumn{1}{l}{}                                   & \multicolumn{1}{l}{}                                    & \begin{tabular}[c]{@{}c@{}}0.14\\ (0.003)\end{tabular} & \multicolumn{1}{l}{}                                    & \multicolumn{1}{l}{}                                   & \multicolumn{1}{l}{}                                    & \multicolumn{1}{l}{}                                   & \multicolumn{1}{l}{}                                    & \multicolumn{1}{l}{}                                   \\
CT.ADQ            &                                                        &                                                         &                                                        & \begin{tabular}[c]{@{}c@{}}-0.14\\ (0.002)\end{tabular} &                                                        &                                                         &                                                        &                                                         &                                                        \\
CT.AR             &                                                        &                                                         &                                                        &                                                         & \begin{tabular}[c]{@{}c@{}}0.17\\ (0.002)\end{tabular} &                                                         &                                                        &                                                         &                                                        \\
S.NOM             &                                                        &                                                         &                                                        &                                                         &                                                        & \begin{tabular}[c]{@{}c@{}}-0.17\\ (0.003)\end{tabular} &                                                        &                                                         &                                                        \\
S.PEQ             &                                                        &                                                         &                                                        &                                                         &                                                        &                                                         & \begin{tabular}[c]{@{}c@{}}0.30\\ (0.002)\end{tabular} &                                                         &                                                        \\
S.NA              &                                                        &                                                         &                                                        &                                                         &                                                        &                                                         &                                                        & \begin{tabular}[c]{@{}c@{}}0.049\\ (0.003)\end{tabular} &                                                        \\
Fav               &                                                        &                                                         &                                                        &                                                         &                                                        &                                                         &                                                        &                                                         & \begin{tabular}[c]{@{}c@{}}0.36\\ (0.002)\end{tabular} \\
\hline
$R^2$             & 0.83                                                   & 0.84                                                    & 0.83                                                   & 0.84                                                    & 0.84                                                   & 0.84                                                    & 0.88                                                   & 0.82                                                    & 0.90                                                   \\
VIF               & 3.00                                                   & 2.01                                                    & 2.49                                                   & 1.04                                                    & 1.41                                                   & 2.15                                                    & 1.70                                                   & 2.62                                                    & 1.71                    \\
\hline \hline
\end{tabular}
\caption{{\it Multivariate regression in PCS class:}  For all models the dependent variable is the fraction of contracts by single-bidder in the PCS class, and one of the independent variables is always the fraction of contracts in which the relation buyer-supplier has an {\bf RAD $\geq$ 0.5}. The second independent variable is a control variable related to procedure character ({\bf PC}), contract type ({\bf CT}), supplier's size ({\bf S}), or the risk factor favoritism ({\bf Fav}). Observations are weighted by the number of PCS contracts per year. The table shows the standardized coefficients, their standard error (in parentheses), the $R^2$, and the mean variance inflation factor (VIF). Only models with VIF $<$ 5 are shown. All the reported parameters are highly significant with $p_v\ll 0.01$. {\it Note:} A \lq\lq complete'' model with the most significant independent variables in each model was performed, but the multicollinearity between the independent variables was very high. We do not report it.}
\end{table*}